\documentclass[twocolumn,secnumarabic,amsmath,amssymb,reprint,nobibnotes, aps, prb]{revtex4-1}

\pdfoutput=1

\usepackage{graphicx}
\usepackage{dcolumn}
\usepackage{bm}
\usepackage{color}
\usepackage{float}
\usepackage{epsfig}
\usepackage[a4paper,left=1.5cm,right=1.5cm, top=1.5cm,bottom=2cm]{geometry}

\newcommand{\Sec}[1]{Sec.\,\ref{#1}}

\newcommand{\be}{\begin{equation}}
\newcommand{\ee}{\end{equation}}
\newcommand{\bea}{\begin{eqnarray}}
\newcommand{\eea}{\end{eqnarray}}
\newcommand{\Fig}[1]{Fig.\,\ref{#1}}
\newcommand{\Eq}[1]{Eq.\,(\ref{#1})}
\newcommand{\etal}{{\it et al. }}
\newcommand{\la}{\langle}
\newcommand{\ra}{\rangle}

\newcommand{\Tr }{\textrm{Tr}}

\begin{document}

\preprint{AIP/123-QED}

\title{Diagrammatic quantum Monte Carlo toward the calculation of transport properties in disordered semiconductors}

\author{Yu-Chen Wang}
\affiliation{
State Key Laboratory of Physical Chemistry of Solid Surfaces, iCHEM, Fujian Provincial Key Laboratory of Theoretical and Computational Chemistry, and  College of Chemistry and Chemical Engineering, Xiamen University, Xiamen 361005, People's Republic of China
}

\author{Yi Zhao}
\email[E-mail: ]{yizhao@xmu.edu.cn}
\affiliation{
State Key Laboratory of Physical Chemistry of Solid Surfaces, iCHEM, Fujian Provincial Key Laboratory of Theoretical and Computational Chemistry, and  College of Chemistry and Chemical Engineering, Xiamen University, Xiamen 361005, People's Republic of China
}

\begin{abstract}
A new diagrammatic quantum Monte Carlo approach is proposed to deal with the imaginary time propagator involving both dynamic disorder (\textit{i.e.}, electron-phonon interactions) and static disorder of local or nonlocal nature in a unified and numerically exact way.
The establishment of the whole framework relies on a general reciprocal-space expression and a generalized Wick's theorem for the static disorder.
Since the numerical cost is independent of the system size, various physical quantities such as the thermally averaged coherence, Matsubara one-particle Green's function and current autocorrelation function can be efficiently evaluated in the thermodynamic limit (infinite in the system size).
The validity and performance of the proposed approach are systematically examined in a broad parameter regimes.
This approach, combined with proper numerical analytic continuation methods and first-principles calculations, is expected to be a versatile tool toward the calculation of various transport properties like mobilities in realistic semiconductors involving multiple electronic energy bands, high-frequency optical and low-frequency acoustic phonons, different forms of dynamic and static disorders, anisotropy, \textit{etc}.

\end{abstract}


\maketitle

\section{INTRODUCTION}
Structural disorders at the microscopic level generally exist in materials such as semiconductors and light-harvesting systems.
These disorders may substantially alter the electronic structure of the system and ulteriorly influence the transport properties of the carriers (electron, hole, or exciton) therein.

According to the time scale of the influence, the disorders can be categorized into two types.
The first type called the static disorder comes from obvious local structural imperfections caused by impurities, vacancies, irregular stacking arrangements, structural mismatching, \textit{etc}., and is reflected in the electronic state energies and electronic couplings as certain static randomness over a long timescale.
The second type manifests time-evolving changes in the electronic structure at a relatively short timescale (from femtoseconds to picoseconds), and thereby is referred to as the dynamic disorder.
The dynamic one, which originates from the continuous motion of the nuclei, is also recognized as the electron-phonon interactions (or exciton-phonon interactions for the excitation energy transfer process).

The transport properties in many disordered semiconductors and light-harvesting systems often exhibit unexpected and sometimes exhilarating behaviors.
For instance, recent two-dimensional electronic spectroscopic experiments on the pigment-protein complexes of photosynthetic organisms\cite{Brixner-N-2005-625,Engel-N-2007-782,Lee-S-2007-1462,
Collini-N-2010-644,Panitchayangkoon-PNASU-2010-12766,Romero-NP-2014-676,
Fuller-NC-2014-706} and conjugated polymers\cite{Collini-S-2009-369} have revealed long-lasting oscillating cross peaks at both cryogenic and physiological temperatures, implying that quantum coherence may play a significant role in the excitation energy transfer process.
These observations challenge the traditional view that the exciton dynamics in such disordered systems should follow incoherent hopping motion.
As an another example, it has been known for a long time that the carrier mobilities in some organic semiconductors decline as the temperature arises, exhibiting a typical band-like behavior\cite{Karl-SM-2003-649,Jurchescu-APL-2004-3061,
Podzorov-PRL-2005-226601,Ostroverkhova-APL-2006-162101}.
However, the magnitude of these mobilities is close to the Mott-Ioffe-Regel limit, which indicates a strong localization nature of the charge carrier and contradicts the delocalized band picture\cite{Cheng-JCP-2003-3764}.
Similar band-type tendency of the mobility has also been found in the titanium dioxide of different crystallographic forms\cite{Breckenridge-PR-1953-793,Austin-AP-1969-41,
Tang-JAP-1994-2042,Forro-JAP-1994-633,Yagi-PRB-1996-7945,
Bak-JPCS-2003-1069}, but several experimental\cite{Setvin-PRL-2014-86402,Yang-PRB-2013-125201} and theoretical\cite{Deak-PRB-2011-155207,Deak-PRB-2012-195206,
DiValentin-JPCC-2009-20543} evidences have shown that the charge carrier may form localized small polarons.

Over the decades, great efforts have been devoted to uncover these fascinating mysteries, which has also boosted the development of many insightful theoretical models and powerful numerical approaches.
In the field of quantum biology, numerically exact approaches such as the hierarchical equation of motion\cite{Tanimura-JPSJ-1989-101,Ishizaki-PNASU-2009-17255,Tanimura-JCP-2020-20901} and iterative quasi-adiabatic path integral\cite{Makri-JCP-1995-4600,Makri-JCP-1995-4611} have been frequently applied to the exciton dynamics in small-sized systems like the Fenna-Matthews-Olson complex.
For systems as large as the chlorosome from green sulfur bacteria (containing hundreds of thousands of bacteriochlorophylls), impressive studies based on approximate methods like the stochastic Schr\"{o}dinger equation have been carried out\cite{Fujita-JPCL-2012-2357,Fujita-PR-2014-273,Huh-JACS-2014-2048,
Sawaya-NL-2015-1722,Li-JPCB-2020-4026}.
Nevertheless, the dynamical methods for large-sized systems still have remained unsatisfactory.
For example, owing to the use of classical random forces, the stochastic Schr\"{o}dinger equation does not satisfy the detailed balance and may not be able to give a correct description for the quantum coherence.

To explain the peculiar behaviour of the carrier mobility in organic semiconductors, various models and computational schemes have been proposed, which include but not limited to the hopping model with the nuclear tunneling effect\cite{Nan-PRB-2009-115203}, polaron theory\cite{Hannewald-PRB-2004-75212,Wang-JCP-2007-44506}, mixed quantum-classical method\cite{Troisi-PRL-2006-86601} and transient localization scenario\cite{Ciuchi-PRB-2011-81202,Ciuchi-PRB-2012-245201,Fratini-AFM-2016-2292,Fratini-NM-2017-998} based on the Su-Schrieffer-Hegger model, and real-time quantum dynamic simulations\cite{Zhong-NJP-2014-45009,Jiang-NH-2016-53,
Lian-JCP-2019-44115,Wang-JCP-2010-81101,Lian-JCP-2019-44115,
Li-JPCL-2020-4930,Li-NC-2021-4260}.
These pioneering works have greatly promoted our understanding toward the localization-delocalization conflict of the charge carrier.
Nevertheless, each of them has its own limitations.
For example, the hopping model completely neglects the coherence;
the mixed quantum-classical method and transient localization scenario only consider the nonlocal electron-phonon interactions caused by low-frequency intermolecular vibrations;
accurate quantum dynamic simulations are only affordable for relatively small-sized systems and are often limited by the finite-size effect.
In realistic organic semiconductors, the transport mechanism is involved in multiple factors, such as the high-frequency intramolecular and low-frequency intermolecular vibrations, local and nonlocal electron-phonon interactions, static disorder, and complicated transport paths.
The transport mechanism may also be relevant to several electronic energy bands if the unit cell contains multiple components.
Up to now, a general consensus over the transport mechanism in organic semiconductors is yet to be achieved.
As for the transport properties of inorganic semiconductors, the semiclassical Boltzmann equation is generally used in theoretical studies.
However, it is known that many metallic oxides such as the titanium dioxide are characterized by strong Fr\"{o}hlich-type electron-phonon interactions\cite{Hendry-PRB-2004-81101,Persson-APL-2005-231912,
Moser-PRL-2013-196403,Verdi-PRL-2015-176401,Himmetoglu-JPCM-2016-65502,
Franchini-NRM-2021-1}.
This is inconsistent with the assumption of the Boltzmann equation that the electron-phonon interactions can be regarded as perturbation.
Recently, there have been attempts to apply the hopping model to titanium dioxides of different crystallographic forms, but the calculated mobilities are usually several orders of magnitude less than the experimental one \cite{Deskins-PRB-2007-195212,Spreafico-PCCP-2014-26144}.

The aforementioned difficulties arise from the fact that various complex interactions (dynamic and static disorders, electronic couplings, \textit{etc}.) in many disordered materials are at a similar magnitude, precluding a straightforward perturbation treatment for the real-time dynamic simulations.
By contrast, accurate quantum simulations in imaginary time are much easier to carry out due to the absence of the dynamical sign problem.
According to the Matsubara formalism and linear response theory\cite{Mahan--2013-}, transport properties like the optical conductivity and one-particle spectral function can be extracted from the imaginary time propagator via analytic continuation.

In the absence of the dynamical sign problem, the Monte Carlo technique is very suited for the calculations of physical quantities in imaginary time.
Particularly, the advanced diagrammatic quantum Monte Carlo (DQMC) method has gained great success in the bosonic lattice models\cite{Prokofev-JL-1996-911,Beard-PRL-1996-5130}, polaron problems\cite{Prokofev-PRL-1998-2514,Mishchenko-PRB-2000-6317,
Mishchenko-PRL-2003-236401,DeFilippis-PRL-2006-136405,
Marchand-PRL-2010-266605,Goodvin-PRL-2011-76403,
Mishchenko-PRL-2015-146401,
DeFilippis-PRL-2015-86601,Mishchenko-PRB-2018-45141,
Mishchenko-PRL-2019-76601} and quantum impurity models\cite{Gull-RMP-2011-349} during the past two decades.
The basic principle of DQMC is to expand the partition function into the summation of Feynman diagrams of arbitrary order and perform importance samplings over all of the diagrams.
Despite the generality of the pivotal principle, the concrete framework and numerical realization of DQMC differ from model to model.
In the polaron problems, works on the Holstein model\cite{Goodvin-PRL-2011-76403,Mishchenko-PRL-2015-146401}, Su-Schrieffer-Hegger model\cite{Marchand-PRL-2010-266605,DeFilippis-PRL-2015-86601} and Fr\"{o}hlich model\cite{Mishchenko-PRB-2000-6317,Mishchenko-PRL-2003-236401,DeFilippis-PRL-2006-136405,Mishchenko-PRL-2019-76601} have been reported.
It is worth mentioning that by combining DQMC with numerical analytic continuation, the mobility of the Holstein polaron has been obtained for a very broad parameter regime from low to high temperatures and from weak to strong electron-phonon interactions\cite{Mishchenko-PRL-2015-146401}.
Powerful as it is, the DQMC has been only applied to simple theoretical models where the local and nonlocal electron-phonon interactions are not simultaneously presented, and where only single phonon branch is considered.
Furthermore, the static disorder has not yet been incorporated into the existing DQMC framework in the literatures.

Because of the generality of the basic idea, it is expected that under proper formulation and implementation, DQMC can in principle apply to realistic materials with multifarious intricate factors beyond the model systems.
In this work, we will provide a new formulation of DQMC toward the calculation of transport properties in realistic disordered semiconductors and light-harvesting materials that simultaneously involve multiple electronic dispersions, high-frequency optical and low-frequency acoustic vibrations, and different kinds of dynamic and static disorders in a unified and numerically exact way.
As will be seen, the extension for the dynamic disorder is straightforward, but the incorporation of the static disorder is much more involved.

The paper is organized as follows.
In \Sec{sec2}, detailed formulation is presented to arrive at a new DQMC approach.
In \Sec{sec3}, the performance of the proposed approach in different parameter regimes are systematically examined.
In \Sec{sec4}, further discussion of the this approach and concluding remarks are given.
Throughout the paper, we set $\hbar=k_B=1$ for the sake of convenience.

\section{Methodology\label{sec2}}
In this work, we focus on the imaginary time propagator $e^{-\beta\hat{H}}$, from which various transport properties near the thermal equilibrium can be extracted, where $\hat{H}$ is the total Hamiltonian of the system and $\beta=1/T$ is the inverse of the temperature.
The DQMC method to be introduced relies on the diagrammatic expansion of the canonical partition function $Z=\Tr\{e^{-\beta\hat{H}}\}$ and the importance sampling over all of the diagrams through a Monte Carlo procedure.

Before presenting the formulation of the new DQMC framework, it is necessary to detailedly introduce the comprehensive model and the mathematical techniques we will use.
Concretely speaking, in \Sec{sec2.1}, the total Hamiltonian and the current operator in reciprocal space are first presented, and then the origin, the derivation, and several examples of the static disorder Hamiltonian is discussed at some length.
In \Sec{sec2.2}, we first briefly summarize the Wick's theorem for phonons, and then propose and prove the generalized Wick's theorem for static disorder.
In \Sec{sec2.3}, the diagrammatic expansion of the partition function is detailedly presented.

\subsection{Hamiltonian}\label{sec2.1}
For the sake of convenience, here we take the carrier transport in semiconductors as an example, but the whole DQMC methodology can also be applied to the excitation energy transfer process occurring in regularly arranged light-harvesting materials such as the chlorosomes.

\subsubsection{Hamiltonian in reciprocal space}
Consider a three-dimensional semiconductor material with the periodic boundary condition.
In the absence of disorder in the morphology, the semiconductor is in the crystal form and is translational symmetric.
Under this situation, the electronic structure and phonon spectrum are well-described by the band theory.
By including various extrinsic disordered factors such as impurities, vacancies and irregular stacking arrangements, the translational symmetry is broken and the electronic energy band is interrupted.
With this in mind, it is convenient to divide the total Hamiltonian of the disordered semiconductor into four parts as
\begin{equation}
\label{htot}
\hat{H}=\hat{H}_{el}+\hat{H}_{ph}+\hat{H}_{el-ph}+\hat{H}_{sd},
\end{equation}
where $\hat{H}_{el}$, $\hat{H}_{ph}$ and $\hat{H}_{el-ph}$ are the primitive electronic Hamiltonian, phonon Hamiltonian and electron-phonon interactions, respectively, and $\hat{H}_{sd}$ contains the static disorder in electronic state energies and electronic couplings induced by extrinsic disordered factors.
Technically speaking, $\hat{H}_{sd}$ is also one part of the electronic Hamiltonian.
We assume that the crystal lattice parameters along the directions of the three crystal axes are $a$, $b$, and $c$, respectively, and the total number of unit cells within the semiconductor is $N_a\times N_b\times N_c$.
For the sake of convenience, we set $N_a=N_b=N_c=N$ hereafter, but the DQMC formalism being introduced can be straightforwardly extended to situations where the three integers are different.

Owing to the translational symmetry in the absence of the static disorder, $\hat{H}_{el}$ and $\hat{H}_{ph}$ are both diagonal in reciprocal space (also called the Bloch representation).
Concretely speaking, the electronic Hamiltonian is given by
\begin{equation}
\label{hel}
\hat{H}_{el}=\sum_{n\mathbf{k}}\epsilon_{n\mathbf{k}}
\hat{c}_{n\mathbf{k}}^\dagger\hat{c}_{n\mathbf{k}}.
\end{equation}
Here, $n$ corresponds to the nth energy band.
The three-dimensional vector $\mathbf{k}=(\frac{2\pi n_a}{aN},\frac{2\pi n_b}{bN},\frac{2\pi n_c}{cN})$ is the crystal momentum of the electron, where the possible values of $n_a$, $n_b$ and $n_c$ are $-\frac{N-1}{2},-\frac{N-3}{2},\cdots,\frac{N-1}{2}$ when $N$ is odd and $-\frac{N-2}{2},-\frac{N-4}{2},\cdots,\frac{N}{2}$ when $N$ is even.
$\epsilon_{n\mathbf{k}}$ is the energy of the Bloch orbital at the nth energy band with crystal momentum $\mathbf{k}$, and $\hat{c}_{n\mathbf{k}}^\dagger$ and $\hat{c}_{n\mathbf{k}}$ are the corresponding electron creation and annihilation operators, respectively.
Note that $\epsilon_{n\mathbf{k}}$ is symmetric with respect to the $\Gamma$ point in the Brillouin zone, $\epsilon_{n\mathbf{k}}=\epsilon_{n,-\mathbf{k}}$.

The phonon Hamiltonian is written as
\begin{equation}
\label{hph}
\hat{H}_{ph}=\sum_{\nu\mathbf{q}}\omega_{\nu\mathbf{q}}
\left(\hat{b}_{\nu\mathbf{q}}^\dagger\hat{b}_{\nu\mathbf{q}}+\frac{1}{2}\right).
\end{equation}
Here, $\nu$ represents the $\nu$th phonon branch.
$\mathbf{q}$ is the crystal momentum of the phonon, and its possible value is identical to that of $\mathbf{k}$.
$\omega_{\nu\mathbf{q}}$ is the frequency of the phonon at the $\nu$th branch with crystal momentum $\mathbf{q}$, and $\hat{b}_{\nu\mathbf{q}}^\dagger$ and $\hat{b}_{\nu\mathbf{q}}$ are the corresponding phonon creation and annihilation operators, respectively.
The phonon frequency is also symmetric with respect to the $\Gamma$ point, $\omega_{\nu\mathbf{q}}=\omega_{\nu,-\mathbf{q}}$.

The electron-phonon interaction has the following form
\begin{equation}
\label{helph}
\hat{H}_{el-ph}=\sum_{\nu\mathbf{q}}\hat{A}_{\nu\mathbf{q}}^\dagger\otimes\hat{B}_{\nu\mathbf{q}}
\end{equation}
with
\begin{equation}
\label{avq}
\hat{A}_{\nu\mathbf{q}}^\dagger=N^{-\frac{3}{2}}\sum_{nm\mathbf{k}}
g_{nm\mathbf{kq}\nu}\hat{c}_{n,\mathbf{k+q}}^\dagger\hat{c}_{m\mathbf{k}}
\end{equation}
and
\begin{equation}
\label{bvq}
\hat{B}_{\nu\mathbf{q}}=\hat{b}_{\nu\mathbf{q}}
+\hat{b}_{\nu,-\mathbf{q}}^\dagger,
\end{equation}
where the complex number $g_{nm\mathbf{kq}\nu}$ is the electron-phonon interaction parameter obeying $(g_{nm\mathbf{kq}\nu})^*=g_{mn,\mathbf{k+q,-q},\nu}$.
\Eq{helph} indicates that for the electronic crystal momentum to be increased by $\mathbf{q}$, one must annihilate (create) a phonon with the same (opposite) crystal momentum.
Thereby, the electron-phonon interactions conserve the total crystal momentum, which is a natural result of the translational symmetry.
Furthermore, it is easy to prove that $\hat{A}_{\nu,-\mathbf{q}}^\dagger=\hat{A}_{\nu\mathbf{q}}$ and
$\hat{B}_{\nu,-\mathbf{q}}=\hat{B}_{\nu\mathbf{q}}^\dagger$.
Being the given general form, the local and nonlocal electron-phonon interactions are included in \Eq{helph} on an equal footing\cite{Giustino-RMP-2017-105003}.

Eqs.\,(\ref{hel})-(\ref{helph}) are widely used in the literatures for the description of the carrier dynamics in periodic systems.
To be able to describe extrinsic disordered factors within the same theoretical framework, we adopt the static disorder Hamiltonian with the following form
\begin{equation}
\label{hsd-k}
\hat{H}_{sd}=\sum_{\mu\mathbf{q}}D_{\mu\mathbf{q}}
\hat{C}_{\mu\mathbf{q}}^\dagger
\end{equation}
with
\begin{equation}
\label{hsd-cq}
\hat{C}_{\mu\mathbf{q}}^\dagger=N^{-\frac{3}{2}}\sum_{nm\mathbf{k}}
f_{nm\mathbf{kq}\mu}\hat{c}_{n,\mathbf{k+q}}^\dagger
\hat{c}_{m\mathbf{k}}.
\end{equation}
Here, resembling the concept of the phonon spectrum, we have introduced two indexes $\mu$ and $\mathbf{q}$ to distinguish different random variables of the static disorder (termed as the disorder variables in the following), where $\mu$ represents the $\mu$th static disorder branch (that is, the $\mu$th type of static disorder), and $\mathbf{q}$ is a three-dimensional vector, the physical meaning of which is identical to that of the phonon crystal momentum.
$D_{\mu\mathbf{q}}$ is a standard complex Gaussian random variable obeying
\begin{equation}
\label{sd-dq}
\left\{
\begin{aligned}
&(D_{\mu\mathbf{q}})^*=D_{\mu,-\mathbf{q}}, \\
&\mathcal{M}\left\{D_{\mu\mathbf{q}}\right\}=0, \\
&\mathcal{M}\left\{D_{\mu\mathbf{q}}D_{\mu'\mathbf{q}'}\right\}
=\delta_{\mu\mu'}\delta_{\mathbf{q,-q}'}, \\
\end{aligned}
\right.
\end{equation}
where $\mathcal{M}\left\{\cdot\right\}$ represents to take the average.
Such kind of random variables can be generated via
\begin{equation}
\label{dq-xqyq}
\left\{
\begin{aligned}
&D_{\mu\mathbf{q}}=\frac{1}{\sqrt{2}}(x_{\mu\mathbf{q}}+iy_{\mu\mathbf{q}}), \\
&D_{\mu,-\mathbf{q}}=\frac{1}{\sqrt{2}}(x_{\mu\mathbf{q}}-iy_{\mu\mathbf{q}}), \\
\end{aligned}
\right.
\end{equation}
where $x_{\mu\mathbf{q}}$ and $y_{\mu\mathbf{q}}$ are independent real Gaussian variables obeying $\mathcal{M}\left\{x_{\mu\mathbf{q}}x_{\mu'\mathbf{q}'}\right\}
=\mathcal{M}\left\{y_{\mu\mathbf{q}}y_{\mu'\mathbf{q}'}\right\}
=\delta_{\mu\mu'}\delta_{\mathbf{q,q}'}$, and their index $\mathbf{q}$ is restricted to half of the first Brillouin zone.
$f_{nm\mathbf{kq}\mu}$ is a deterministic complex number and has the property $(f_{nm\mathbf{kq}\mu})^*=f_{mn,\mathbf{k+q,-q},\mu}$.

Apparently, \Eq{hsd-k} is in a form very different from that of the static disorder Hamiltonian commonly adopted in the literatures.
In \Sec{sec2.1.2}, we will give a detailed derivation of this expression and prove that any kinds of Gaussian static disorder in real space can be transformed into \Eq{hsd-k} if only the covariance of the static disorder is translational symmetric.
Thereby, Eqs.\,(\ref{hsd-k})-(\ref{sd-dq}) can be regarded as the general expressions of the Gaussian static disorder in reciprocal space.
Note that the random nature of the static disorder is condensed into $D_{\mu\mathbf{q}}$, whereas the magnitude and correlation of the static disorder is fully characterized by $f_{nm\mathbf{kq}\mu}$.
Comparing Eqs.\,(\ref{hsd-k})-(\ref{hsd-cq}) with Eqs.\,(\ref{helph})-(\ref{avq}), it is obvious that the static disorder has a very similar form to the electron-phonon interaction in reciprocal space.
Concretely, $f_{nm\mathbf{kq}\mu}$ and $D_{\mu\mathbf{q}}$ correspond to the electron-phonon interaction parameter $g_{nm\mathbf{kq}\nu}$ and the phonon coordinate $\hat{B}_{\nu\mathbf{q}}$, respectively.
As such, one may consider $D_{\mu\mathbf{q}}$ as the ``coordinate'' of the static disorder, and regard $f_{nm\mathbf{kq}\mu}$ as the static disorder parameters.
It is worth noting that different from $\hat{B}_{\nu\mathbf{q}}$, $D_{\mu\mathbf{q}}$ is a complex random variable instead of an operator.

Many transport properties such as the carrier mobility, group velocity and mean free path are tightly relevant to the current operator $\hat{\mathbf{J}}$, which is a three-dimensional vector,
$\hat{\mathbf{J}}\equiv(\hat{J}_a,\hat{J}_b,\hat{J}_c)$.
The current operator is defined as the time derivative of the polarization operator $\mathbf{P}$, $\hat{\mathbf{J}}=i[\hat{H},\hat{\mathbf{P}}]$.
For the Hamiltonian \Eq{htot} being considered here, it can be decomposed into three components, $
\hat{\mathbf{J}}=\hat{\mathbf{J}}_{el}+\hat{\mathbf{J}}_{el-ph}+\hat{\mathbf{J}}_{sd}$,
where $\hat{\mathbf{J}}_{el}$, $\hat{\mathbf{J}}_{el-ph}$, and $\hat{\mathbf{J}}_{sd}$ originate from the electronic kinetic energy, the electron-phonon interactions, and the static disorder, respectively.
In reciprocal space, $\hat{\mathbf{J}}_{el}$ is given by
\begin{equation}
\label{jel}
\hat{\mathbf{J}}_{el}=e\sum_{nm}\sum_{\mathbf{k}}\mathbf{v}_{nm\mathbf{k}}
\hat{c}_{n\mathbf{k}}^\dagger\hat{c}_{m\mathbf{k}},
\end{equation}
where $e$ is the charge of the electron, and the vector $\mathbf{v}_{nm\mathbf{k}}\equiv(v_{nm\mathbf{k}}^a,v_{nm\mathbf{k}}^b,v_{nm\mathbf{k}}^c)$ is the element of the velocity matrix associated with crystal momentum $\mathbf{k}$.
For the diagonal element, $\mathbf{v}_{nn\mathbf{k}}=\nabla_{\mathbf{k}}\epsilon_{n\mathbf{k}}$.
One can see from \Eq{jel} that $\hat{\mathbf{J}}_{el}$ is partially diagonal in reciprocal space in the sense that there is no cross terms between orbitals with different crystal momentums.

The component from the electron-phonon interactions reads
\begin{equation}
\label{jelph}
\hat{\mathbf{J}}_{el-ph}=
eN^{-\frac{3}{2}}\sum_{nm\mathbf{k}}\sum_{\nu\mathbf{q}}
\mathbf{u}_{nm\mathbf{kq}\nu}\hat{c}_{n,\mathbf{k+q}}^\dagger\hat{c}_{m\mathbf{k}}
\otimes\hat{B}_{\nu\mathbf{q}},
\end{equation}
where $\mathbf{u}_{nm\mathbf{kq}\nu}\equiv(u_{nm\mathbf{kq}\nu}^{a},u_{nm\mathbf{kq}\nu}^{b},u_{nm\mathbf{kq}\nu}^{c})$ is also a vector.
Comparing \Eq{jelph} with Eqs.\,(\ref{helph})-(\ref{bvq}), one can find that $\hat{\mathbf{J}}_{el-ph}$ has a very similar form to $\hat{H}_{el-ph}$.
Actually, just like the connection between $\mathbf{v}_{nn\mathbf{k}}$ and $\epsilon_{n\mathbf{k}}$, a simple relationship $\mathbf{u}_{nn\mathbf{kq}\nu}=\nabla_{\mathbf{k}}g_{nn\mathbf{kq}\nu}$ approximately holds true if there is only one electronic band.
In Appendix A, we briefly discuss the origin of \Eq{jelph}, and suggest a first-principles calculation scheme of $\mathbf{u}_{nn\mathbf{kq}\nu}$ on the basis of the Wannier function.

Likewise, the last component originating from the static disorder is given by
\begin{equation}
\label{jsd}
\hat{\mathbf{J}}_{sd}=
eN^{-\frac{3}{2}}\sum_{nm\mathbf{k}}\sum_{\mu\mathbf{q}}
\tilde{\mathbf{u}}_{nm\mathbf{kq}\mu}
D_{\mu\mathbf{q}}
\hat{c}_{n,\mathbf{k+q}}^\dagger\hat{c}_{m\mathbf{k}},
\end{equation}
where $\tilde{\mathbf{u}}_{nm\mathbf{kq}\mu}\equiv(\tilde{u}_{nm\mathbf{kq}\mu}^{a},\tilde{u}_{nm\mathbf{kq}\mu}^{b},\tilde{u}_{nm\mathbf{kq}\mu}^{c})$ is a three-dimensional vector.
Appendix B provides a brief derivation of \Eq{jsd} and discusses the relation between $\tilde{\mathbf{u}}_{nm\mathbf{kq}\mu}$ and $f_{nm\mathbf{kq}\mu}$.

In \Sec{sec3}, the current operator will be involved in the calculations of the imaginary time and imaginary frequency current autocorrelation functions defined as
\begin{equation}
\label{gt}
G_{\alpha\beta}(\tau)=\frac{1}{Z}\Tr\left\{e^{\tau\hat{H}}\hat{J}_\alpha e^{-\tau\hat{H}}\hat{J}_\beta e^{-\beta\hat{H}}\right\}
\end{equation}
and
\begin{equation}
\label{gw}
\widetilde{G}_{\alpha\beta}(i\omega_n)=\int_0^\beta\mathrm{d}\tau e^{i\omega_n\tau}G_{\alpha\beta}(\tau),
\end{equation}
respectively, where $\omega_n=2\pi nT$ is the Matsubara frequency, and $n$ is a nonnegative integer.
From $G_{\alpha\alpha}(\tau)$ or $\widetilde{G}_{\alpha\alpha}(i\omega_n)$, the optical conductivity and the mobility along the $\alpha$ direction can be extracted via numerical analytic continuation.

\subsubsection{Derivation of the reciprocal-space static disorder}\label{sec2.1.2}
In this subsection, we provide a derivation of Eqs.\,(\ref{hsd-k})-(\ref{hsd-cq}) from its real-space correspondence.
To this end, we first need to adopt a specific real-space representation.
It is known that for the electronic degrees of freedom, the definition of real space is not unique, and one may choose a proper localized basis set according to the system under investigation.
For instance, the frontier molecular orbitals of the monomers are usually used as the basis set for organic materials, whereas for inorganic materials the maximally localized Wannier orbitals are frequently adopted\cite{Marzari-PRB-1997-12847,Marzari-RMP-2012-1419}.

Despite the non-unique definition of such a representation, the real-space localized orbitals can be generally expressed as the linear combinations of the Bloch orbitals as
\begin{equation}
\label{el-r2k}
\left\{
\begin{aligned}
&\hat{c}_{\bar{n}\mathbf{R}}^\dagger=N^{-\frac{3}{2}}\sum_{n\mathbf{k}}e^{-i\mathbf{k\cdot R}}
(\hat{U}_\mathbf{k})_{n\bar{n}}\hat{c}_{n\mathbf{k}}^\dagger, \\
&\hat{c}_{\bar{n}\mathbf{R}}=N^{-\frac{3}{2}}\sum_{n\mathbf{k}}e^{i\mathbf{k\cdot R}}
(\hat{U}_\mathbf{k}^\dagger)_{\bar{n}n}\hat{c}_{n\mathbf{k}}, \\
\end{aligned}
\right.
\end{equation}
where $\mathbf{R}=(m_aa,m_bb,m_cc)$ with $m_a,m_b,m_c=1,\cdots,N$ is a lattice vector, $\hat{c}_{\bar{n}\mathbf{R}}^\dagger$ and $\hat{c}_{\bar{n}\mathbf{R}}$ are the creation and annihilation operators of the $\bar{n}$th localized orbitals in the unit cell centered at $\mathbf{R}$, and $\hat{U}_\mathbf{k}$ is the unitary transformation operator associated with $\mathbf{k}$.
The short line over $\bar{n}$ is used to remind that $\bar{n}$ is a real-space index.

The static disorder in real space can be generally written as
\begin{equation}
\label{hsd-r1}
\hat{H}_{sd}=\sum_{\bar{n}\bar{m}\bar{\mathbf{R}}_e}\sum_{\mathbf{R}}
\gamma_{\bar{n}\bar{m}\bar{\mathbf{R}}_e\mathbf{R}}
\left(\hat{c}_{\bar{n}\mathbf{R}}^\dagger
\hat{c}_{\bar{m},\mathbf{R}+\bar{\mathbf{R}}_e}
+\hat{c}_{\bar{m},\mathbf{R}+\bar{\mathbf{R}}_e}^\dagger
\hat{c}_{\bar{n}\mathbf{R}}\right).
\end{equation}
Here, $\bar{\mathbf{R}}_e$ is a lattice vector. $\gamma_{\bar{n}\bar{m}\bar{\mathbf{R}}_e\mathbf{R}}$ is a real random variable obeying certain statistical properties.
When $\bar{n}=\bar{m}$ and $\bar{\mathbf{R}}_e=\mathbf{0}$, $\gamma_{\bar{n}\bar{m}\bar{\mathbf{R}}_e\mathbf{R}}$ corresponds to the static disorder in the orbital energy (local static disorder), otherwise it corresponds to that in the electronic coupling (nonlocal static disorder).
To simplify the notation, we introduce a compact index $\bar{\mu}\equiv(\bar{n},\bar{m},\bar{\mathbf{R}}_e)$ and the corresponding summation $\sum_{\bar{\mu}}\equiv\sum_{\bar{n}\bar{m}\bar{\mathbf{R}}_e}$.
Note that the short bars over these indexes are used to remind that they are real-space indexes.
As such, \Eq{hsd-r1} is simplified to
\begin{equation}
\label{hsd-r2}
\hat{H}_{sd}=\sum_{\bar{\mu}}\sum_{\mathbf{R}}
\gamma_{\bar{\mu}\mathbf{R}}
\left(\hat{c}_{\bar{n}\mathbf{R}}^\dagger
\hat{c}_{\bar{m},\mathbf{R}+\bar{\mathbf{R}}_e}
+\hat{c}_{\bar{m},\mathbf{R}+\bar{\mathbf{R}}_e}^\dagger
\hat{c}_{\bar{n}\mathbf{R}}\right).
\end{equation}

Up to now, the statistical properties of the static disorder have not been given yet.
We further assume that $\gamma_{\bar{\mu}\mathbf{R}}$ is a Gaussian random variable with a zero average, and it satisfies the following statistical property
\begin{equation}
\label{sd-cov}
\mathcal{M}\left\{\gamma_{\bar{\mu}\mathbf{R}}
\gamma_{\bar{\mu}',\mathbf{R}+\Delta\mathbf{R}}\right\}
=\Gamma_{\bar{\mu}\bar{\mu}'\Delta\mathbf{R}},
\end{equation}
where $\Gamma_{\bar{\mu}\bar{\mu}'\Delta\mathbf{R}}$ is the covariance of the static disorder, which determines the correlation between different disorder variables $\gamma_{\bar{\mu}\mathbf{R}}$.
In traditional theoretical studies, it is often assumed that different disorder variables are independent with each other.
However, in realistic materials, there may be nonnegligible correlations between disorder variables that are close in space location.
It is worth noting that $\Gamma_{\bar{\mu}\bar{\mu}'\Delta\mathbf{R}}$ is independent of $\mathbf{R}$ because we have assumed that the covariance of the static disorder is translational symmetric.
This is in consistent with the fact that the morphological disorders in different locations of a realistic semiconductor usually exhibit similar statistical properties at a macroscopic level.

Since the static disorder technically is part of the electronic Hamiltonian, one can directly transform \Eq{hsd-r2} into reciprocal space via \Eq{el-r2k}.
Before doing so, we first show how to disentangle different disorder variables $\gamma_{\bar{\mu}\mathbf{R}} $ and  acquire the ``eigenstates'' of the static disorder by diagonalizing the covariance matrix.
To this end, we introduce a new set of random variables
\begin{equation}
\gamma_{\bar{\mu}\mathbf{q}}=N^{-\frac{3}{2}}\sum_{\mathbf{R}}
e^{-i\mathbf{q\cdot R}}\gamma_{\bar{\mu}\mathbf{R}}.
\end{equation}
Using the property $N^{-3}\sum_{\mathbf{R}}e^{-i(\mathbf{q}+\mathbf{q}')\cdot\mathbf{R}}
=\delta_{\mathbf{q},-\mathbf{q}'}$ and invoking \Eq{sd-cov}, it is easy to prove that the new random variables satisfy the following statistical property
\begin{equation}
\label{sd-cov-mid}
\mathcal{M}\left\{\gamma_{\bar{\mu}\mathbf{q}}
\gamma_{\bar{\mu}'\mathbf{q}'}\right\}
=\delta_{\mathbf{q},-\mathbf{q}'}
(\hat{\Gamma}_{\mathbf{q}})_{\bar{\mu}\bar{\mu}'},
\end{equation}
where the matrix element $(\hat{\Gamma}_{\mathbf{q}})_{\bar{\mu}\bar{\mu}'}
=\sum_{\Delta\mathbf{R}}e^{i\mathbf{q}\cdot
\Delta\mathbf{R}}\Gamma_{\bar{\mu}\bar{\mu}'\Delta\mathbf{R}}$.
It is seen that $\gamma_{\bar{\mu}\mathbf{q}}$ with different crystal momentum $\mathbf{q}$ are independent with each other.
Furthermore, from \Eq{sd-cov}, we have $\Gamma_{\bar{\mu}\bar{\mu}'\Delta\mathbf{R}}
=\Gamma_{\bar{\mu}'\bar{\mu},-\Delta\mathbf{R}}$.
Using this property, it is easy to prove that $(\hat{\Gamma}_{\mathbf{q}})_{\bar{\mu}\bar{\mu}'}
=(\hat{\Gamma}_{\mathbf{q}})_{\bar{\mu}'\bar{\mu}}^*$.
Therefore, $\hat{\Gamma}_{\mathbf{q}}$ is hermitian, and there exists a unitary matrix $\hat{\Phi}_{\mathbf{q}}$ that can diagonalize $\hat{\Gamma}_{\mathbf{q}}$, $\hat{\Phi}_{\mathbf{q}}\hat{\Gamma}_{\mathbf{q}}
\hat{\Phi}_{\mathbf{q}}^\dagger=\widetilde{\Gamma}_{\mathbf{q}}$, where $(\widetilde{\Gamma}_{\mathbf{q}})_{\mu\mu'}
=\delta_{\mu\mu'}\lambda_{\mu\mathbf{q}}$.
Note that we have removed the short bar over the index $\bar{\mu}$ to remind that the covariance matrix is diagonal in this new representation.
The nonnegative definiteness of the covariance matrix guarantees that $\lambda_{\mu\mathbf{q}}\ge0$.
In addition, it can also be proved that $\hat{\Phi}_{-\mathbf{q}}^T=\hat{\Phi}_{\mathbf{q}}=\hat{\Phi}_{\mathbf{q}}^\dagger$, $\lambda_{\mu\mathbf{q}}=\lambda_{\mu,-\mathbf{q}}$.

Utilizing the above unitary matrix, we can further define a new set of random variables,
$\tilde{\gamma}_{\mu\mathbf{q}}=
\sum_{\bar{\mu}}(\hat{\Phi}_{\mathbf{q}})_{\mu\bar{\mu}}
\gamma_{\bar{\mu}\mathbf{q}}$.
It is straightforward to show that $\tilde{\gamma}_{\mu\mathbf{q}}^*=\tilde{\gamma}_{\mu,-\mathbf{q}}$ and $\mathcal{M}\left\{\tilde{\gamma}_{\mu\mathbf{q}}
\tilde{\gamma}_{\mu'\mathbf{q}'}\right\}
=\delta_{\mu\mu'}\delta_{\mathbf{q,-q}'}\lambda_{\mu\mathbf{q}}$.
Thereby, $\tilde{\gamma}_{\mu\mathbf{q}}$ with different indexes are completely independent with other, and they constitute the spectrum of the static disorder.
To go further, we set $\tilde{\gamma}_{\mu\mathbf{q}}=
\sqrt{\lambda_{\mu\mathbf{q}}}D_{\mu\mathbf{q}}$, where $D_{\mu\mathbf{q}}$ are complex Gaussian random variables obeying \Eq{sd-dq}.
Then, starting from $\tilde{\gamma}_{\mu\mathbf{q}}$, and inversely applying all of the above transformations, we obtain
\begin{equation}
\label{gam-dq}
\begin{split}
\gamma_{\bar{\mu}\mathbf{R}}&=
N^{-\frac{3}{2}}\sum_{\mathbf{q}}e^{i\mathbf{q\cdot R}}
\sum_{\mu}(\hat{\Phi}_{\mathbf{q}}^\dagger)_{\bar{\mu}\mu}
\tilde{\gamma}_{\mu\mathbf{q}} \\
&=N^{-\frac{3}{2}}\sum_{\mathbf{q}}e^{i\mathbf{q\cdot R}}
\sum_{\mu}(\hat{\Phi}_{\mathbf{q}}^\dagger)_{\bar{\mu}\mu}
\sqrt{\lambda_{\mu\mathbf{q}}}D_{\mu\mathbf{q}}.
\end{split}
\end{equation}
\Eq{gam-dq} exhibits the relation between $D_{\mu\mathbf{q}}$ and the real-space disorder variable $\gamma_{\bar{\mu}\mathbf{R}}$.
On the other hand, we can also start from \Eq{gam-dq} and use \Eq{sd-dq} to derive the statistical properties of $\gamma_{\bar{\mu}\mathbf{R}}$
\begin{equation}
\begin{split}
\mathcal{M}\left\{\gamma_{\bar{\mu}\mathbf{R}}
\gamma_{\bar{\mu}',\mathbf{R}+\Delta\mathbf{R}}\right\}
&=N^{-3}\sum_{\mathbf{q}}e^{-i\mathbf{q}\cdot\Delta\mathbf{R}}
(\hat{\Phi}_{\mathbf{q}}^\dagger\widetilde{\Gamma}_{\mathbf{q}}
\hat{\Phi}_{\mathbf{q}})_{\bar{\mu}\bar{\mu}'} \\
&=N^{-3}\sum_{\mathbf{q}}e^{-i\mathbf{q}\cdot\Delta\mathbf{R}}
(\hat{\Gamma}_{\mathbf{q}})_{\bar{\mu}\bar{\mu}'} \\
&=\Gamma_{\bar{\mu}\bar{\mu}'\Delta\mathbf{R}},
\end{split}
\end{equation}
where in the last equality we have used the transformation between  $(\hat{\Gamma}_{\mathbf{q}})_{\bar{\mu}\bar{\mu}'}$ and $\Gamma_{\bar{\mu}\bar{\mu}'\Delta\mathbf{R}}$.

So far, we have successfully disentangle different disorder variables $\gamma_{\bar{\mu}\mathbf{R}}$.
Substituting \Eq{el-r2k} and \Eq{gam-dq} into \Eq{hsd-r2}, we finally arrive at Eqs.\,(\ref{hsd-k})-(\ref{hsd-cq}) with
\begin{equation}
\label{sd-r2k}
\begin{split}
f_{nm\mathbf{kq}\mu}=&\sqrt{\lambda_{\mu\mathbf{q}}}
\sum_{\bar{\mu}}(\hat{\Phi}_{\mathbf{q}}^\dagger)_{\bar{\mu}\mu}
\left[e^{i\mathbf{k}\cdot\bar{\mathbf{R}}_e}
(\hat{U}_\mathbf{k+q})_{n\bar{n}}
(\hat{U}_\mathbf{k}^\dagger)_{\bar{m}m}\right. \\
&\left.+e^{-i(\mathbf{k+q})\cdot\bar{\mathbf{R}}_e}
(\hat{U}_\mathbf{k+q})_{n\bar{m}}
(\hat{U}_\mathbf{k}^\dagger)_{\bar{n}m}\right].
\end{split}
\end{equation}
Note that $\bar{\mu}\equiv(\bar{n},\bar{m},\bar{\mathbf{R}}_e)$ is a simplified notation.
\Eq{sd-r2k} provides a direct way to transform the static disorder from real space to reciprocal space.
In another word, by assigning the concrete function form of $f_{nm\mathbf{kq}\mu}$, \Eq{hsd-k} can represent any Gaussian static disorder with a translational symmetric covariance.

\subsubsection{Examples of the static disorder}
Here we give some examples of the static disorder to show the generality of \Eq{hsd-k}.
For the sake of simplicity, we only consider the situation of single electronic energy band and single type of static disorder.
As such, the real-space and reciprocal-space expression of the static disorder are simplified to
\begin{equation}
\label{hsd-r-exam}
\hat{H}_{sd}=\sum_{\mathbf{R}}
\gamma_{\mathbf{R}}\left(
\hat{c}_{\mathbf{R}}^\dagger\hat{c}_{\mathbf{R}+\bar{\mathbf{R}}_0}
+\hat{c}_{\mathbf{R}+\bar{\mathbf{R}}_0}^\dagger\hat{c}_{\mathbf{R}}\right)
\end{equation}
and
\begin{equation}
\hat{H}_{sd}=N^{-\frac{3}{2}}\sum_{\mathbf{q}}\sum_{\mathbf{k}}
f_{\mathbf{kq}} D_{\mathbf{q}}
\hat{c}_{\mathbf{k+q}}^\dagger\hat{c}_{\mathbf{k}},
\end{equation}
respectively.
Here, $\bar{\mathbf{R}}_0$ is a specific lattice vector.
$\bar{\mathbf{R}}_0=\mathbf{0}$ and $\bar{\mathbf{R}}_0\ne\mathbf{0}$ correspond to local and nonlocal static disorders, respectively.
For the sake of convenience, we omit the indexes $\bar{\mu}$ and $\mu$ in $\gamma_{\bar{\mu}\mathbf{R}}$ and $f_{\mathbf{kq}\mu}$, respectively.
The unitary matrices $\hat{U}_\mathbf{k+q}$, $\hat{U}_\mathbf{k}^\dagger$ and $\hat{\Phi}_\mathbf{q}^\dagger$ all reduce to the identity matrix, and \Eq{sd-r2k} is simplified to
\begin{equation}
\label{sd-fkq-1}
f_{\mathbf{kq}}=\sqrt{\lambda_{\mathbf{q}}}
\left[e^{i\mathbf{k}\cdot\bar{\mathbf{R}}_0}
+e^{-i(\mathbf{k+q})\cdot\bar{\mathbf{R}}_0}\right],
\end{equation}
where
\begin{equation}
\label{sd-gamq-1}
\lambda_{\mathbf{q}}=\sum_{\Delta\mathbf{R}}
e^{i\mathbf{q}\cdot\Delta\mathbf{R}}
\mathcal{M}\left\{\gamma_{\mathbf{R}}
\gamma_{\mathbf{R}+\Delta\mathbf{R}}\right\}
\end{equation}
is the eigenvalue of the covariance matrix of the static disorder.
In the following, we discuss the concrete form of $f_{\mathbf{kq}}$ under different situations.

We first consider the case $\bar{\mathbf{R}}_0=\mathbf{0}$, which corresponds to the static disorder in the orbital energy.
From \Eq{sd-fkq-1}, it is clear that $f_\mathbf{kq}=2\sqrt{\lambda_{\mathbf{q}}}$ is only dependent on $\mathbf{q}$ but not on $\mathbf{k}$.
Assume that the disorder variables $\gamma_{\mathbf{R}}$ are uncorrelated, $\mathcal{M}\left\{\gamma_{\mathbf{R}}
\gamma_{\mathbf{R}+\Delta\mathbf{R}}\right\}
=\sigma\delta_{\mathbf{0},\Delta\mathbf{R}}$, where $\sigma$ characterizes the magnitude of the static disorder.
Substituting $\lambda_\mathbf{q}=\sigma$ and $\bar{\mathbf{R}}_0=\mathbf{0}$ into \Eq{sd-fkq-1}, we obtain
\begin{equation}
\label{sd-fkq-eloc}
f_{\mathbf{kq}}=2\sqrt{\sigma}.
\end{equation}
Therefore, uncorrelated local static disorder corresponds to a constant $f_{\mathbf{kq}}$.

Then, consider the local static disorder with the following correlation property
\begin{equation}
\label{sd-cov-gauss}
\mathcal{M}\left\{\gamma_{\mathbf{R}}\gamma_{\mathbf{R}+\Delta\mathbf{R}}\right\}
=\sigma e^{-\frac{|\Delta\mathbf{R}|^2}{2L^2}},
\end{equation}
where $\sigma$ and $L$ characterize the magnitude and correlation length of the static disorder, respectively.
When the relative distance $|\Delta\mathbf{R}|$ exceeds $L$, the correlation between different disorder variables will rapidly decline as $|\Delta\mathbf{R}|$ further increases.
Substituting \Eq{sd-cov-gauss} into \Eq{sd-gamq-1} and \Eq{sd-fkq-1}, we obtain
\begin{equation}
f_{\mathbf{kq}}=2\left(\sigma\sum_{\Delta\mathbf{R}}
e^{i\mathbf{q\cdot\Delta R}}e^{-\frac{|\Delta\mathbf{R}|^2}{2L^2}}\right)^{\frac{1}{2}}.
\end{equation}
As compared with \Eq{sd-fkq-eloc}, it can be seen that the dependence of $f_{\mathbf{kq}}$ on $\mathbf{q}$ reflects the real-space correlation of the static disorder.

Now turn to the situation that $\bar{\mathbf{R}}_0$ is a nonzero vector, which corresponds to the static disorder in electronic couplings.
We first consider the uncorrelated disorder variables,
$\mathcal{M}\left\{\gamma_{\mathbf{R}}\gamma_{\mathbf{R}'}\right\}
=\delta_{\mathbf{RR}'}\sigma$.
Substituting the eigenvalue $\lambda_\mathbf{q}=\sigma$ into \Eq{sd-fkq-1}, we obtain
\begin{equation}
f_{\mathbf{kq}}=\sqrt{\sigma}\left(e^{i\mathbf{k}\cdot\bar{\mathbf{R}}_0}
+e^{-i(\mathbf{k+q})\cdot\bar{\mathbf{R}}_0}\right).
\end{equation}
On can see that $f_{\mathbf{kq}}$ exhibits dependency on both $\mathbf{k}$ and $\mathbf{q}$, which is a character of the nonlocal static disorder.
Likewise, consider the nonlocal static disorder with the following correlation property
\begin{equation}
\mathcal{M}\left\{\gamma_{\mathbf{R}}\gamma_{\mathbf{R}+\Delta\mathbf{R}}\right\}
=\sigma e^{-\frac{|\Delta\mathbf{R}|^2}{2L^2}}.
\end{equation}
After the same procedure, we obtain
\begin{equation}
\begin{split}
f_{\mathbf{kq}}=&\left(\sigma\sum_{\Delta\mathbf{R}}
e^{i\mathbf{q\cdot\Delta R}}e^{-\frac{|\Delta\mathbf{R}|^2}{2L^2}}\right)^{\frac{1}{2}} \\
&\times\left(e^{i\mathbf{k}\cdot\mathbf{R}_0}
+e^{-i(\mathbf{k+q})\cdot\mathbf{R}_0}\right).
\end{split}
\end{equation}

As can be seen, Eqs.\,(\ref{hsd-k})-(\ref{sd-dq}) provide an elegant way for the description of the static disorder in reciprocal space.
The resemblance between the static disorder and the electron-phonon interactions also paves the way for incorporating the former one into the DQMC framework, as will be seen in \Sec{sec2.3}.

\subsection{Wick's theorem}\label{sec2.2}
The establishment of the DQMC framework relies on two mathematical techniques.
The first one is the Wick's theorem for phonons being briefly summarized in \Sec{sec2.2.1}.
The second one is the generalized Wick's theorem for static disorder being proposed and proven (for the first time to our knowledge) in \Sec{sec2.2.2}.

\subsubsection{Wick's theorem for phonons}\label{sec2.2.1}
In the field of condensed matter physics, one frequently encounters the thermal average over a multiple product of phonon coordinates expressed as $\la\mathcal{T}_+\hat{B}_{\nu_1\mathbf{q}_1}(\tau_1)
\cdots\hat{B}_{\nu_{2\kappa}\mathbf{q}_{2\kappa}}(\tau_{2\kappa})\ra_{ph}$, where $\kappa$ is a positive integer, $\hat{B}_{\nu\mathbf{q}}(\tau)=e^{\tau\hat{H}_{ph}}\hat{B}_{\nu\mathbf{q}}e^{-\tau\hat{H}_{ph}}$ is the phonon coordinate at imaginary time $\tau$, $\mathcal{T}_+$ denotes the chronological time-ordering operation, $\la\hat{O}\ra_{ph}\equiv\Tr\{\hat{O}e^{-\hat{H}_{ph}}/Z_{ph}\}$ for an arbitrary phonon opeartor $\hat{O}$, and $Z_{ph}=\Tr\{e^{-\hat{H}_{ph}}\}$ is the phonon canonical partition function.
With the help of Wick's theorem\cite{Mahan--2013-}, the phonon thermal average can be rewritten as
\begin{equation}
\label{wick}
\begin{split}
&\la\mathcal{T}_+\hat{B}_{\nu_1\mathbf{q}_1}(\tau_1)
\cdots\hat{B}_{\nu_{2\kappa}\mathbf{q}_{2\kappa}}(\tau_{2\kappa})\ra_{ph} \\
=&\sum_{\substack{\mathrm{all~possible}\\\mathrm{pairings}}}
\la\mathcal{T}_+\hat{B}_{\nu_{i_1}\mathbf{q}_{i_1}}(\tau_{i_1})
\hat{B}_{\nu_{i_2}\mathbf{q}_{i_2}}(\tau_{i_2})\ra_{ph}\times\cdots \\
&\quad\times\la\mathcal{T}_+\hat{B}_{\nu_{i_{2\kappa-1}}\mathbf{q}_{i_{2\kappa-1}}}(\tau_{i_{2\kappa-1}})
\hat{B}_{\nu_{i_{2\kappa}}\mathbf{q}_{i_{2\kappa}}}(\tau_{i_{2\kappa}})\ra_{ph}
\end{split}
\end{equation}
where $(i_1,i_2;\cdots;i_{2\kappa-1},i_{2\kappa})$ represents a possible pairwise combination of $(1,2,\cdots,2\kappa)$, and there are totally $\frac{(2\kappa)!}{2^\kappa\kappa!}$ terms in the summation in \Eq{wick}.
Thereby, the thermal average over the multiple product of phonon coordinates is reduced to the summation over the multiple product of various binary thermal averages.
From Wick's theorem, we can also know that \Eq{wick} should be zero if the number of $\hat{B}_{\nu\mathbf{q}}$ is odd.
In the eigenstate representation of the phonon, it is easy to show that
\begin{equation}
\label{bt1bt2}
\la\mathcal{T}_+\hat{B}_{\nu\mathbf{q}}(\tau)
\hat{B}_{{\nu'}\mathbf{q}'}(\tau')\ra_{ph}
=\delta_{\nu\nu'}\delta_{\mathbf{q,-q}'}\alpha_{\nu\mathbf{q}}(|\tau-\tau'|),
\end{equation}
where
\begin{equation}
\label{alpha-vq}
\alpha_{\nu\mathbf{q}}(\tau)=n_{\nu\mathbf{q}}e^{\tau\omega_{\nu\mathbf{q}}}
+(n_{\nu\mathbf{q}}+1)e^{-\tau\omega_{\nu\mathbf{q}}}
\end{equation}
is the free phonon propagator at finite temperatures and $n_{\nu\mathbf{q}}=1/(e^{\beta\omega_{\nu\mathbf{q}}}-1)$ is the thermal average occupation number.

\subsubsection{Generalized Wick's theorem for static disorder}\label{sec2.2.2}
When introducing the static disorder to the DQMC framework, one will have to deal with the following complex product of disorder variables
\begin{equation}
\label{wick-sd-1}
\begin{split}
\mathcal{Q}_{2\kappa,\mathbf{k}_0}=&
N^{-3\kappa}\sum_{\mu_1\mathbf{q}_1}\cdots
\sum_{\mu_{2\kappa}\mathbf{q}_{2\kappa}}
\delta_{\mathbf{q}_1+\cdots+\mathbf{q}_{2\kappa},\mathbf{k}_0}
D_{\mu_1\mathbf{q}_1}\cdots \\
&\times D_{\mu_{2\kappa}\mathbf{q}_{2\kappa}}
Y_{2\kappa}(\mu_1,\mathbf{q}_1;
\cdots;\mu_{2\kappa},\mathbf{q}_{2\kappa}).
\end{split}
\end{equation}
Here, $\kappa$ is a positive integer.
$\mathbf{k}_0$ is a crystal momentum in the first Brillouin zone.
$D_{\mu\mathbf{q}}$ is a complex Gaussian random variable obeying \Eq{sd-dq}.
$Y_{2\kappa}$ is a bounded continuous complex function of $\mathbf{q}_1,\cdots,\mathbf{q}_{2\kappa}$, and is not always zero in the definition domain.
\Eq{wick-sd-1} resembles very much the phonon thermal average \Eq{wick}.
In fact, based on the central limit theorem, a generalized Wick's theorem can be proven to greatly simplify \Eq{wick-sd-1}.
The pivotal idea relies on the fact that most of the terms in the summation will cancel out with each other due to the stochastic nature of $D_{\mu\mathbf{q}}$, and only those terms where all $D_{\mu\mathbf{q}}$ are in pairwise combination with each other contribute to \Eq{wick-sd-1}.

We first briefly introduce the central limit theorem.
Consider $L$ independent random variables $x_1,\cdots,x_L$, where the average value and standard deviation of the $n$th random variable $x_n$ are $\bar{x}_n$ and $\sigma_n$, respectively.
We set $\sigma_L=\sqrt{\sum_{n=1}^L\sigma_n^2}$.
According to the Lyapunov's central limit theorem, if there exists a positive number $\eta$ such that
\begin{equation}
\label{wick-sd-lyapunov}
\frac{\sum_{n=1}^{L}\mathcal{M}\{|x_n-\bar{x}_n|^{2+\eta}\}}{(\sigma_L)^{2+\eta}}\rightarrow0
\end{equation}
when $L\rightarrow\infty$, then the summation of the random variables $\sum_{n=1}^{L}x_n$ is subject to the Gaussian distribution with an average of $\sum_{n=1}^{L}\bar{x}_n$ and a standard deviation of $\sigma_L$.
In other words, we have
\begin{equation}
\label{wick-sd-central}
\sum_{n=1}^{L}x_n\sim\sum_{n=1}^{L}
\bar{x}_n\pm\bar{\sigma}L^{\frac{1}{2}},
\end{equation}
where $A\sim B$ represents that $A$ and $B$ are at the same magnitude, and $\bar{\sigma}=\sqrt{\sum_{n=1}^{L}\sigma_n^2/L}$.

Now we show how to utilize \Eq{wick-sd-central} and mathematical induction to prove the following relation
\begin{equation}
\label{wick-t1}
\begin{split}
&\sum_{\mu_1\mathbf{q}_1}\cdots\sum_{\mu_\kappa\mathbf{q}_\kappa}
D_{\mu_1\mathbf{q}_1}\cdots D_{\mu_\kappa\mathbf{q}_\kappa} \\
&\qquad\times Y_\kappa(\mu_1,\mathbf{q}_1;\cdots;\mu_\kappa,\mathbf{q}_\kappa)
\sim N^{\frac{3\kappa}{2}}.
\end{split}
\end{equation}
For the sake of simplicity, we first assume that $Y_\kappa$ is a real-variable function and $D_{\mu\mathbf{q}}$ are standard real Gaussian random variables.
The results will be generalized to the complex-number case later.
For $\kappa=1$, the left of \Eq{wick-t1} becomes $\sum_{\mu\mathbf{q}}D_{\mu\mathbf{q}}Y_1(\mu,\mathbf{q})$.
Since $D_{\mu\mathbf{q}}Y_1(\mu,\mathbf{q})$ can be regarded as a Gaussian random variable with a zero average value and a standard deviation of $|Y_1(\mu,\mathbf{q})|$, we can use Lyapunov's central limit theorem to evaluate the magnitude of $\sum_{\mu\mathbf{q}}D_{\mu\mathbf{q}}Y_1(\mu,\mathbf{q})$.
We might set $\eta=2$, then the numerator in \Eq{wick-sd-lyapunov} is
\begin{equation}
\label{wick-sd-lya-1}
\begin{split}
\sum_{n=1}^{L}\mathcal{M}\{|x_n-\bar{x}_n|^{2+\eta}\}
&=\sum_{\mu\mathbf{q}}|Y_1(\mu,\mathbf{q})|^4
\mathcal{M}\{|D_{\mu\mathbf{q}}|^4\} \\
&=3\sum_{\mu\mathbf{q}}|Y_1(\mu,\mathbf{q})|^4 \\
&\le3|Y_\mathrm{max}|^4N_{sd}N^3\sim N^3,
\end{split}
\end{equation}
where $N_{sd}$ is the total number of the index $\mu$ (that is, the total number of static disorder types), and $|Y_\mathrm{max}|$ is the maximal absolute value of $Y$ (remember that $Y$ is bounded in its definition domain).
Furthermore, due to the continuity of $Y_1$, we can always find a continuous region $R_0$ in which $Y_1$ is always nonzero.
Assume that the minimal absolute value of $Y$ in $R_0$ is $|Y_\mathrm{min,R_0}|$.
For the denominator in \Eq{wick-sd-lyapunov}, we have
\begin{equation}
\label{wick-sd-lya-2}
\begin{split}
(\sigma_L)^{2+\eta}&=
\left[\sum_{\mu\mathbf{q}}|Y_1(\mu,\mathbf{q})|^2\right]^{2} \\
&\ge\left[\sum_{(\mu,\mathbf{q})\in R_0}|Y_1(\mu,\mathbf{q})|^2\right]^{2} \\
&\ge\left[|Y_\mathrm{min,R_0}|^2r_0N_{sd}N^3\right]^{2}
\sim N^{6},
\end{split}
\end{equation}
where $r_0$ is the ratio between the area of $R_0$ and the total area of the definition domain of $Y_1(\mu,\mathbf{q})$.
Combining \Eq{wick-sd-lya-1} and \Eq{wick-sd-lya-2}, the condition \Eq{wick-sd-lyapunov} is satisfied, and one can use Lyapunov's central limit theorem to evaluate the magnitude of $\sum_{\mu\mathbf{q}}D_{\mu\mathbf{q}}Y_1(\mu,\mathbf{q})$.
Keeping in mind that the total number of $\mathbf{q}$ point is $N^3$ and the fact that the average value of $D_{\mu\mathbf{q}}Y_1(\mu,\mathbf{q})$ is zero, from \Eq{wick-sd-central} we have
\begin{equation}
\sum_{\mu\mathbf{q}}D_{\mu\mathbf{q}} Y_1(\mu,\mathbf{q})
\sim\pm \left[\frac{\sum_{\mu\mathbf{q}}|Y_1(\mu,\mathbf{q})|^2}
{N_{sd}N^3}\right]^{\frac{1}{2}}N_{sd}^{\frac{1}{2}}
N^{\frac{3}{2}}
\sim N^{\frac{3}{2}},
\end{equation}
which proves that \Eq{wick-t1} is valid for $\kappa=1$.
In the next step, we assume that \Eq{wick-t1} is valid for a positive integer $\kappa$, and inspect the case of $\kappa+1$.
Setting
\begin{equation}
\begin{split}
\mathcal{K}_{\kappa,\mu\mathbf{q}}=&
\sum_{\mu_1\mathbf{q}_1}\cdots\sum_{\mu_{\kappa}\mathbf{q}_{\kappa}}
D_{\mu_1\mathbf{q}_1}\cdots D_{\mu_{\kappa}\mathbf{q}_{\kappa}} \\
&\times Y_{\kappa+1}(\mu_1,\mathbf{q}_1;\cdots;
\mu_{\kappa},\mathbf{q}_{\kappa};\mu,\mathbf{q}),
\end{split}
\end{equation}
then
\begin{equation}
\label{wick-t2}
\begin{split}
&\sum_{\mu_1\mathbf{q}_1}\cdots\sum_{\mu_{\kappa+1}\mathbf{q}_{\kappa+1}}
D_{\mu_1\mathbf{q}_1}\cdots D_{\mu_{\kappa+1}\mathbf{q}_{\kappa+1}} \\
&\times Y_{\kappa+1}(\mu_1,\mathbf{q}_1;\cdots;
\mu_{\kappa+1},\mathbf{q}_{\kappa+1})
=\sum_{\mu\mathbf{q}}
\mathcal{K}_{\kappa,\mu\mathbf{q}}
D_{\mu\mathbf{q}}.
\end{split}
\end{equation}
It is easy to know that $\mathcal{K}_{\kappa,\mu\mathbf{q}}$ is also a bounded continuous function of $\mathbf{q}$, and it is not always zero in the definition domain.
Since $\mathcal{K}_{\kappa,\mu\mathbf{q}}\sim N^{\frac{3\kappa}{2}}$, we can regard $\mathcal{K}_{\kappa,\mu\mathbf{q}}D_{\mu\mathbf{q}}$ as a Gaussian random variable with a zero average and a standard deviation at the magnitude of $N^{\frac{3\kappa}{2}}$.
Applying \Eq{wick-sd-central} again, we obtain $\sum_{\mu\mathbf{q}}\mathcal{K}_{\kappa,\mu\mathbf{q}}
D_{\mu\mathbf{q}}\sim N^{\frac{3(\kappa+1)}{2}}$.
Hence, \Eq{wick-t1} is valid for any positive integer $\kappa$.
Finally, in the complex-number case, we can divide \Eq{wick-t1} into the real and imaginary parts  and complete the prove separately through the same procedure.

Now we utilize \Eq{wick-t1} to prove the generalized Wick's theorem for static disorder.
First consider the case $\mathbf{k}_0\neq\mathbf{0}$.
Inspecting \Eq{wick-sd-1}, we can find that due to the existence of $\delta_{\mathbf{q}_1+\cdots+\mathbf{q}_{2\kappa},\mathbf{k}_0}$,
one of the $2\kappa$ $\mathbf{q}$ indexes in \Eq{wick-sd-1} is eliminated, and \Eq{wick-sd-1} only involves the summation over $2\kappa-1$ $\mathbf{q}$ indexes.
From \Eq{wick-t1}, we readily obtain
\begin{equation}
\mathcal{Q}_{2\kappa,\mathbf{k}_0}\sim N^{-\frac{3}{2}}\quad\mathrm{for}~\mathbf{k}_0\neq\mathbf{0}.
\end{equation}
The same argument also applies to the case where \Eq{wick-sd-1} only involves an odd number of $(\mu,\mathbf{q})$ index sets.

Then we consider the case $\mathbf{k}_0=\mathbf{0}$.
Divide $\mathcal{Q}_{2\kappa,\mathbf{0}}$ into two parts, $\mathcal{Q}_{2\kappa,\mathbf{0}}=\mathcal{Q}_{2\kappa}^\mathrm{unpair}
+\mathcal{Q}_{2\kappa}^\mathrm{pair}$, where the terms involving at least one unpaired $D_{\mu\mathbf{q}}$ are included in $\mathcal{Q}_{2\kappa}^\mathrm{unpair}$, whereas those with all the $D_{\mu\mathbf{q}}$ in pairwise combinations are included in $\mathcal{Q}_{2\kappa}^\mathrm{pair}$ (we say that $D_{\mu_i\mathbf{q}_i}$ and $D_{\mu_j\mathbf{q}_j}$ are paired if and only if $\mu_i=\mu_j$ and $\mathbf{q}_i=-\mathbf{q}_j$).
Following the same argument, it is easy to show that $\mathcal{Q}_{2\kappa}^\mathrm{unpair}\sim N^{-\frac{3}{2}}$.
However, the situation of $\mathcal{Q}_{2\kappa}^\mathrm{pair}$ is completely different.
Invoking the fact $D_{\mu\mathbf{q}}D_{\mu,-\mathbf{q}}=|D_{\mu\mathbf{q}}|^2$, $\mathcal{Q}_{2\kappa}^\mathrm{pair}$ can be explicitly written as
\begin{equation}
\label{wick-sd-qpair}
\begin{split}
\mathcal{Q}_{2\kappa}^\mathrm{pair}
=&N^{-3\kappa}\sum_{\mu_1\mathbf{q}_1}\cdots
\sum_{\mu_\kappa\mathbf{q}_\kappa}
|D_{\mu_1\mathbf{q}_1}|^2\cdots|D_{\mu_\kappa\mathbf{q}_\kappa}|^2 \\
&\qquad\times\sum_{\substack{\mathrm{all~possible}\\ \mathrm{pairings}}}Y_{2\kappa}(\mathrm{paired}),
\end{split}
\end{equation}
where $Y_{2\kappa}(\mathrm{paired})$ represents that all the variables $(\mu,\mathbf{q})$ of $Y_{2\kappa}$ are in a specific type of pairwise combination.
For instance, when $\kappa=2$, $Y_{4}(\mathrm{paired})$ could be any one of $Y_4(\mu_1,\mathbf{q}_1;\mu_1,-\mathbf{q}_1;
\mu_2,\mathbf{q}_2;\mu_2,-\mathbf{q}_2)$
, $Y_4(\mu_1,\mathbf{q}_1;\mu_2,\mathbf{q}_2;
\mu_1,-\mathbf{q}_1;\mu_2,-\mathbf{q}_2)$
and $Y_4(\mu_1,\mathbf{q}_1;\mu_2,\mathbf{q}_2;
\mu_2,-\mathbf{q}_2;\mu_1,-\mathbf{q}_1)$.
For a general $\kappa$, there are $\frac{(2\kappa)!}{2^\kappa\kappa!}$ kinds of different $Y_{2\kappa}(\mathrm{paired})$ in total.
It is seen that in \Eq{wick-sd-qpair} the Kronecker delta symbol disappears since all paired variables automatically satisfies the condition $\mathbf{q}_1+\cdots+\mathbf{q}_{2\kappa}=\mathbf{0}$.
Furthermore, regarding $|D_{\mu\mathbf{q}}|^2$ as indepedent random variables, \Eq{wick-sd-qpair} can be considered as the summation over $|D_{\mu\mathbf{q}}|^2$ instead of $D_{\mu\mathbf{q}}$.
From \Eq{sd-dq} we know that the average value of $|D_{\mu\mathbf{q}}|^2$ is 1 rather than 0.
As such, applying the Lyapunov's central limit theorem and mathematical induction again, we  have $\mathcal{Q}_{2\kappa}^\mathrm{pair}\sim N^0$.
Combining all the above results, we finally obtain
\begin{widetext}
\begin{equation}
\label{wick-sd}
\begin{split}
\mathcal{Q}_{2\kappa,\mathbf{k}_0}
=&\delta_{\mathbf{k}_0,\mathbf{0}}N^{-3\kappa}
\sum_{\mu_1\mathbf{q}_1}\cdots\sum_{\mu_\kappa\mathbf{q}_\kappa}
|D_{\mu_1\mathbf{q}_1}|^2\cdots|
D_{\mu_\kappa\mathbf{q}_\kappa}|^2
\sum_{\substack{\mathrm{all~possible}\\ \mathrm{pairings}}}
Y_{2\kappa}(\mathrm{paired})
+\mathcal{O}(N^{-\frac{3}{2}}).
\end{split}
\end{equation}
\end{widetext}
\Eq{wick-sd} is the generalized Wick's theorem for static disorder.
The last summation in \Eq{wick-sd} involves $\frac{(2\kappa)!}{2^\kappa\kappa!}$ of different $Y_{2\kappa}(\mathrm{paired})$.
Taking $\kappa=2$ as an example, according to \Eq{wick-sd}, we have
\begin{widetext}
\begin{equation}
\label{wick-sd-kap2}
\begin{split}
&N^{-6}\sum_{\mu_1\mathbf{q}_1}\cdots\sum_{\mu_4\mathbf{q}_{4}}
\delta_{\mathbf{q}_1+\cdots+\mathbf{q}_{4},\mathbf{0}}
D_{\mu_1\mathbf{q}_1}\cdots D_{\mu_4\mathbf{q}_{4}}
Y(\mu_1,\mathbf{q}_1;\mu_2,\mathbf{q}_2;\mu_3,\mathbf{q}_3;\mu_4,\mathbf{q}_4) \\
=&N^{-6}\sum_{\mu_1\mathbf{q}_1}\sum_{\mu_2\mathbf{q}_{2}}
\left[Y(\mu_1,\mathbf{q}_1;\mu_1,-\mathbf{q}_1;\mu_2,\mathbf{q}_2;\mu_2,-\mathbf{q}_2)
+Y(\mu_1,\mathbf{q}_1;\mu_2,\mathbf{q}_2;\mu_1,-\mathbf{q}_1;\mu_2,-\mathbf{q}_2)\right. \\
&\qquad\qquad\quad\left.+Y(\mu_1,\mathbf{q}_1;\mu_2,\mathbf{q}_2;\mu_2,-\mathbf{q}_2;\mu_1,-\mathbf{q}_1)
\right]|D_{\mu_1\mathbf{q}_1}|^2|D_{\mu_2\mathbf{q}_{2}}|^2
+\mathcal{O}(N^{-\frac{3}{2}}).
\end{split}
\end{equation}
\end{widetext}

At this moment, it is worth pointing out that in the generalized Wick's theorem, the contribution from unpaired terms is at the magnitude of $N^{-\frac{3}{2}}$, and it approaches zero only in the thermodynamic limit ($N\rightarrow\infty$).
Whereas in the Wick's theorem for phonons, the contribution from unpaired terms is exactly zero no matter the size of the system.
As compared with Eqs.\,(\ref{wick})-(\ref{alpha-vq}), $|D_{\mu\mathbf{q}}|^2$ may be regarded as the free static disorder propagator, although it does not really depend on time.

\subsection{Diagrammatic expansion of the partition function}\label{sec2.3}
Now we turn to the diagrammatic expansion of the partition function, $Z=\Tr\{e^{-\beta\hat{H}}\}$.
In the new DQMC approach, we consider the single-electron situation.
In other words, for the electronic degree of freedom, we restrict ourselves in the subspace expanded by the single-electron Bloch states $|n\mathbf{k}\ra\equiv\hat{c}_{n\mathbf{k}}^\dagger|\mathrm{vac}\ra$, where $|\mathrm{vac}\ra$ is the vacuum state that the energy bands under consideration are all empty.
This corresponds to the low-density limit frequently encountered in realistic semiconductors.

We divide the total Hamiltonian \Eq{htot} into a plain term $\hat{H}_0=\hat{H}_{el}+\hat{H}_{ph}$ and an interaction term $\hat{V}=\hat{H}_{el-ph}+\hat{H}_{sd}$, where the concrete expressions for $\hat{H}_{el}$, $\hat{H}_{ph}$, $\hat{H}_{el-ph}$ and $\hat{H}_{sd}$ have been given in \Eq{hel}, \Eq{hph}, \Eq{helph} and \Eq{hsd-k}, respectively.
To facilitate the derivation, we can further rewrite the interaction term as
\begin{equation}
\label{vterm}
\hat{V}=\sum_{j\mathbf{q}}\hat{\Lambda}_{j\mathbf{q}}^\dagger
\hat{F}_{j\mathbf{q}}.
\end{equation}
Here, we have introduced several new notations
\begin{equation}
\label{fjq}
\hat{F}_{j\mathbf{q}}=\left\{
\begin{aligned}
&\hat{B}_{j\mathbf{q}}\quad\mathrm{for}~1\le j\le N_{ph}, \\
&D_{j'\mathbf{q}}\quad\mathrm{for}~N_{ph}<j\le N_{ph}+N_{sd}, \\
\end{aligned}
\right.
\end{equation}
\begin{equation}
\label{lambjq}
\hat{\Lambda}_{j\mathbf{q}}^\dagger=N^{-\frac{3}{2}}\sum_{nm\mathbf{k}}
h_{nm\mathbf{kq}j}\hat{c}_{n,\mathbf{k+q}}^\dagger\hat{c}_{m\mathbf{k}},
\end{equation}
with
\begin{equation}
\label{hjq}
h_{nm\mathbf{kq}j}=\left\{
\begin{aligned}
&g_{nm\mathbf{kq}j}\quad\mathrm{for}~1\le j\le N_{ph}, \\
&f_{nm\mathbf{kq}j'}\quad\mathrm{for}~N_{ph}<j\le N_{ph}+N_{sd}, \\
\end{aligned}
\right.
\end{equation}
where $j'=j-N_{ph}$, and $N_{ph}$ and $N_{sd}$ are the total number of phonon branches and static disorder branches, respectively.
Note that we have introduced a unified branch index $j$ to replace $\nu$ and $\mu$.

Then, expand $Z$ into an infinite series in terms of $\hat{V}$ as follows
\begin{equation}
\label{z1}
\begin{split}
Z=&\sum_{\kappa=0}^{\infty}\int_0^\beta\mathrm{d}\tau_{2\kappa}
\int_0^{\tau_{2\kappa}}\mathrm{d}\tau_{2\kappa-1}
\cdots\int_0^{\tau_2}\mathrm{d}\tau_1
\\
&\times\Tr\left\{e^{-\beta\hat{H}_0}\mathcal{T}_+
\hat{V}(\tau_{2\kappa})\cdots\hat{V}(\tau_{1})\right\}
+\mathcal{O}(N^{-\frac{3}{2}}),
\end{split}
\end{equation}
where $\hat{V}(\tau)=e^{\tau\hat{H}_0}\hat{V}e^{-\tau\hat{H}_0}$ is the interaction term at imaginary time $\tau$, the sequential order of imaginary-time variables is $0\le\tau_1<\cdots<\tau_{2\kappa}\le\beta$, and the trace over the electronic part is restricted to the single-electron Bloch states.
It is noted that we only include even-order terms in \Eq{z1}.
As will be seen soon afterwards, according to the two Wick's theorems presented in the last subsection, the contributions from odd-order terms to the partition function are at the magnitude of about $N^{-\frac{3}{2}}$ and have been integrated into $\mathcal{O}(N^{-\frac{3}{2}})$ in \Eq{z1}.

Substituting \Eq{vterm} into \Eq{z1} and using the fact that the electronic operator $\hat{\Lambda}_{j\mathbf{q}}^\dagger$ commutes with the disorder coordinate $\hat{F}_{j\mathbf{q}}$, we obtain
\begin{widetext}
\begin{equation}
\label{z2}
\begin{split}
Z=&\sum_{\kappa=0}^{\infty}\sum_{j_1\mathbf{q}_1}
\cdots\sum_{j_{2\kappa}\mathbf{q}_{2\kappa}}
\int_0^\beta\mathrm{d}\tau_{2\kappa}\cdots\int_0^{\tau_2}\mathrm{d}\tau_1 \Tr\left\{e^{-\beta\hat{H}_{el}}\mathcal{T}_+
\hat{\Lambda}_{j_{2\kappa}\mathbf{q}_{2\kappa}}^\dagger(\tau_{2\kappa})
\cdots\hat{\Lambda}_{j_1\mathbf{q}_1}^\dagger(\tau_{1})\right\} \\
&\times\Tr\left\{e^{-\beta\hat{H}_{ph}}\mathcal{T}_+
\hat{F}_{j_{2\kappa}\mathbf{q}_{2\kappa}}(\tau_{2\kappa})
\cdots\hat{F}_{j_1\mathbf{q}_1}(\tau_{1})\right\}
+\mathcal{O}(N^{-\frac{3}{2}}),
\end{split}
\end{equation}
\end{widetext}
where $\hat{\Lambda}_{j\mathbf{q}}^\dagger(\tau)=e^{\tau\hat{H}_{el}}
\hat{\Lambda}_{j\mathbf{q}}^\dagger e^{-\tau\hat{H}_{el}}$ and $\hat{F}_{j\mathbf{q}}(\tau)=e^{\tau\hat{H}_{ph}}
\hat{F}_{j\mathbf{q}}e^{-\tau\hat{H}_{ph}}$.
It should be noted that all of the operators in the two curly braces have already been arranged in the correct chronological order.

Substituting \Eq{lambjq} into \Eq{z2} and explicitly evaluating the trace over the electronic part, one arrives at
\begin{widetext}
\begin{equation}
\label{z3}
\begin{split}
Z&=\sum_{\kappa=0}^{\infty}\sum_{\mathbf{k}_1}
\sum_{m_1\cdots m_{2\kappa}}
\sum_{j_1\mathbf{q}_1}\cdots\sum_{j_{2\kappa}\mathbf{q}_{2\kappa}}
\int_0^\beta\mathrm{d}\tau_{2\kappa}\cdots\int_0^{\tau_2}\mathrm{d}\tau_1
\delta_{\mathbf{q}_1+\cdots+\mathbf{q}_{2\kappa},\mathbf{0}}
N^{-3\kappa}
\left[\prod_{l=1}^{2\kappa}h_{m_{l+1}m_l\mathbf{k}_l\mathbf{q}_lj_l}\right]
\\
&\qquad\times Z_{ph}\la\mathcal{T}_+
\hat{F}_{j_{2\kappa}\mathbf{q}_{2\kappa}}(\tau_{2\kappa})
\cdots\hat{F}_{j_1\mathbf{q}_1}(\tau_{1})\ra_{ph}
\exp\left\{-\sum_{l=1}^{2\kappa}(\tau_{l}-\tau_{l-1})
\epsilon_{m_l\mathbf{k}_{l}}\right\}
+\mathcal{O}(N^{-\frac{3}{2}}).
\end{split}
\end{equation}
\end{widetext}
Here, for $1\le l\le2\kappa$, $\tau_l$ is the time position of the $l$th interaction vertex, and $\tau_0=\tau_{2\kappa}-\beta$.
$\mathbf{k}_1$ is the initial electronic crystal momentum,
$\mathbf{k}_{l}=\mathbf{k}+\mathbf{q}_1+\cdots+\mathbf{q}_{l-1}$ is the crystal momentum within the time interval $(\tau_{l-1},\tau_{l})$.
$m_l$ is the energy band index within $(\tau_{l-1},\tau_{l})$, and $m_{2\kappa+1}=m_1$.

To go further, we need to simplify the thermal average in the second row of \Eq{z3}.
It is worth pointing out again that the disorder variable $D_{\mu\mathbf{q}}$ is a complex random variable instead of an operator.
Therefore, the phonon coordinate $\hat{B}_{\nu\mathbf{q}}$ commutes with $D_{\mu\mathbf{q}}$.
Combining the Wick's theorem for phonons, \Eq{wick}, and the generalized Wick's theorem for static disorder, \Eq{wick-sd}, we can recast \Eq{z3} into
\begin{widetext}
\begin{equation}
\label{z4}
\begin{split}
Z=&\sum_{\kappa=0}^{\infty}\sum_{\mathbf{k}_1}\sum_{m_1\cdots m_{2\kappa}}
\sum_{j_1\mathbf{q}_1}\cdots\sum_{j_{2\kappa}\mathbf{q}_{2\kappa}}
\int_0^\beta\mathrm{d}\tau_{2\kappa}\cdots\int_0^{\tau_2}\mathrm{d}\tau_1
N^{-3\kappa}Z_{ph}
\left[\prod_{l=1}^{2\kappa}h_{m_{l+1}m_l\mathbf{k}_l\mathbf{q}_lj_l}\right]
\\
&\times\exp\left\{-\sum_{l=1}^{2\kappa}(\tau_{l}-\tau_{l-1})\epsilon_{m_l\mathbf{k}_{l}}\right\}
\sum_{\substack{\mathrm{all~possible}\\ \mathrm{pairings}}}
\delta_{j_{i_1}j_{i_2}}\delta_{\mathbf{q}_{i_1},-\mathbf{q}_{i_2}}\cdots
\delta_{j_{i_{2\kappa-1}}j_{i_{2\kappa}}}\delta_{\mathbf{q}_{i_{2\kappa-1}},-\mathbf{q}_{i_{2\kappa}}}
\\
&\times\xi_{j_{i_1}\mathbf{q}_{i_1}}(\tau_{i_1}-\tau_{i_2})\cdots
\xi_{j_{i_{2\kappa-1}}\mathbf{q}_{i_{2\kappa-1}}}(\tau_{i_{2\kappa-1}}-\tau_{i_{2\kappa}})
+\mathcal{O}(N^{-\frac{3}{2}}),
\end{split}
\end{equation}

\end{widetext}
where the time variable pairs in the last row should follow
$\tau_{i_1}>\tau_{i_2}$, $\cdots$, $\tau_{i_{2\kappa-1}}>\tau_{i_{2\kappa}}$, and
\begin{equation}
\label{fjq}
\xi_{j\mathbf{q}}(\tau)=\left\{
\begin{aligned}
&\alpha_{j\mathbf{q}}(\tau)\quad\mathrm{for}~1\le j\le N_{ph}, \\
&|D_{j'\mathbf{q}}|^2\quad\mathrm{for}~N_{ph}<j\le N_{ph}+N_{sd}, \\
\end{aligned}
\right.
\end{equation}
with $j'=j-N_{ph}$ can be regarded as a compact free disorder propagator.
It should be noted that although the summations in \Eq{z4} seem to be relevant to $2\kappa$ sets of indexes $(j,\mathbf{q})$, only $\kappa$ sets of them are independent with each other due to the existence of the Kronecker delta symbols, and half of them should be left out when doing the summations.

\Eq{z4} is the central result of this work.
To associate with the Monte Carlo technique, we reexpress \Eq{z4} as
\begin{equation}
\label{z-for-mc}
Z=\sum_{\kappa=0}^\infty\sum_{\mathbf{x}_\kappa}
\mathcal{Z}(\mathbf{x}_\kappa).
\end{equation}
Here, the configuration variable $\mathbf{x}_\kappa$ is defined as
\begin{equation}
\label{xk}
\begin{split}
\mathbf{x}_\kappa\equiv&(\mathbf{k}_1;m_1,\cdots,m_{2\kappa};
j_{i_1},\mathbf{q}_{i_1},\tau_{i_1},\tau_{i_2};j_{i_3}, \\
&\,\,\mathbf{q}_{i_3},\tau_{i_3},\tau_{i_4};
\cdots;
j_{i_{2\kappa-1}},\mathbf{q}_{i_{2\kappa-1}},\tau_{i_{2\kappa-1}},\tau_{i_{2\kappa}}),
\end{split}
\end{equation}
where $(i_1,i_2;\cdots;i_{2\kappa-1},i_{2\kappa})$ represents a possible pairwise combination of $(1,2,\cdots,2\kappa)$, and the corresponding time variable pairs obey $\tau_{i_1}>\tau_{i_2}$, $\cdots$, $\tau_{i_{2\kappa-1}}>\tau_{i_{2\kappa}}$.
The ordered sequence $(j_1\mathbf{q}_1,j_2\mathbf{q}_2,\cdots,j_{2\kappa}\mathbf{q}_{2\kappa})$ can be regarded as the chronological rearrangement of the sequence $(j_{i_1}\mathbf{q}_{i_1},j_{i_2}\mathbf{q}_{i_2},\cdots,j_{i_{2\kappa}}\mathbf{q}_{i_{2\kappa}})$, where $j_{i_{2n}}=j_{i_{2n-1}}$ and  $\mathbf{q}_{i_{2n}}=-\mathbf{q}_{i_{2n-1}}$ with $1\le n\le\kappa$ (a requirement of the Kronecker delta symbol in \Eq{z4}).
Note that we have denoted the order of the configuration $\mathbf{x}_\kappa$ explicitly in the subscript.
In \Eq{z-for-mc}, the summation over $\mathbf{x}_\kappa$ represents
\begin{equation}
\label{xk-sum}
\begin{split}
\sum_{\mathbf{x}_\kappa}\equiv&
\sum_{\mathbf{k}_1}\sum_{m_1\cdots m_{2\kappa}}
\int_0^\beta\int_0^{\tau_{2\kappa}}\cdots\int_0^{\tau_2}\\
&\times\sum_{\substack{\mathrm{all~possible}\\ \mathrm{pairings}}}
\sum_{j_{i_1}\mathbf{q}_{i_1}}\sum_{j_{i_3}\mathbf{q}_{i_3}}\cdots
\sum_{j_{i_{2\kappa-1}}\mathbf{q}_{i_{2\kappa-1}}},
\end{split}
\end{equation}
and the function $\mathcal{Z}(\mathbf{x}_\kappa)$ is
\begin{equation}
\label{zx-kappa1}
\mathcal{Z}(\mathbf{x}_\kappa)=N^{-3\kappa}Z_{ph}z(\mathbf{x}_\kappa)\mathrm{d}\tau_1\cdots\mathrm{d}\tau_{2\kappa},
\end{equation}
where
\begin{equation}
\label{zx-kappa2}
\begin{split}
z(\mathbf{x}_{\kappa})
=&\left[\prod_{l=1}^{2\kappa}h_{m_{l+1}m_l\mathbf{k}_l\mathbf{q}_lj_l}\right]
\exp\left\{-\sum_{l=1}^{2\kappa}(\tau_{l}-\tau_{l-1})
\epsilon_{m_l\mathbf{k}_{l}}\right\}
\\
&\times\xi_{j_{i_1}\mathbf{q}_{i_1}}(\tau_{i_1}-\tau_{i_2})
\xi_{j_{i_3}\mathbf{q}_{i_3}}(\tau_{i_3}-\tau_{i_4})\cdots \\
&\times
\xi_{j_{i_{2\kappa-1}}\mathbf{q}_{i_{2\kappa-1}}}
(\tau_{i_{2\kappa-1}}-\tau_{i_{2\kappa}}).
\end{split}
\end{equation}

There is a one-to-one correspondence between $z(\mathbf{x}_\kappa)$ and the Feynman diagram.
With the help of the diagrams, one can easily obtain the concrete expression of
$z(\mathbf{x}_\kappa)$ at any configurations.
Based on Eqs.\,(\ref{z-for-mc})-(\ref{zx-kappa2}), the implementation and numerical realization of DQMC is straightforward and similar to the other variants in the literatures\cite{Prokofev-JL-1996-911,Prokofev-PRL-1998-2514,Mishchenko-PRB-2000-6317,
Gull-RMP-2011-349}
Therefore, we will not present the details here.
For the completeness and self-consistency of this paper, we provide several relevant contents in the Supplemental Material, including a brief summary of the Monte Carlo technique,  the correspondence between $z(\mathbf{x}_\kappa)$ and the diagrams, the necessary configuration updating procedures required for the Monte Carlo stochastic sampling, and several efficient estimators for the calculations of different physical quantities such as the thermally averaged coherence, one-particle Green's function, and imaginary time current autocorrelation function.

\section{RESULTS AND DISCUSSION\label{sec3}}

\subsection{Model Hamiltonian}

In this section, we systematically examine the validity and performance of the proposed method.
To this end, we adopt a single-band molecular chain model involving various kinds of disorders.
Results for the two-dimensional case can be found in the Supplementary Material.
In real space, the four components of the total Hamiltonian \Eq{htot} are given by
\begin{equation}
\hat{H}_{el}=V\sum_n(\hat{c}_{n+1}^\dagger\hat{c}_n
+\mathrm{H.c.}),
\end{equation}
\begin{equation}
\begin{split}
\hat{H}_{ph}&=\frac{1}{2}\sum_n\sum_j\left[
\hat{P}_{nj}^2+\omega_j^2\hat{Q}_{nj}^2\right]+
\frac{1}{2}\sum_n\hat{p}_n^2 \\
&+\frac{1}{2}\omega_\mathrm{off}^2\sum_n
\left[(1-2b_0)\hat{x}_n^2+b_0(\hat{x}_n-\hat{x}_{n+1})^2
\right], \\
\end{split}
\end{equation}
\begin{equation}
\label{helph-1d}
\begin{split}
\hat{H}_{el-ph}&=\sum_{n}\sum_{j}g_j\hat{Q}_{nj}
\hat{c}_n^\dagger\hat{c}_n \\
&+\frac{g_\mathrm{off}}{\sqrt{2}}\sum_n(\hat{x}_n-\hat{x}_{n+1})
(\hat{c}_{n+1}^\dagger\hat{c}_n+\mathrm{H.c.}),
\end{split}
\end{equation}
\begin{equation}
\label{hsd-1d}
\hat{H}_{sd}=\sum_n\gamma_n\hat{c}_n^\dagger\hat{c}_n
+\sum_n\tilde{\gamma}_n
(\hat{c}_{n+1}^\dagger\hat{c}_n+\mathrm{H.c.}),
\end{equation}
separately, and the current operator is
\begin{equation}
\label{J-1d}
\begin{split}
\hat{J}=&-iea\sum_n\left[V
+\frac{g_\mathrm{off}}{\sqrt{2}}(\hat{x}_n-\hat{x}_{n+1})
+\tilde{\gamma}_n\right] \\
&\qquad\qquad\times(\hat{c}_{n+1}^\dagger\hat{c}_n
-\hat{c}_n^\dagger\hat{c}_{n+1}).
\end{split}
\end{equation}
Here, the index $n$ represents the $n$th site.
$\hat{c}_n^\dagger$ and $\hat{c}_n$ are the electron creation and annihilation operators for site $n$, respectively, and $V$ is the nearest neighbouring electronic coupling.
H.c. represents to take the Hermitian conjugate.
$\hat{P}_{nj}$ and $\hat{Q}_{nj}$ are the momentum and position operators of the $j$th intramolecular vibrational mode at site $n$ with the frequency $\omega_j$.
$\hat{p}_n$ and $\hat{x}_n$ are the real-space momentum and position operators of the acoustic phonon coordinates (intermolecular vibrations) at site $n$, $\omega_\mathrm{off}$ is the characteristic frequency of the acoustic phonon, and $b_0$ determines the width of the phonon dispersion, $\omega_q=\omega_\mathrm{off}\sqrt{1-2b_0\cos{aq}}$, where $a$ is the lattice constant of the molecular chain.
In the following, we fix $\omega_\mathrm{off}=1$ and $b_0=0.1$, which corresponds to a width of about 0.2.

The first and second terms in \Eq{helph-1d} are the local and nonlocal electron-phonon interactions, respectively.
The local electron-phonon interaction parameter $g_j$ can be conveniently described by the spectral density function defined as
\begin{equation}
J(\omega)=\frac{\pi}{2}\sum_j
\frac{g_j^2}{\omega_j}\delta(\omega-\omega_j).
\end{equation}
Here we adopt the super-Ohmic spectral density function with an exponential cutoff,
\begin{equation}
J(\omega)=\frac{\pi\lambda_\mathrm{loc}}{2}
\left(\frac{\omega}{\omega_c}\right)^3e^{-\omega/\omega_c},
\end{equation}
where $\lambda_\mathrm{loc}=\frac{1}{\pi}\int \frac{J(\omega)}{\omega}\mathrm{d}\omega$ characterizes the strength of the local electron-phonon interactions, and $\omega_c$ is the cutoff frequency.
$g_\mathrm{off}$ is the nonlocal electron-phonon interaction parameter.
Note that we adopt an antisymmetric form $\hat{x}_n-\hat{x}_{n+1}$ in the second row of \Eq{helph-1d}, which is typical in conjugated polymers.
Similar to the local case, we define $\lambda_\mathrm{non}=\frac{g_\mathrm{off}^2}{2\omega_\mathrm{off}^2}$ to characterize the strength of the nonlocal one.
In \Eq{hsd-1d}, $\gamma_n$ and $\tilde{\gamma}_n$ are real Gaussian random variables obeying $\la\gamma_n\ra=\la\tilde{\gamma}_n\ra=0$, $\la\gamma_n^2\ra=\Delta_\mathrm{loc}^2$ and $\la\tilde{\gamma}_n^2\ra=\Delta_\mathrm{non}^2$, where $\Delta_\mathrm{loc}$ and $\Delta_\mathrm{non}$ characterize the strengths of the local and nonlocal static disorders, respectively.
Finally, in \Eq{J-1d}, $e$ is the charge of the electron. We set $e=a=1$ in what follows.

\subsection{Numerical Results}
To eliminate any finite-size effects, all of the DQMC calculations are performed in the thermodynamic limit ($N\rightarrow\infty$).
Two methods, the stochastic Liouville-von Neumann equation\cite{Stockburger-PRL-2002-170407,Moix-PRB-2012-115412} (SLN) and the polaron theory, are used to benchmark the results of DQMC.
SLN is a numerically exact method that incorporates the influence of electron-phonon interactions on the system via stochastic fields, but the numerical cost grows cubically with respect to the system size, and it may be hard to converge at low temperatures or strong electron-phonon interactions.
To afford the numerical cost, we adopt a finite-size molecular chain consisting of dozens of sites in the SLN calculations.
On the other hand, the polaron theory treats nonlocal interactions as perturbation, and only includes the lowest-order contributions of $V$, $\lambda_\mathrm{non}$ and $\Delta_\mathrm{non}$.
As such, the polaron theory is accurate only in the limit $V,\lambda_\mathrm{non},\Delta_\mathrm{non}\ll T$.

\begin{figure}[htbp]
\begin{center}
\includegraphics[width=3.4in]{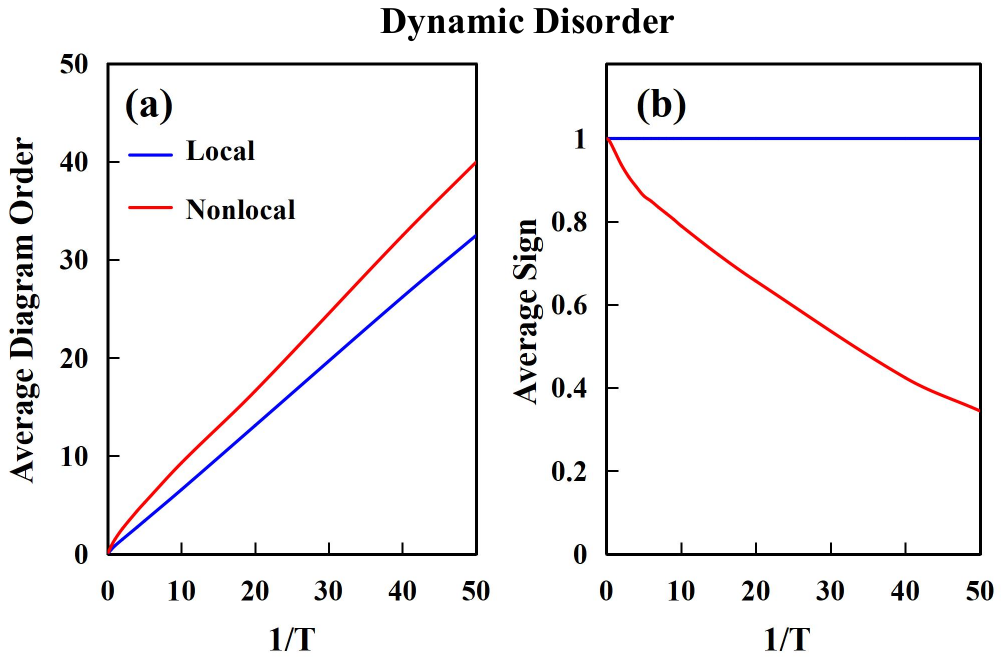}
\caption{Variation of the (a) average diagram order $\la\kappa\ra$ and (b) average sign versus $1/T$ with the presence of local (blue line) or nonlocal (red line) dynamic disorders. The electron-phonon interaction strengths are $\lambda_\mathrm{loc}=1$, $\lambda_\mathrm{non}=0$ for the local case and $\lambda_\mathrm{loc}=0$, $\lambda_\mathrm{non}=1$ for the nonlocal case, other parameters are $V=\omega_c=1$ and $\Delta_\mathrm{loc}=\Delta_\mathrm{non}=0$.}
\label{rfig1}
\end{center}
\end{figure}

We start from the case of dynamic disorder.
It it noted that the performance of DQMC is tightly connected to the average diagram order $\la\kappa\ra$ and the average diagram sign, the latter one is defined by
\begin{equation}
\la\mathrm{Sign}\ra=\sum_{\kappa=0}^\infty\sum_{\mathbf{x}_\kappa}
\frac{|\mathcal{Z}(\mathbf{x}_\kappa)|}
{\mathcal{Z}(\mathbf{x}_\kappa)}.
\end{equation}
The numerical cost for calculating the weight of a concrete diagram is linear in the diagram order, and so as those for calculating various estimators.
As such, it is expected that the overall numerical cost should be at least linear in the average diagram order.
On the other hand, it is well known that in a Monte Carlo simulation the stochastic error can be magnified by the inverse of $\la\mathrm{Sign}\ra$.
Therefore, the error may become uncontrollable if $\la\mathrm{Sign}\ra\ll1$.

\begin{figure}[htbp]
\begin{center}
\includegraphics[width=2.8in]{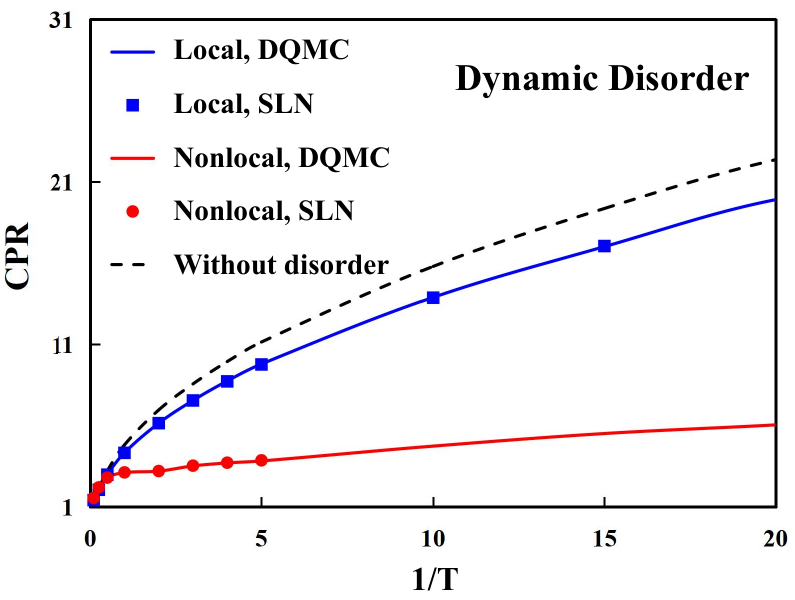}
\caption{Variation of CPR versus $1/T$ with the presence of local (blue line) or nonlocal (red line) dynamic disorders. The electron-phonon interaction strengths are $\lambda_\mathrm{loc}=1$, $\lambda_\mathrm{non}=0$ for the local case and $\lambda_\mathrm{loc}=0$, $\lambda_\mathrm{non}=1$ for the nonlocal case. Other parameters are $V=\omega_c=1$ and $\Delta_\mathrm{loc}=\Delta_\mathrm{non}=0$. Also shown are CPR without the presence of disorder (dashed line) and the results obtained by SLN (blue square and red circle).}
\label{rfig2}
\end{center}
\end{figure}

\Fig{rfig1} (a) and (b) present the average diagram order and average sign, respectively, at different temperatures with the presence of local or nonlocal electron-phonon interactions.
It is seen that for both types of dynamic disorders, the average diagram order increases linearly with $1/T$, and the order of the nonlocal one is slightly larger than that of the local one at the same temperature.
From \Fig{rfig1} (b), it is found that for local dynamic disorder the average sign is always unity in the whole temperature regime, whereas that for the nonlocal one slowly decreases from unity as the temperature decreases.
Inspecting \Eq{zx-kappa2}, it is clear that the phase factor of a diagram is exclusively determined by the electron-phonon interaction parameters $g_{nm\mathbf{kq}\nu}$.
For the local dynamic disorder, $g_{nm\mathbf{kq}\nu}$ is usually independent of the electronic crystal momentum $\mathbf{k}$.
Since $(g_{nm\mathbf{kq}\nu})^*=g_{nm,\mathbf{k}+\mathbf{q},-\mathbf{q},\nu}$, and the interaction vertexes with the opposite $\mathbf{q}$ always exist in pairs, we can expect that there is usually no sign problem for the local dynamic disorder.
On the contrary, $g_{nm\mathbf{kq}\nu}$ can exhibit fruitful dependence on both $\mathbf{k}$ and $\mathbf{q}$ for the nonlocal one, and the sign problem may show up in some parameter regimes.
In spite of this, the average sign is still larger than 0.3 even if the temperature is extremely low.

Disorders of different types alter the electronic properties of the system in varying degrees.
To show this, we calculate the coherence participation ratio (CPR), which measures the magnitude of real-space coherence and is therefore a good indicator for characterizing the localization degree of the electron.
CPR is defined as
\begin{equation}
\mathrm{CPR}=\left(\sum_n|\rho_{nn}|^2\right)
\frac{(\sum_n\sum_l|\rho_{n,n+l}|)^2}
{\sum_n\sum_l|\rho_{n,n+l}|^2},
\end{equation}
where $\rho_{n,n+l}=\Tr\{\hat{c}_{n}^\dagger\hat{c}_{n+l}e^{-\beta\hat{H}}/Z\}$ is the thermally averaged coherence between site $n$ and site $n+l$.
For a completely delocalized state, CPR equals to the total site number $N$, whereas for a completely localized state it equals to one.

\begin{figure}[htbp]
\begin{center}
\includegraphics[width=3.4in]{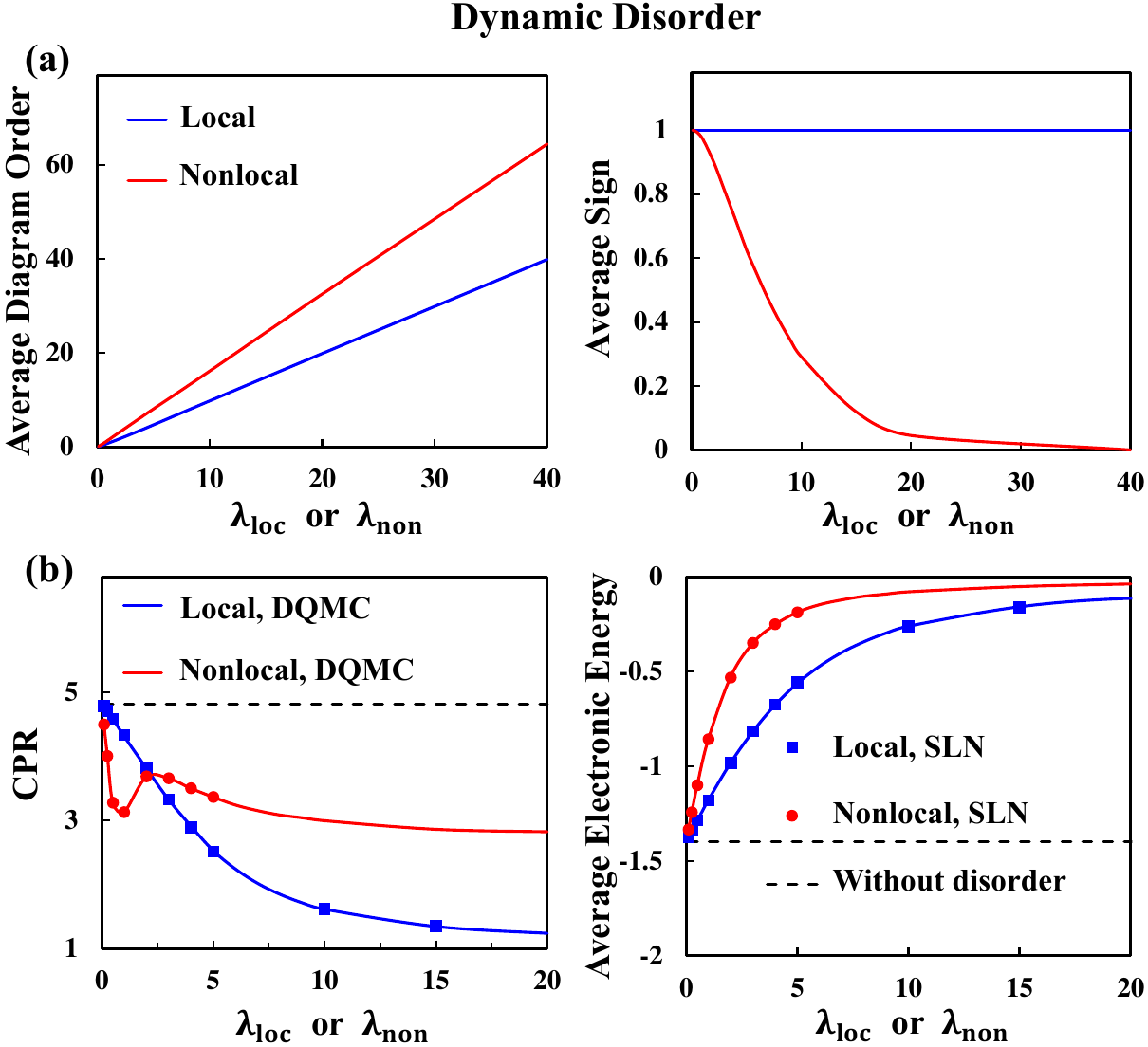}
\caption{Variation of the (a) average diagram order $\la\kappa\ra$ and average sign and (b) CPR and average electronic energy $\la E_{el}\ra$ versus electron-phonon interaction strengths with the presence of local (blue line) or nonlocal (red line) dynamic disorders. Also shown in (b) are CPR and the average electronic energy without the presence of disorder (dashed line) and the results obtained by SLN (blue square and red circle). Other parameters are $T=V=\omega_c=1$ and $\Delta_\mathrm{loc}=\Delta_\mathrm{non}=0$.}
\label{rfig3}
\end{center}
\end{figure}

\begin{figure}[htbp]
\begin{center}
\includegraphics[width=3.0in]{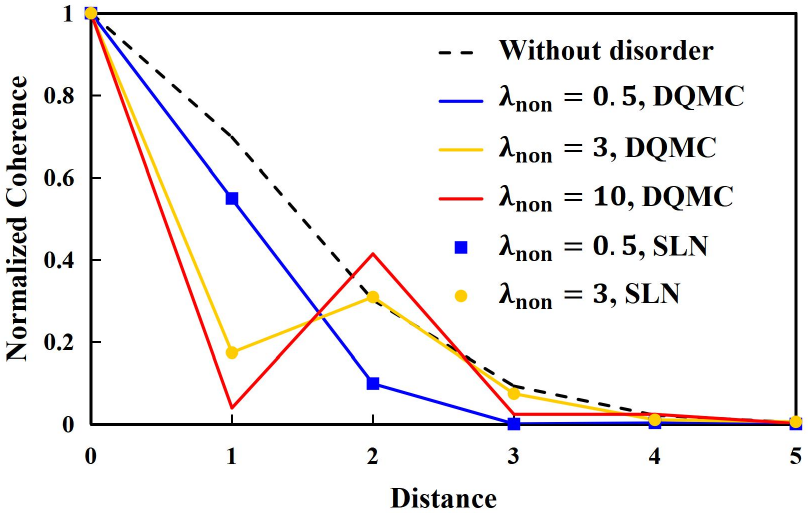}
\caption{Variation of the normalized coherence $|\tilde{\rho}_{n,n+l}|$ versus the distance $l$ with the presence of nonlocal dynamic disorder. Also shown are the coherence without the presence of disorder (dashed line) and the results obtained by SLN (blue square and yellow circle). Other parameters are $T=V=1$ and $\lambda_\mathrm{loc}=\Delta_\mathrm{loc}=\Delta_\mathrm{non}=0$. }
\label{rfig4}
\end{center}
\end{figure}

As can be seen in \Fig{rfig2}, without the presence of any disorder, CPR quickly arises as the temperature decreases.
Under the influence of local dynamic disorder, the magnitude of CPR is slightly decreased, but it still exhibits a large value at low temperatures, indicating that moderate local dynamic disorder ($\lambda_\mathrm{loc}\approx V$) only has limited impact on the system.
However, when the nonlocal dynamic disorder takes part in, CPR is substantially suppressed to a small value for the whole temperature range.
This implies that the nonlocal one has a much more significant decoherence effect than the local one, which is consistent with the physical picture of the transient localization scenario\cite{Ciuchi-PRB-2011-81202,Ciuchi-PRB-2012-245201,Fratini-AFM-2016-2292,Fratini-NM-2017-998}.
In addition, the results obtained by SLN shown in \Fig{rfig2} coincide with those calculated by DQMC since they are both numerically exact.
We note that it becomes very hard to acquire a convergent result for SLN as $\lambda_\mathrm{loc}>15$ or $\lambda_\mathrm{non}>5$.

Next, we analyze the situation of varying the dynamic disorder strength at a fixed temperature.
The results are presented in \Fig{rfig3} (a).
Similarly, the average diagram order is linearly dependent on $\lambda_\mathrm{loc}$ and $\lambda_\mathrm{non}$.
The average sign for the local dynamic disorder is still unity despite the concrete value of $\lambda_\mathrm{loc}$, whereas that for the nonlocal one continuously declines as $\lambda_\mathrm{non}$ increases, and it becomes smaller than 0.05 when $\lambda_\mathrm{non}$ exceeds 20.
As a result, it is not easy to obtain convergent results for DQMC when $\lambda_\mathrm{non}>20$.

To see the impacts of strong dynamic disorder on the system, in \Fig{rfig3} (b) we present the CPR and the average electronic energy defined as $\la E_{el}\ra=\Tr\{\hat{H}_{el}e^{-\beta\hat{H}}/Z\}$ for different values of $\lambda_\mathrm{loc}$ and $\lambda_\mathrm{non}$.
As can be seen, by varying $\lambda_\mathrm{loc}$ from weak to strong regimes, the CPR gradually decreases and finally approaches 1.
Simultaneously, the average electronic energy continuously increases and eventually approaches zero.
This is not surprising but can be well explained by the energy band narrowing effect induced  by the local electron-phonon interactions.
On the other hand, under the influence of the nonlocal dynamic disorder, the CPR shows much more fruitful behaviours.
As $\lambda_\mathrm{non}$ increases from zero, the CPR sharply declines in the beginning and quickly reaches a minimal value at around $\lambda_\mathrm{non}=V$.
After that, the CPR briefly arises and then decreases again at a slow rate.
This peculiar behaviour implies that the nonlocal dynamic disorder may play a dual role in the decoherence.
Furthermore, from the right panel of \Fig{rfig3} (b), it can be seen that the nonlocal dynamic disorder also has a stronger impact on the average electronic energy than the local one.

\begin{figure}[htbp]
\begin{center}
\includegraphics[width=3.4in]{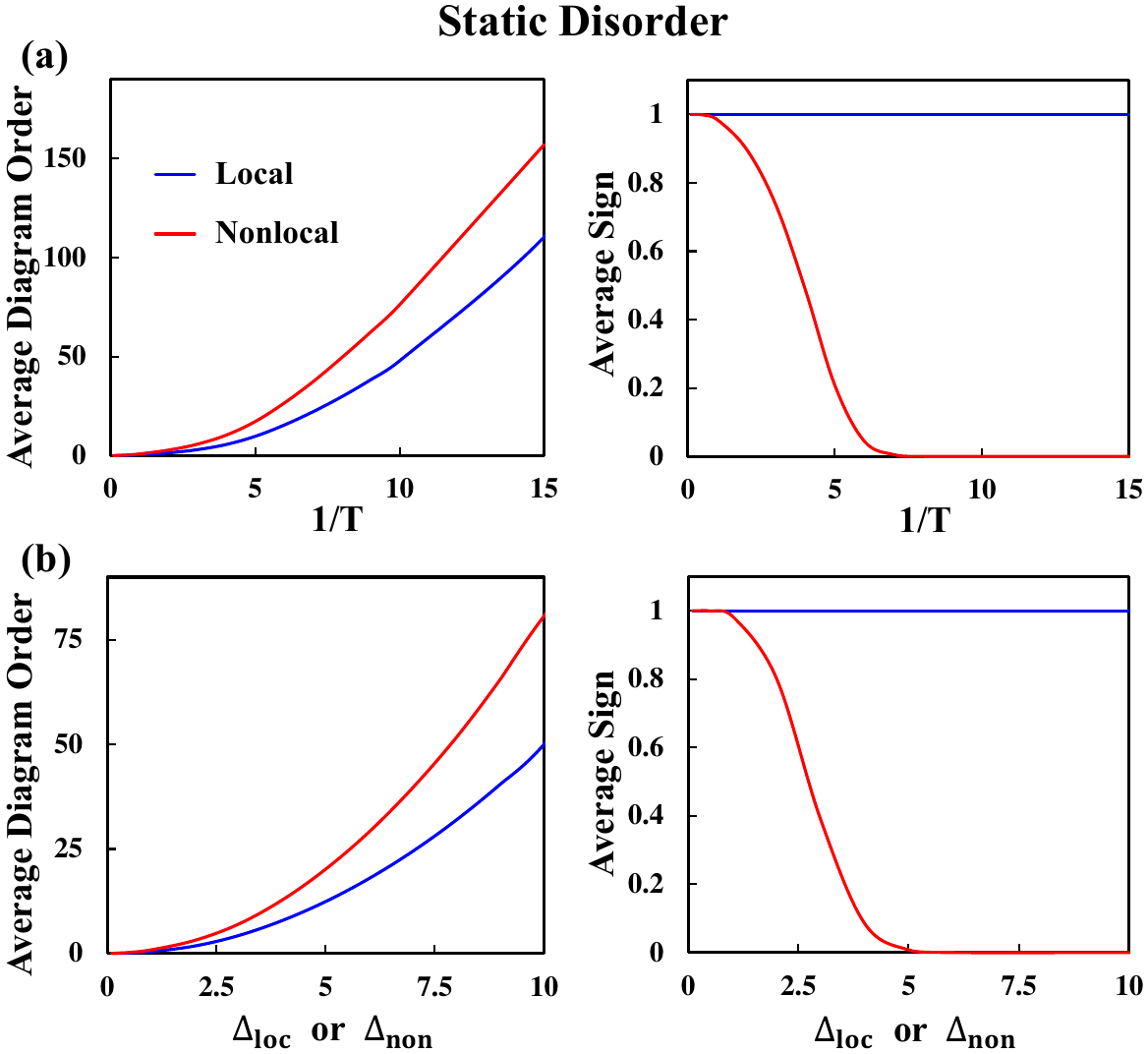}
\caption{Variation of the average diagram order $\la\kappa\ra$ (left panel) and average sign (right panel) versus (a) $1/T$ and (b) static disorder strengths with the presence of local (blue line) or nonlocal (red line) static disorders. In (a), the static disorder strengths are $\Delta_\mathrm{loc}=1$, $\Delta_\mathrm{non}=0$ for the local case and $\Delta_\mathrm{loc}=0$, $\Delta_\mathrm{non}=1$ for the nonlocal case. In (b), we set $T=1$. Other parameters are $V=1$ and $\lambda_\mathrm{loc}=\lambda_\mathrm{non}=0$. }
\label{rfig5}
\end{center}
\end{figure}

To further elucidate the effects of the nonlocal dynamic disorder on the coherence, we present the normalized real-space coherence $|\tilde{\rho}_{n,n+l}|=|\rho_{n,n+l}|/|\rho_{nn}|$ with different strengths of nonlocal electron-phonon interactions in \Fig{rfig4}.
It is found that although the presence of weak nonlocal dynamic disorder ($\lambda_\mathrm{non}=0.5$) substantially suppresses the coherence, the next nearest-neighbouring coherence $|\tilde{\rho}_{n,n+2}|$ is resurrected as $\lambda_\mathrm{non}$ exceeds $V$, and it even surpasses that without disorder when $\lambda_\mathrm{non}$ is larger than 3.
Similar resurgence phenomenon is also observed for $|\tilde{\rho}_{n,n+3}|$ and $|\tilde{\rho}_{n,n+4}|$.
Combining the results shown in \Fig{rfig2}, we conclude that when the nonlocal dynamic disorder is weak it strongly suppresses the coherence, whereas it turns to help regenerate the coherence when it becomes strong.
In addition, it is noted that the nearest-neighbouring coherence $|\tilde{\rho}_{n,n+1}|$ is always suppressed no matter the strength of the disorder, which is found to be relevant to the anti-symmetric form of the nonlocal electron-phonon interaction.

\begin{figure}[htbp]
\begin{center}
\includegraphics[width=3.1in]{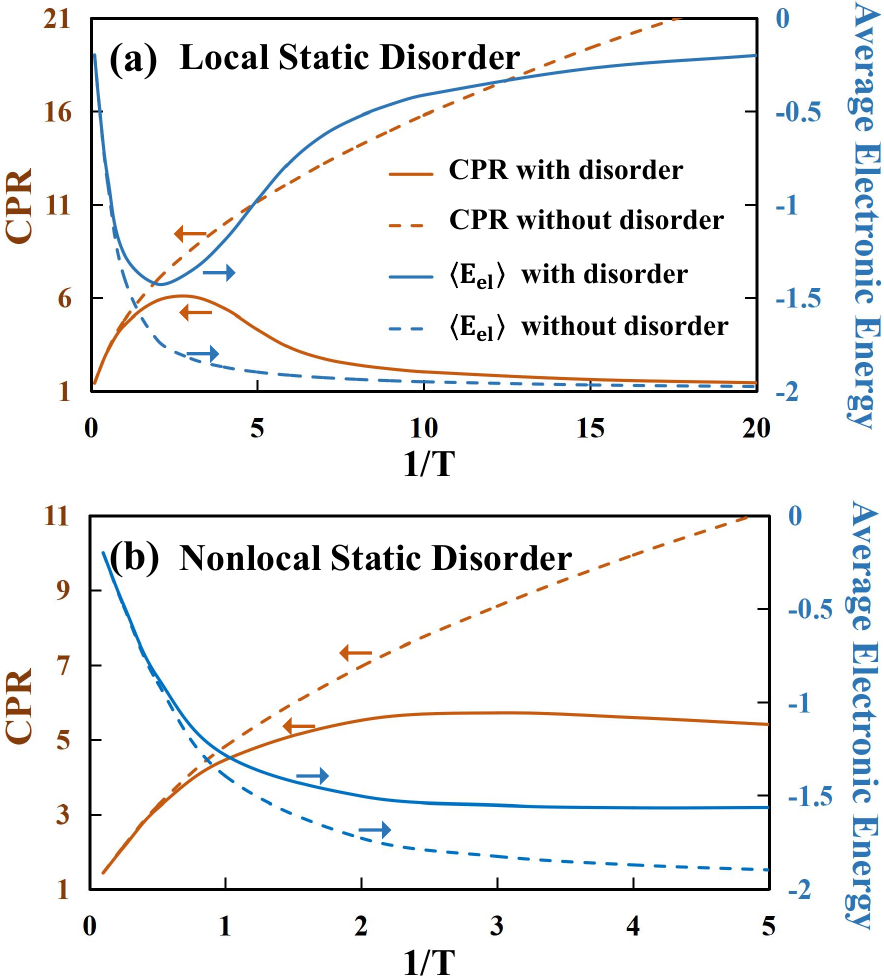}
\caption{Variation of CPR and the average electronic energy $\la E_{el}\ra$ versus $1/T$ with the presence of (a) local or (b) nonlocal static disorders. Also shown are the results without the presence of disorder (dashed line). The disorder strengths are $\Delta_\mathrm{loc}=1$, $\Delta_\mathrm{non}=0$ in (a) and $\Delta_\mathrm{loc}=0$, $\Delta_\mathrm{non}=1$ in (b). Other parameters are $V=1$ and $\lambda_\mathrm{loc}=\lambda_\mathrm{non}=0$.}
\label{rfig6}
\end{center}
\end{figure}

Now we turn to the static disorder.
\Fig{rfig5} presents the average diagram order and average sign for varying temperatures and varying static disorder strengths.
Different from the linear relation of the dynamic one, it is found that the average diagram order increases quadratically as $1/T$, $\Delta_\mathrm{loc}$ or $\Delta_\mathrm{non}$ increases.
Since the diagram order partially reflects the importance of the interaction, one can expect that the static disorder will dominate the transport properties at low temperatures, whereas the dynamic one will overwhelm at high temperatures.
Furthermore, similar to the dynamic disorder, the average sign for a local static disorder is always unity, but a nonlocal one can have a severe sign problem when $\Delta_\mathrm{non}/T$ is larger than 5, which is much worse than the case of the dynamic disorder.
Consequently, it is hard to obtain convergent results when  $\Delta_\mathrm{non}/T>5$ for the current version of DQMC.

\begin{figure*}[htbp]
\begin{center}
\includegraphics[width=6.5in]{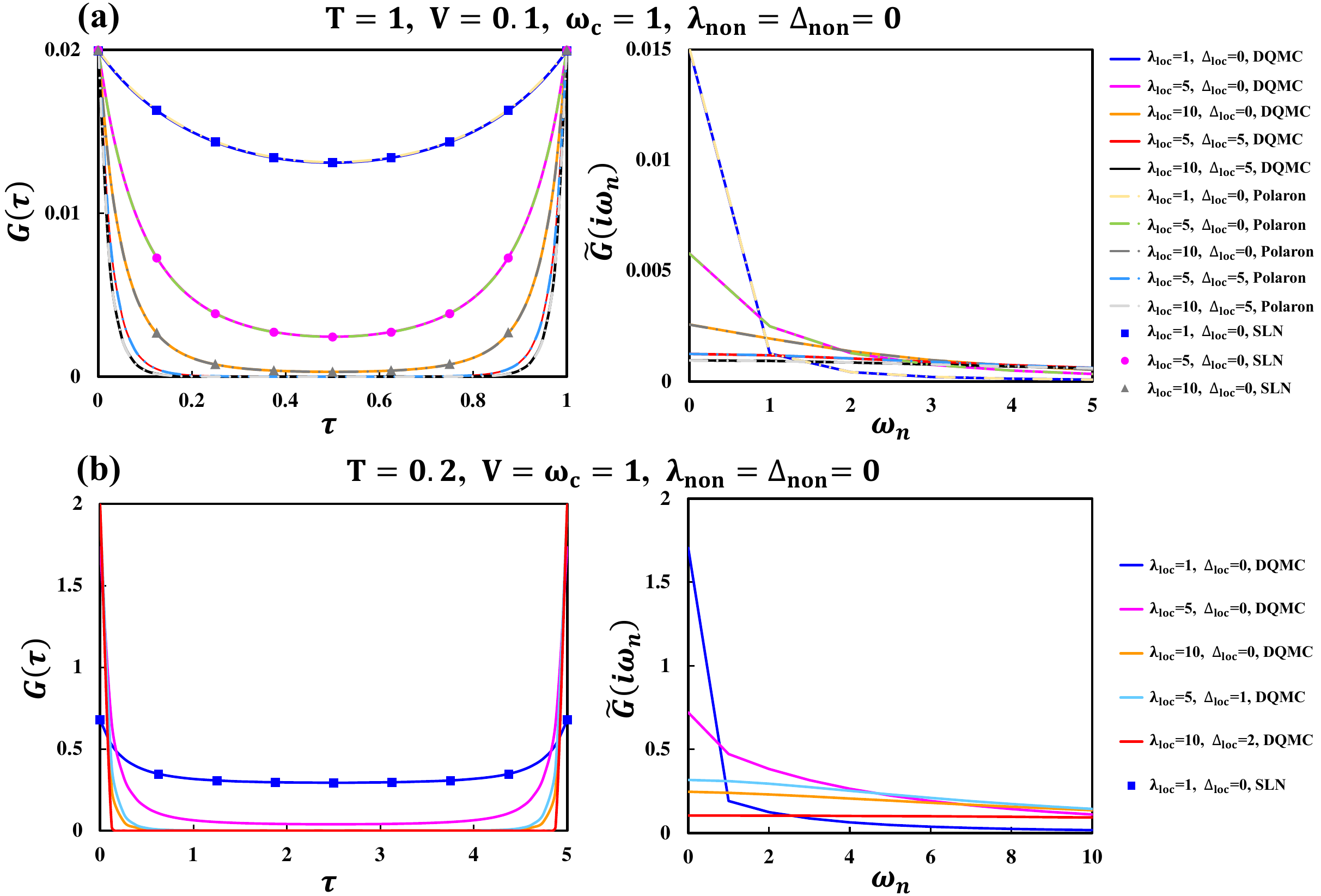}
\caption{Imaginary time (left panel) and imaginary frequency (right panel) current autocorrelation functions with the presence of local dynamic disorder and local static disorder for (a) small and (b) large electronic couplings.}
\label{rfig7}
\end{center}
\end{figure*}

\begin{figure*}[htbp]
\begin{center}
\includegraphics[width=6.5in]{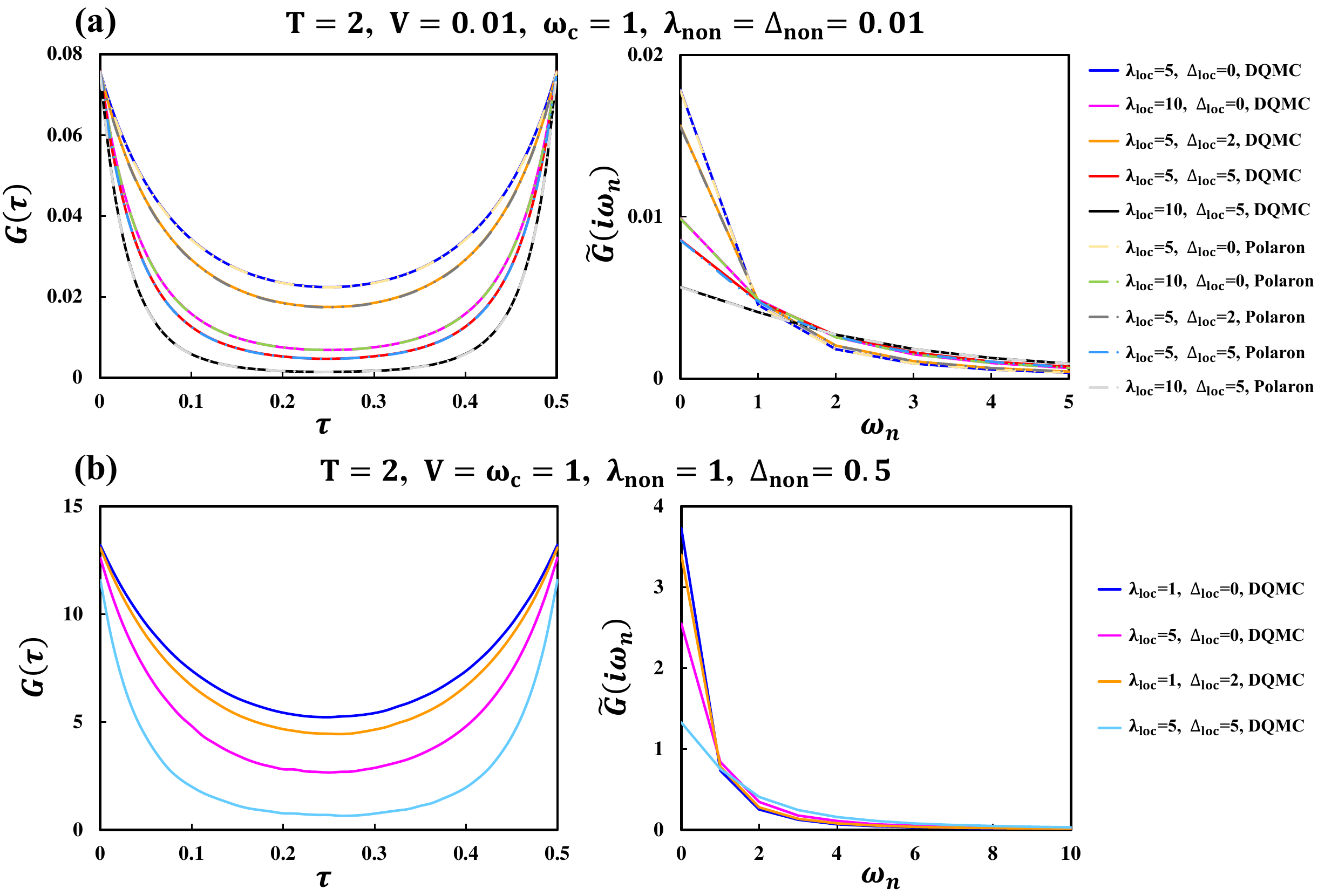}
\caption{Imaginary time (left panel) and imaginary frequency (right panel) current autocorrelation functions with the presence of various types of disorders for (a) small and (b) large electronic couplings.}
\label{rfig8}
\end{center}
\end{figure*}

The static disorder can strongly influence the electronic properties of the system at low temperatures.
\Fig{rfig6} presents the CPR and average electronic energy at different temperatures with the presence of local or nonlocal static disorders.
In the local case, from \Fig{rfig6} (a) it can be seen that the CPR and average electronic energy start to deviate those without the disorder when $\Delta_\mathrm{non}/T$ is larger than 1.
As $\Delta_\mathrm{non}/T$ exceeds 2, the CPR gradually approaches a maximal value of about six and then turns to decline.
In the meantime, the average electronic energy exhibits a similar inversion behaviour.
At extremely low temperatures (where the ground state is mostly occupied), the CPR and average electronic energy finally approach one and zero, respectively.
It is well known that with the presence of static disorder the ground state of a one-dimensional chain is exponentially localized, the so-called Anderson localization.
Our results successfully capture this feature.
As for the results of the nonlocal static disorder shown in \Fig{rfig6} (b), we find that the influence of the disorder also becomes important when $\Delta_\mathrm{non}/T>1$, but the effects are less significant than those in the local case, and the inversion behaviour do not obviously appears within the range $\Delta_\mathrm{non}/T<5$.
Convergent results are not available for lower temperatures.

One of the most important transport properties is the optical conductivity, which can be obtained by applying analytic continuation to the imaginary time or imaginary frequency current autocorrelation function.
\Fig{rfig7} (a) presents the current autocorrelation function for small electronic couplings with the presence of local dynamic and local static disorders.
As can be seen, the imaginary-time one, $G(\tau)$, is a symmetric function with respect to $\tau$.
When there is no disorder, $G(\tau)$ is a constant function; as the disorder is turned on, $G(\tau)$ becomes cupulate, and the stronger the disorder, the more concave in the middle of $G(\tau)$.
In the strong disorder limit, $G(0)$ can even be larger than $G(\beta/2)$ by several orders of magnitude.
On the contrary, the imaginary-frequency current autocorrelation function $\widetilde{G}(i\omega_n)$ becomes less and less structured as the disorder strength increases.
Similar behaviours are found for the case of large electronic couplings at low temperatures shown in \Fig{rfig7} (b).
We also present the results of the polaron theory and SLN in \Fig{rfig7}.
In the small electronic coupling limit, the polaron theory is quite accurate, and the results perfectly coincide with those obtained by DQMC.
As for SLN, it is found to be hard to obtain convergent results for large $\lambda_\mathrm{loc}$ at low temperatures.
\Fig{rfig8} presents the current autocorrelation functions with the inclusion of nonlocal disorders.
Again, one can find that the results of the polaron theory and DQMC are consistent with each other in the limit of weak nonlocal interactions.

In the last part of this section, we examine the reliability of
combining DQMC and numerical analytic continuation to extract charge carrier mobilities in different transport regimes from imaginary-time data.
The imaginary time and imaginary frequency current autocorrelation function is related to the real part of the optical conductivity $\sigma_{\alpha}(\omega)$ via\cite{Mahan--2013-}
\begin{equation}
\label{gt-sig}
G_{\alpha\alpha}(\tau)=\frac{1}{\pi}\int_{-\infty}^{+\infty}\mathrm{d}\omega
\frac{\omega e^{-\tau\omega}}{1-e^{-\beta\omega}}\sigma_{\alpha}(\omega)
\end{equation}
and
\begin{equation}
\label{gw-sig}
\widetilde{G}_{\alpha\alpha}(i\omega_n)
=\frac{1}{\pi}\int_{-\infty}^{+\infty}\mathrm{d}\omega
\frac{\omega^2}{\omega^2+\omega_n^2}\sigma_{\alpha}(\omega),
\end{equation}
respectively.
Inversely, $\sigma_{\alpha}(\omega)$ can be obtained by applying numerical analytic continuation to $G_{\alpha\alpha}(\tau)$ or $\widetilde{G}_{\alpha\alpha}(i\omega_n)$.
The mobility of the charge carrier along the $\alpha$ direction can be extracted from the optical conductivity via
$\mu_{\alpha}=\lim_{\omega\rightarrow0}\sigma_{\alpha}(\omega)/|e|$.

We choose the state-of-the-art stochastic optimization method\cite{Mishchenko-PRB-2000-6317,Mishchenko--2012-} (SOM) to perform numerical analytic continuation on $\widetilde{G}(i\omega_n)$.
It is worth mentioning that based on the Holstein model, Mishchenko \etal have successfully obtained the mobilities in several regions including the band conduction region\cite{Mishchenko-PRL-2015-146401}.
Their work on the Su-Schrieffer-Hegger model\cite{DeFilippis-PRL-2015-86601}. also covers the regime of transient localization.
As such, in the following we mainly focus on the strong disorder regimes involving several types of disorders.

\begin{figure}[htbp]
\begin{center}
\includegraphics[width=2.8in]{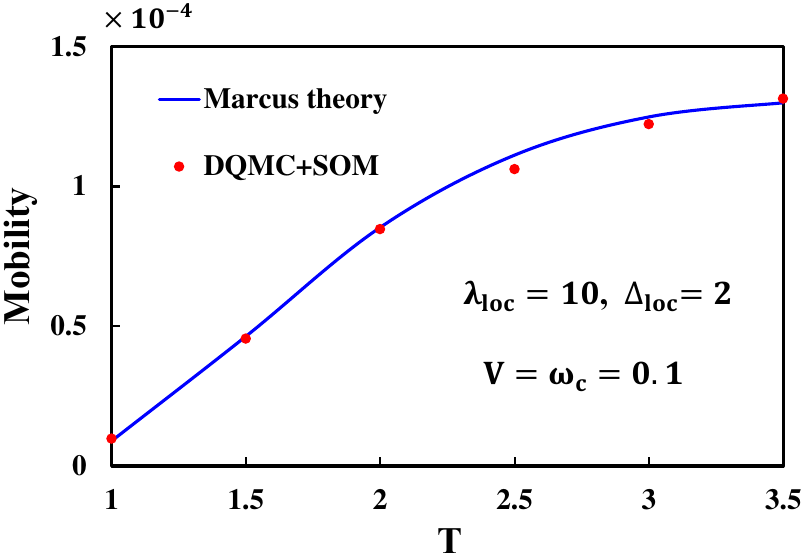}
\caption{Mobility in the thermally activated hopping region calculated by the Marcus theory \Eq{marcus} (solid blue line) and by the combination of DQMC and SOM (red circle). The Hamiltonian parameters are $V=\omega_c=0.1$, $\lambda_\mathrm{loc}=10$, $\Delta_\mathrm{loc}=2$, and $\lambda_\mathrm{non}=\Delta_\mathrm{non}=0$.}
\label{rfig9}
\end{center}
\end{figure}

\begin{figure}[htbp]
\begin{center}
\includegraphics[width=2.8in]{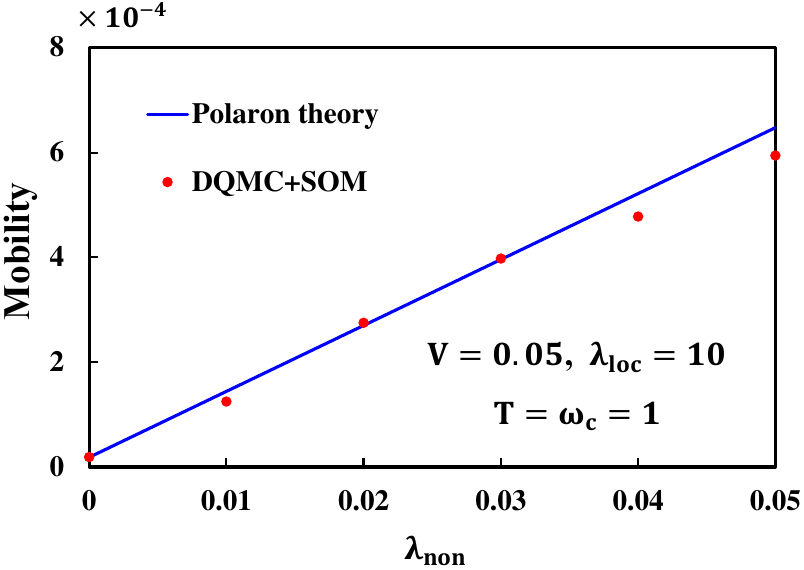}
\caption{Mobility in the phonon-assisted transport region calculated by the polaron theory \Eq{mu-polaron} (solid blue line) and by the combination of DQMC and SOM (red circle). The temperature is set as 1, and the Hamiltonian parameters are $V=0.05$, $\lambda_\mathrm{loc}=10$, $\omega_c=1$, and $\Delta_\mathrm{loc}=\Delta_\mathrm{non}=0$.}
\label{rfig10}
\end{center}
\end{figure}

The first example is the thermally activated hopping transport shown in \Fig{rfig9}.
In this region, the Marcus theory is applicable, and the mobility can be calculated by
\begin{equation}
\label{marcus}
\begin{split}
\mu=&\frac{1}{z_\mathrm{sd}k_BT}\int\mathrm{d}E_1\int\mathrm{d}E_2
P_\mathrm{sd}(E_1)P_\mathrm{sd}(E_2)
e^{-E_1/k_BT} \\
&\times|V|^2\sqrt{\frac{\pi}{k_BT\lambda}}
e^{-(E_2-E_1+\lambda)^2/(4k_{B}T\lambda)},
\end{split}
\end{equation}
where $\lambda=2\lambda_\mathrm{loc}$ is the reorganization energy for the electron transfer,
\begin{equation}
P_\mathrm{sd}(E)=\frac{1}{\sqrt{2\pi}\Delta_\mathrm{loc}}
e^{-E^2/(2\Delta_\mathrm{loc}^2)}
\end{equation}
is the probability distribution function of the static disorder, and $z_\mathrm{sd}=\int\mathrm{d}EP_\mathrm{sd}(E)e^{-E/k_BT}$.
As can be seen in \Fig{rfig9}, the mobilities calculated by DQMC and SOM coincide very well with the Marcus theory despite very small deviations at some temperatures.
This confirms the validity of the analytic continuation strategy in the hopping region with the presence of both dynamic and static disorders.

The second one is the phonon-assisted transport presented in \Fig{rfig10}.
In this region, the mobility can be obtained by
\begin{equation}
\label{mu-polaron}
\mu=k_\mathrm{pol}/k_BT,
\end{equation}
where $k_\mathrm{pol}$ is the rate constant calculated via the real-time version of the polaron theory\cite{Hannewald-PRB-2004-75212a}.
From \Fig{rfig10}, it is seen that as the nonlocal electron-phonon interaction strength increases, the mobility continuously increases, which is a character of the phonon-assisted transport.
The comparison with the polaron theory further validates the results of DQMC and SOM.

\begin{figure}[htbp]
\begin{center}
\includegraphics[width=2.8in]{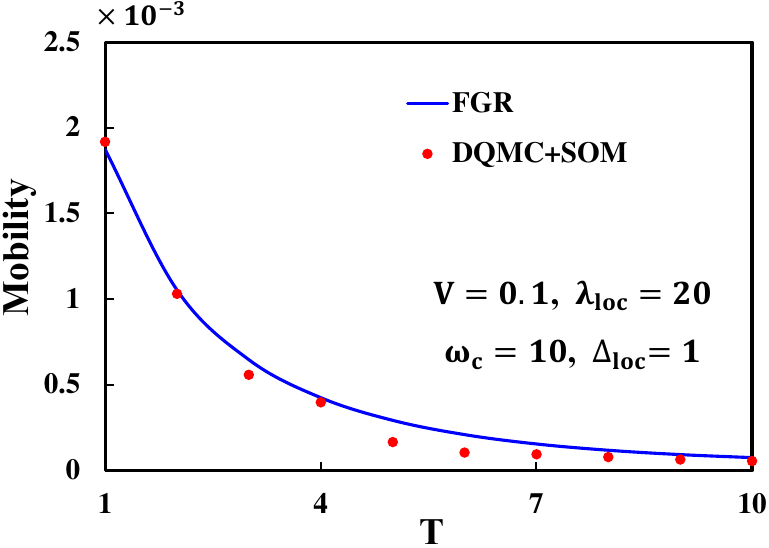}
\caption{Mobility in the nuclear tunneling region calculated by FGR \Eq{fgr} (solid blue line) and by the combination of DQMC and SOM (red circle). The Hamiltonian parameters are $V=0.1$, $\lambda_\mathrm{loc}=20$, $\omega_c=10$, $\Delta_\mathrm{loc}=1$, and $\lambda_\mathrm{non}=\Delta_\mathrm{non}=0$.}
\label{rfig11}
\end{center}
\end{figure}

The last one is the nuclear tunneling transport mechanism.
To explain the conflict between the localization nature of the carrier and the band-like behaviour of the mobility in many organic semiconductors, Nan and coworkers\cite{Nan-PRB-2009-115203} have proposed that the carrier still follows hopping motion but the hopping rate decreases with the increase in the temperature due to the nuclear tunneling effect of high-frequency intramolecular vibrations.
By using the Fermi's golden rule (FGR) to calculate the hopping rate, they successfully obtained the band-like mobilities for several organic semiconductors.
\Fig{rfig11} presents the mobilities in the nuclear tunneling region calculated by our method and by FGR.
We set $\omega_c=10$ to mimic high-frequency intramolecular vibrations.
The results of FGR are calculated by

\begin{equation}
\label{fgr}
\begin{split}
\mu=&\frac{1}{z_\mathrm{sd}k_BT}\int\mathrm{d}E_1\int\mathrm{d}E_2
P_\mathrm{sd}(E_1)P_\mathrm{sd}(E_2) \\
&\qquad\qquad\times
e^{-E_1/k_BT}k_\mathrm{FGR}(E_2-E_1),
\end{split}
\end{equation}
where
\begin{equation}
\begin{split}
k_\mathrm{FGR}(E)=&|V|^2\int_{-\infty}^{+\infty}\mathrm{d}t
\exp\{iEt-\sum_jS_j[(2n_j+1) \\
&\qquad\qquad-n_je^{-i\omega_jt}
-(n_j+1)e^{i\omega_jt}]\},
\end{split}
\end{equation}
and $S_j=g_j^2/2\omega_j^3$ is the Huang-Rhys factor.
It can be seen in \Fig{rfig11} that the results of DQMC and SOM are quantitatively in line with those of FGR at both low and high temperatures.
Nevertheless, significant deviations (with a relative error larger than 30\%) appear for the temperatures ranging from 5 to 8.
This indicates that one should be cautious when applying the numerical analytic continuation in this parameter regime.

\section{FURTHER DISCUSSION AND CONCLUDING REMARKS}\label{sec4}

The DQMC approach presented here naturally inherits several merits of many other variants in the literatures.
First, it is a numerically exact method since we did not introduce any approximation to the framework.
Second, this approach can be used to obtain transport properties in the thermodynamic limit (infinitely large-sized systems) and therefore is not bothered by any finite-size effects.
Carefully inspecting the acceptance ratios and estimators presented in Supporting Information, one can see that the numerical cost of the Monte Carlo procedure is not relevant to the size of the system but is only determined by the temperature and the Hamiltonian parameters, \textit{ie}., the electron-phonon interactions, phonon frequencies, static disorder, and electronic dispersions.
Third, the diagrammatic expansion technique treats the imaginary time as a continuous variable and dispenses with the systematic error caused by the discretization of the time axis (as such, it is more often called the continuous-time QMC in the field of quantum impurities\cite{Gull-RMP-2011-349}).

In addition to the aforementioned advantages, the present DQMC approach also benefits from the generality of the Hamiltonian we adopted.
To give a theoretical description as accurate as possible for realistic disordered semiconductor materials, we have included multiple electronic energy bands and phonon branches with arbitrary dispersions, and various kinds of dynamic and static disorders in the Hamiltonian \Eq{htot}.
We have also explicitly included a general disorder-induced current operator ($\hat{\mathbf{J}}_{el-ph}+\hat{\mathbf{J}}_{sd}$) in our framework.
As such, the proposed DQMC approach is versatile and applicable to diverse semiconductor materials involving various intricate factors such as the local and nonlocal electron-phonon interactions, high-frequency optical and low-frequency acoustic phonons, static disorder with complicated covariance properties, complex unit cells with multiple components, and anisotropy.

When applying the DQMC approach to a realistic semiconductor materials, the most expensive part should be the first-principles calculation of the Hamiltonian parameters.
For this task, there have already been many powerful state-of-the-art first-principles methods.
For example, the electronic energy bands are routinely obtained by the density functional theory, whereas the phonon dispersions and electron-phonon interaction parameters for a sparse q grid can be efficiently calculated by the density functional perturbation theory\cite{Baroni-RMP-2001-515}.
Nevertheless, in practical calculations, the k and q grids in the Brillouin zone should be dense enough so as to eliminate the finite-size effect.
This can be a formidable task for direct \textit{ab initio} calculations.
To circumvent this problem, it is useful to adopt efficient interpolation schemes\cite{Giustino-PRB-2007-165108,Verdi-PRL-2015-176401} to obtain Hamiltonian parameters at any points in the Brillouin zone based on the original data in a relatively coarse grid.
This strategy also applies to the parameterization of $\hat{\mathbf{J}}_{el}$ in \Eq{jel}\cite{Yates-PRB-2007-195121}.
As for the parameters $\mathbf{u}_{nm\mathbf{kq}\nu}$ and  $\tilde{\mathbf{u}}_{nm\mathbf{kq}\mu}$ given in \Eq{jelph} and \Eq{jsd}, respectively, there is no computational method currently available to our knowledge.
In Appendix A, we establish a quantitative relationship between $\hat{\mathbf{J}}_{el-ph}$ and the electron-phonon interactions with the aid of the localized Wannier functions\cite{Marzari-RMP-2012-1419}.
This relationship can serve as an efficient computational scheme for $\mathbf{u}_{nm\mathbf{kq}\nu}$.
A similar computational scheme for $\hat{\mathbf{J}}_{sd}$ is discussed in Appendix B.

It is worth noting that in the theoretical framework presented in this work, the static disorder is introduced as the certain randomness in the electronic-state energies and electronic couplings on the basis of a perfect periodic crystal.
Thereby, this description is suitable for semiconductors with regularly arranged microscopic structures containing small or moderate static disorder, but may not apply to very disordered materials such as amorphous polymers.

Finally, although we take the semiconductor material as an example, the current approach can also be used to study the transport properties of the exciton in natural and artificial light-harvesting systems as long as the system being studied consists of regularly stacking units.
Typical examples of the suitable systems are small-molecule optoelectronic materials like the metal phthalocyanines\cite{Feng-JCP-2020-34116,Feng-JPCA-2021-2932} and self-assembled tubular molecular aggregates such as the chlorosome from green sulfur bacteria\cite{Fujita-JPCL-2012-2357,Fujita-PR-2014-273,Huh-JACS-2014-2048,
Sawaya-NL-2015-1722,Li-JPCB-2020-4026} and synthetic nanotubes\cite{Barclay-CR-2014-10217,Vlaming-JPCB-2009-2273,
Eisele-NC-2012-655,Doria-AN-2018-4556}.

In summary, starting from a general Hamiltonian expressed in reciprocal space and utilizing the standard diagrammatic expansion technique to the imaginary time propagator, we have proposed a new DQMC approach which is suitable for the accurate calculation of various transport properties in realistic disordered semiconductors and light-harvesting materials.
This approach deals with the dynamic and static disorders in a unified and numerically exact way and is applicable to material systems containing multiple electronic bands and phonon branches with general dispersions.
It is expected that the DQMC approach proposed here will become a promising theoretical tool for elucidating various fascinating transport properties experimentally observed in disordered semiconductors and light-harvesting materials.

\section*{ACKNOWLEDGEMENTS}
This work is supported by the National Science Foundation of China (Grant Nos. 22033006 and 21833006). Y.-C. Wang acknowledges the support from China Postdoctoral Science Foundation (No. 2021M702734).

\section*{Supplementary Material}
The supplementary material includes a brief summary of the Monte Carlo technique; discussion about the correspondence between the expression \Eq{zx-kappa2} and the diagrams; detailed introduction of the necessary configuration updating procedures required for the Monte Carlo stochastic sampling and several efficient estimators for the calculations of different physical quantities; supplementary results for the one-dimensional and two-dimensional models.

\section*{Data Availability}
The data that support the findings of this study are available from the corresponding author upon reasonable
request.

\appendix
\section{Current operator from the electron-phonon interactions}
In this appendix, we provide a scheme for calculating the parameters $\mathbf{u}_{nm\mathbf{kq}\nu}$ in \Eq{jelph} from electron-phonon interaction parameters with the aid of the localized Wannier functions.

We start from the definition of the phonon-induced current operator
\begin{equation}
\label{jelph-def}
\hat{\mathbf{J}}_{el-ph}=i[\hat{H}_{el-ph},\hat{\mathbf{P}}],
\end{equation}
where $\hat{\mathbf{P}}$ is the polarization operator.
Assume that the transformation between the Wannier orbitals and the Bloch orbitals is given by \Eq{el-r2k}.
In real space, the polarization operator can be decomposed into two parts as
\begin{equation}
\label{p}
\hat{\mathbf{P}}=\hat{\mathbf{P}}^{(1)}+\hat{\mathbf{P}}^{(2)}
\end{equation}
with
\begin{equation}
\label{p1}
\hat{\mathbf{P}}^{(1)}=
e\sum_{\bar{n}\mathbf{R}}\mathbf{r}_{\bar{n}\mathbf{R}}
\hat{c}_{\bar{n}\mathbf{R}}^\dagger\hat{c}_{\bar{n}\mathbf{R}}
\end{equation}
and
\begin{equation}
\label{p2}
\begin{split}
\hat{\mathbf{P}}^{(2)}=&
e\sum_{\bar{n}}\sum_{\Delta\mathbf{R}\ne\mathbf{0}}
\mathbf{r}_{\bar{n}\bar{n}\Delta\mathbf{R}}
\sum_{\mathbf{R}}
\hat{c}_{\bar{n},\mathbf{R}+\Delta\mathbf{R}}^\dagger
\hat{c}_{\bar{n}\mathbf{R}} \\
&+e\sum_{\bar{n}\ne\bar{m}}\sum_{\Delta\mathbf{R}}
\mathbf{r}_{\bar{n}\bar{m}\Delta\mathbf{R}}
\sum_{\mathbf{R}}
\hat{c}_{\bar{n},\mathbf{R}+\Delta\mathbf{R}}^\dagger
\hat{c}_{\bar{m}\mathbf{R}}
\end{split}.
\end{equation}
Here, $\mathbf{r}_{\bar{n}\mathbf{R}}=\int\mathbf{r}
|\psi_{\bar{n}\mathbf{R}}(\mathbf{r})|^2\mathrm{d}\mathbf{r}
=\mathbf{r}_{\bar{n}}+\mathbf{R}$ with $\mathbf{r}_{\bar{n}}=\int\mathbf{r}
|\psi_{\bar{n}\mathbf{0}}(\mathbf{r})|^2\mathrm{d}\mathbf{r}$ and
$\mathbf{r}_{\bar{n}\bar{m}\Delta\mathbf{R}}=\int\mathbf{r}
\psi_{\bar{n},\mathbf{R}+\Delta\mathbf{R}}(\mathbf{r})
\psi_{\bar{m}\mathbf{R}}(\mathbf{r})\mathrm{d}\mathbf{r}$  are diagonal and off-diagonal elements of the position operator, respectively, and $\psi_{\bar{n}\mathbf{R}}(\mathbf{r})$ is the wavefunction of the $\bar{n}$th Wannier orbital in the unit cell centered at $\mathbf{R}$.
It is noted that $\mathbf{r}_{\bar{n}\bar{m}\Delta\mathbf{R}}$ is independent of $\mathbf{R}$, which is due to the orthogonality of the Wannier functions.
$\hat{\mathbf{P}}^{(1)}$ and $\hat{\mathbf{P}}^{(2)}$ can be regarded as the local and nonlocal components of the polarization operator, respectively.
Due to the localization nature of the Wannier orbital, $\hat{\mathbf{P}}^{(2)}$ is usually much smaller than $\hat{\mathbf{P}}^{(1)}$ and thereby is frequently neglected in the literatures (which is just the tight-binding approximation).

The electron-phonon interactions in real space reads
\begin{equation}
\label{helph-r}
\begin{split}
\hat{H}_{el-ph}=&\sum_{\bar{n}\bar{m}\mathbf{R}_e}\sum_{\bar{\nu}\mathbf{R}_p}
\sum_{\mathbf{R}}g_{\bar{n}\bar{m}\mathbf{R}_e\mathbf{R}_p\bar{\nu}} \\
&\times
\hat{Q}_{\bar{\nu},\mathbf{R}+\mathbf{R}_p}\otimes
\hat{c}_{\bar{n}\mathbf{R}}^\dagger\hat{c}_{\bar{m},\mathbf{R}+\mathbf{R}_e}.
\end{split}
\end{equation}
Here, $\mathbf{R}_e$ and $\mathbf{R}_p$ are three-dimensional vectors, the possible values of which are the same as that of $\mathbf{R}$.
The real-space vibrational coordinate $\hat{Q}_{\bar{\nu}\mathbf{R}}$ can be generally expressed as the linear combination of the reciprocal-space coordinate $\hat{B}_{\nu\mathbf{q}}$ as
\begin{equation}
\label{ph-r2k}
\hat{Q}_{\bar{v}\mathbf{R}}=N^{-\frac{3}{2}}\sum_{\mathbf{q}}e^{i\mathbf{q\cdot R}}
\sum_{\nu}\frac{(\hat{\Theta}_\mathbf{q}^\dagger)_{\bar{\nu}\nu}}{\sqrt{2\omega_{\nu\mathbf{q}}}}
\hat{B}_{\nu\mathbf{q}},
\end{equation}
where $\hat{\Theta}_\mathbf{q}$ is the unitary transformation operator associated with the phonon crystal momentum $\mathbf{q}$.
The parameter $g_{\bar{n}\bar{m}\mathbf{R}_e\mathbf{R}_p\bar{\nu}}$ corresponds to the electron-phonon interaction between $|\bar{n}\mathbf{R}\ra\equiv\hat{c}_{\bar{n}\mathbf{R}}^\dagger|\mathrm{vac}\ra$ and $|\bar{m},\mathbf{R}+\mathbf{R}_e\ra\equiv\hat{c}_{\bar{m},\mathbf{R}+\mathbf{R}_e}^\dagger|\mathrm{vac}\ra$ caused by $\hat{Q}_{\bar{\nu},\mathbf{R}+\mathbf{R}_p}$, and it satisfies the property
$g_{\bar{n}\bar{m}\mathbf{R}_e\mathbf{R}_p\bar{\nu}}
=g_{\bar{m}\bar{n},-\mathbf{R}_e,\mathbf{R}_p-\mathbf{R}_e,\bar{\nu}}$.
Substituting Eqs.\,(\ref{el-r2k}) and (\ref{ph-r2k}) into \Eq{helph-r}, one can recover \Eq{helph} with the following relation
\begin{equation}
\label{g-r2k}
\begin{split}
g_{nm\mathbf{kq}\nu}=&\frac{1}{\sqrt{2\omega_{\nu\mathbf{q}}}}\sum_{\mathbf{R}_e\mathbf{R}_p}
e^{i\mathbf{k}\cdot\mathbf{R}_e+i\mathbf{q}\cdot\mathbf{R}_p}
\sum_{\bar{n}\bar{m}}(\hat{U}_{\mathbf{k+q}})_{n\bar{n}} \\
&\times(\hat{U}_{\mathbf{k}}^\dagger)_{\bar{m}m}
\sum_{\bar{\nu}}(\hat{\Theta}_{\mathbf{q}}^\dagger)_{\bar{\nu}\nu}
g_{\bar{n}\bar{m}\mathbf{R}_e\mathbf{R}_p\bar{\nu}},
\end{split}
\end{equation}
where we have invoked the property $\sum_{\mathbf{R}}e^{i\mathbf{k}\cdot\mathbf{R}}
=\delta_{\mathbf{k}\mathbf{0}}N^{3}$.
Likewise, applying the inverse transformation leads to
\begin{equation}
\label{g-k2r}
\begin{split}
g_{\bar{n}\bar{m}\mathbf{R}_e\mathbf{R}_p\bar{\nu}}
=&N^{-6}\sum_{\mathbf{kq}}e^{-i\mathbf{k}\cdot\mathbf{R}_e-i\mathbf{q}\cdot\mathbf{R}_p}
\sum_{nm}(\hat{U}_{\mathbf{k+q}}^\dagger)_{\bar{n}n} \\
&\times(\hat{U}_{\mathbf{k}})_{m\bar{m}}
\sum_{\nu}\sqrt{2\omega_{\nu\mathbf{q}}}(\hat{\Theta}_{\mathbf{q}})_{\nu\bar{\nu}}
g_{nm\mathbf{kq}\nu}.
\end{split}
\end{equation}
\Eq{g-r2k} and \Eq{g-k2r} as reciprocal relations have been utilized in the Wannier-Fourier interpolation scheme to calculate the electron-phonon interactions in a dense reciprocal-space grid based on the original \textit{ab initio} data in a coarse grid\cite{Giustino-PRB-2007-165108}.
In the following, we provide a similar relation between $g_{\bar{n}\bar{m}\mathbf{R}_e\mathbf{R}_p\bar{\nu}}$ and $\mathbf{u}_{nm\mathbf{kq}\nu}$.

For the convenience of derivation, we divide $\hat{\mathbf{J}}_{el-ph}$ into two parts, $\hat{\mathbf{J}}_{el-ph}=\hat{\mathbf{J}}_{el-ph}^{(1)}
+\hat{\mathbf{J}}_{el-ph}^{(2)}$, where
\begin{equation}
\label{jelph-l}
\hat{\mathbf{J}}_{el-ph}^{(l)}=i[\hat{H}_{el-ph},\hat{\mathbf{P}}^{(l)}]
\end{equation}
with $l=1,2$.
Substituting \Eq{p1} and \Eq{helph-r} into \Eq{jelph-l} and using the property
\begin{equation}
[\hat{c}_i^\dagger\hat{c}_j,\hat{c}_k^\dagger\hat{c}_l]=
\delta_{jk}\hat{c}_i^\dagger\hat{c}_l-\delta_{il}\hat{c}_k^\dagger\hat{c}_j,
\end{equation}
it is straightforward to obtain the following real-space expression for the first part
\begin{equation}
\label{j1-r}
\begin{split}
\hat{\mathbf{J}}_{el-ph}^{(1)}&=
ie\sum_{\bar{n}\bar{m}\mathbf{R}_e}\sum_{\bar{\nu}\mathbf{R}_p}
\sum_{\mathbf{R}}(\mathbf{r}_{\bar{m}}-\mathbf{r}_{\bar{n}}+\mathbf{R}_e)
g_{\bar{n}\bar{m}\mathbf{R}_e\mathbf{R}_p\bar{\nu}} \\
&\qquad\quad\times
\hat{Q}_{\bar{\nu},\mathbf{R}+\mathbf{R}_p}\otimes
\hat{c}_{\bar{n}\mathbf{R}}^\dagger\hat{c}_{\bar{m},\mathbf{R}+\mathbf{R}_e}.
\end{split}
\end{equation}
It is noted that \Eq{j1-r} is very similar to \Eq{helph-r}.
The only differences are the presence of the prefactor $ie$ and the relative distance $(\mathbf{r}_{\bar{m}}-\mathbf{r}_{\bar{n}}+\mathbf{R}_e)$.
Applying the Wannier-to-Bloch transformation, Eqs.\,(\ref{el-r2k}) and (\ref{ph-r2k}), to \Eq{j1-r}, we arrive at the following reciprocal-space expression
\begin{equation}
\label{j1-k}
\hat{\mathbf{J}}_{el-ph}^{(1)}=
eN^{-\frac{3}{2}}\sum_{nm\mathbf{k}}\sum_{\nu\mathbf{q}}
\mathbf{u}_{nm\mathbf{kq}\nu}^{(1)}
\hat{c}_{n,\mathbf{k+q}}^\dagger\hat{c}_{m\mathbf{k}}
\otimes\hat{B}_{\nu\mathbf{q}},
\end{equation}
where
\begin{equation}
\label{u1}
\begin{split}
\mathbf{u}_{nm\mathbf{kq}\nu}^{(1)}
=&\frac{i}{\sqrt{2\omega_{\nu\mathbf{q}}}}\sum_{\mathbf{R}_e\mathbf{R}_p}
e^{i\mathbf{k}\cdot\mathbf{R}_e+i\mathbf{q}\cdot\mathbf{R}_p}
\sum_{\bar{n}\bar{m}}(\mathbf{r}_{\bar{m}}-\mathbf{r}_{\bar{n}}+\mathbf{R}_e)
\\
&\times(\hat{U}_{\mathbf{k+q}})_{n\bar{n}}
(\hat{U}_{\mathbf{k}}^\dagger)_{\bar{m}m}
\sum_{\bar{\nu}}(\hat{\Theta}_{\mathbf{q}}^\dagger)_{\bar{\nu}\nu}
g_{\bar{n}\bar{m}\mathbf{R}_e\mathbf{R}_p\bar{\nu}}
\end{split}
\end{equation}
is the main component of $\mathbf{u}_{nm\mathbf{kq}\nu}$.
Comparing \Eq{u1} with \Eq{g-k2r}, we can immediately find that
$\mathbf{u}_{nm\mathbf{kq}\nu}^{(1)}=\nabla_\mathbf{k}g_{nm\mathbf{kq}\nu}$ if there is only one single electronic band.
In the multiple-band situation, this relation is no longer valid.

Following the same straightforward but tedious procedure, the second part of $\hat{\mathbf{J}}_{el-ph}$ can be cast into
\begin{equation}
\begin{split}
\hat{\mathbf{J}}_{el-ph}^{(2)}&
=eN^{-\frac{3}{2}}\sum_{nm\mathbf{k}}\sum_{\nu\mathbf{q}}
\mathbf{u}_{nm\mathbf{kq}\nu}^{(2)}
\hat{c}_{n,\mathbf{k+q}}^\dagger\hat{c}_{m\mathbf{k}}
\otimes\hat{B}_{\nu\mathbf{q}},
\end{split}
\end{equation}
where
\begin{widetext}
\begin{equation}
\label{u2}
\begin{split}
&\mathbf{u}_{nm\mathbf{kq}\nu}^{(2)}
=\frac{i}{\sqrt{2\omega_{\nu\mathbf{q}}}}
\sum_{\mathbf{R}_e\mathbf{R}_p}
e^{i\mathbf{k}\cdot\mathbf{R}_e
+i\mathbf{q}\cdot\mathbf{R}_p}\sum_{\bar{n}\bar{m}}(\hat{U}_{\mathbf{k+q}})_{n\bar{n}}
(\hat{U}_{\mathbf{k}}^\dagger)_{\bar{m}m}
\sum_{\bar{\nu}}
(\hat{\Theta}_{\mathbf{q}}^\dagger)_{\bar{\nu}\nu}
\left(\sum_{\bar{n}'\ne\bar{m}}\mathbf{r}_{\bar{n}'\bar{m}\mathbf{0}}
g_{\bar{n}\bar{n}'\mathbf{R}_e\mathbf{R}_p\bar{\nu}}\right. \\
&\left.-\sum_{\bar{n}'\ne\bar{n}}
\mathbf{r}_{\bar{n}\bar{n}'\mathbf{0}}
g_{\bar{n}'\bar{m}\mathbf{R}_e\mathbf{R}_p\bar{\nu}}
+\sum_{\Delta\mathbf{R}\ne\mathbf{0}}\sum_{\bar{n}'}
e^{-i\mathbf{k}\cdot\Delta\mathbf{R}}
\mathbf{r}_{\bar{n}'\bar{m}\Delta\mathbf{R}}
g_{\bar{n}\bar{n}'\mathbf{R}_e\mathbf{R}_p\bar{\nu}}
-\sum_{\Delta\mathbf{R}\ne\mathbf{0}}\sum_{\bar{n}'}
e^{-i(\mathbf{k}+\mathbf{q})\cdot\Delta\mathbf{R}}
\mathbf{r}_{\bar{n}\bar{n}'\Delta\mathbf{R}}
g_{\bar{n}'\bar{m}\mathbf{R}_e\mathbf{R}_p\bar{\nu}}\right)
\end{split}
\end{equation}
\end{widetext}
is the other component of $\mathbf{u}_{nm\mathbf{kq}\nu}$.

\Eq{g-k2r} together with \Eq{u1} and \Eq{u2} provide a feasible way to compute $\mathbf{u}_{nm\mathbf{kq}\nu}$ from $g_{nm\mathbf{kq}\nu}$.
The concrete steps can be summarized as follows.
First, perform first-principles calculations to acquire the electronic and phonon dispersions and the electron-phonon interaction parameters $g_{nm\mathbf{kq}\nu}$ in a uniform reciprocal-space grid.
The transformation matrix $\hat{\Theta}_\mathbf{q}$ between the real-space and reciprocal-space coordinates is simultaneously obtained during the calculation of the phonon normal modes.
Second, use computational tools like the Wannier90 program\cite{Mostofi-CPC-2008-685} to construct Bloch-to-Wannier transformation matrix $\hat{U}_\mathbf{k}$ and acquire $\mathbf{r}_{\bar{n}}$ and $\mathbf{r}_{\bar{n}\bar{m}\Delta\mathbf{R}}$.
Third, calculate $g_{\bar{n}\bar{m}\mathbf{R}_e\mathbf{R}_p\bar{\nu}}$ according to \Eq{g-k2r}.
Fourth, calculate $\mathbf{u}_{nm\mathbf{kq}\nu}^{(1)}$ and $\mathbf{u}_{nm\mathbf{kq}\nu}^{(2)}$ according to \Eq{u1} and \Eq{u2}, respectively.

Beneficial from the exponential localization of the maximally localized Wannier functions\cite{Marzari-RMP-2012-1419}, the above computational scheme can be used to calculate $\mathbf{u}_{nm\mathbf{kq}\nu}$ at any points in the Brillouin zone for nonpolar materials by following the conventional Wannier-Fourier interpolation strategy\cite{Giustino-PRB-2007-165108,Yates-PRB-2007-195121,Marzari-RMP-2012-1419}.
As for polar materials like the titanium dioxide, special techniques toward the calculation of the Fr\"{o}hlich vertex should be addressed\cite{Verdi-PRL-2015-176401}.

\section{Current operator from the static disorder}
The derivation of \Eq{jsd} follows the same procedure presented in Appendix A.
Divide $\hat{\mathbf{J}}_{sd}$ into two parts, $\hat{\mathbf{J}}_{sd}=\hat{\mathbf{J}}_{sd}^{(1)}+\hat{\mathbf{J}}_{sd}^{(2)}$, where $\hat{\mathbf{J}}_{sd}^{(l)}=i[\hat{H}_{sd},\hat{\mathbf{P}}^{(l)}]$ with $l=1,2$.
Using \Eq{hsd-r2} and \Eq{p1}, the first term in real space is given by
\begin{equation}
\label{jsd-1}
\begin{split}
\hat{\mathbf{J}}_{sd}=&ie\sum_{\bar{\mu}}\sum_{\mathbf{R}}
\gamma_{\bar{\mu}\mathbf{R}}
(\mathbf{r}_{\bar{m}}-\mathbf{r}_{\bar{n}}+\mathbf{R}_e) \\
&\times\left(\hat{c}_{\bar{n}\mathbf{R}}^\dagger
\hat{c}_{\bar{m},\mathbf{R}+\bar{\mathbf{R}}_e}
-\hat{c}_{\bar{m},\mathbf{R}+\bar{\mathbf{R}}_e}^\dagger
\hat{c}_{\bar{n}\mathbf{R}}\right).
\end{split}
\end{equation}
Remember that $\bar{\mu}\equiv(\bar{n},\bar{m},\bar{\mathbf{R}}_e)$ is an integrated index.
Applying the transformations \Eq{el-r2k} and \Eq{gam-dq} to \Eq{jsd-1} and simplifying the expression, we arrive at
\begin{equation}
\label{jsd-1-k}
\hat{\mathbf{J}}_{sd}^{(1)}=
eN^{-\frac{3}{2}}\sum_{nm\mathbf{k}}\sum_{\nu\mathbf{q}}
\tilde{\mathbf{u}}_{nm\mathbf{kq}\mu}^{(1)}
D_{\mu\mathbf{q}}
\hat{c}_{n,\mathbf{k+q}}^\dagger\hat{c}_{m\mathbf{k}},
\end{equation}
where
\begin{widetext}
\begin{equation}
\label{usd-r2k}
\begin{split}
&\tilde{\mathbf{u}}_{nm\mathbf{kq}\mu}^{(1)}
=i\sqrt{\lambda_{\mu\mathbf{q}}}
\sum_{\bar{\mu}}(\mathbf{r}_{\bar{m}}-\mathbf{r}_{\bar{n}}+\mathbf{R}_e)
(\hat{\Phi}_{\mathbf{q}}^\dagger)_{\bar{\mu}\mu}
\left[e^{i\mathbf{k}\cdot\bar{\mathbf{R}}_e}
(\hat{U}_\mathbf{k+q})_{n\bar{n}}
(\hat{U}_\mathbf{k}^\dagger)_{\bar{m}m}
-e^{-i(\mathbf{k+q})\cdot\bar{\mathbf{R}}_e}
(\hat{U}_\mathbf{k+q})_{n\bar{m}}
(\hat{U}_\mathbf{k}^\dagger)_{\bar{n}m}\right].
\end{split}
\end{equation}
\end{widetext}
One can easily find the resemblance between \Eq{sd-r2k} and \Eq{usd-r2k}.
The other component, $\hat{\mathbf{J}}_{sd}^{(2)}$, can be obtained by the same procedure.


\begin{thebibliography}{87}%
\makeatletter
\providecommand \@ifxundefined [1]{%
 \@ifx{#1\undefined}
}%
\providecommand \@ifnum [1]{%
 \ifnum #1\expandafter \@firstoftwo
 \else \expandafter \@secondoftwo
 \fi
}%
\providecommand \@ifx [1]{%
 \ifx #1\expandafter \@firstoftwo
 \else \expandafter \@secondoftwo
 \fi
}%
\providecommand \natexlab [1]{#1}%
\providecommand \enquote  [1]{``#1''}%
\providecommand \bibnamefont  [1]{#1}%
\providecommand \bibfnamefont [1]{#1}%
\providecommand \citenamefont [1]{#1}%
\providecommand \href@noop [0]{\@secondoftwo}%
\providecommand \href [0]{\begingroup \@sanitize@url \@href}%
\providecommand \@href[1]{\@@startlink{#1}\@@href}%
\providecommand \@@href[1]{\endgroup#1\@@endlink}%
\providecommand \@sanitize@url [0]{\catcode `\\12\catcode `\$12\catcode
  `\&12\catcode `\#12\catcode `\^12\catcode `\_12\catcode `\%12\relax}%
\providecommand \@@startlink[1]{}%
\providecommand \@@endlink[0]{}%
\providecommand \url  [0]{\begingroup\@sanitize@url \@url }%
\providecommand \@url [1]{\endgroup\@href {#1}{\urlprefix }}%
\providecommand \urlprefix  [0]{URL }%
\providecommand \Eprint [0]{\href }%
\providecommand \doibase [0]{http://dx.doi.org/}%
\providecommand \selectlanguage [0]{\@gobble}%
\providecommand \bibinfo  [0]{\@secondoftwo}%
\providecommand \bibfield  [0]{\@secondoftwo}%
\providecommand \translation [1]{[#1]}%
\providecommand \BibitemOpen [0]{}%
\providecommand \bibitemStop [0]{}%
\providecommand \bibitemNoStop [0]{.\EOS\space}%
\providecommand \EOS [0]{\spacefactor3000\relax}%
\providecommand \BibitemShut  [1]{\csname bibitem#1\endcsname}%
\let\auto@bib@innerbib\@empty
\bibitem [{\citenamefont {Brixner}\ \emph {et~al.}(2005)\citenamefont
  {Brixner}, \citenamefont {Stenger}, \citenamefont {Vaswani}, \citenamefont
  {Cho}, \citenamefont {Blankenship},\ and\ \citenamefont
  {Fleming}}]{Brixner-N-2005-625}%
  \BibitemOpen
  \bibfield  {author} {\bibinfo {author} {\bibfnamefont {T.}~\bibnamefont
  {Brixner}}, \bibinfo {author} {\bibfnamefont {J.}~\bibnamefont {Stenger}},
  \bibinfo {author} {\bibfnamefont {H.~M.}\ \bibnamefont {Vaswani}}, \bibinfo
  {author} {\bibfnamefont {M.}~\bibnamefont {Cho}}, \bibinfo {author}
  {\bibfnamefont {R.~E.}\ \bibnamefont {Blankenship}}, \ and\ \bibinfo {author}
  {\bibfnamefont {G.~R.}\ \bibnamefont {Fleming}},\ }\href@noop {} {\bibfield
  {journal} {\bibinfo  {journal} {Nature}\ }\textbf {\bibinfo {volume} {434}},\
  \bibinfo {pages} {625} (\bibinfo {year} {2005})}\BibitemShut {NoStop}%
\bibitem [{\citenamefont {Engel}\ \emph {et~al.}(2007)\citenamefont {Engel},
  \citenamefont {Calhoun}, \citenamefont {Read}, \citenamefont {Ahn},
  \citenamefont {Man{\v{c}}al}, \citenamefont {Cheng}, \citenamefont
  {Blankenship},\ and\ \citenamefont {Fleming}}]{Engel-N-2007-782}%
  \BibitemOpen
  \bibfield  {author} {\bibinfo {author} {\bibfnamefont {G.~S.}\ \bibnamefont
  {Engel}}, \bibinfo {author} {\bibfnamefont {T.~R.}\ \bibnamefont {Calhoun}},
  \bibinfo {author} {\bibfnamefont {E.~L.}\ \bibnamefont {Read}}, \bibinfo
  {author} {\bibfnamefont {T.-K.}\ \bibnamefont {Ahn}}, \bibinfo {author}
  {\bibfnamefont {T.}~\bibnamefont {Man{\v{c}}al}}, \bibinfo {author}
  {\bibfnamefont {Y.-C.}\ \bibnamefont {Cheng}}, \bibinfo {author}
  {\bibfnamefont {R.~E.}\ \bibnamefont {Blankenship}}, \ and\ \bibinfo {author}
  {\bibfnamefont {G.~R.}\ \bibnamefont {Fleming}},\ }\href@noop {} {\bibfield
  {journal} {\bibinfo  {journal} {Nature}\ }\textbf {\bibinfo {volume} {446}},\
  \bibinfo {pages} {782} (\bibinfo {year} {2007})}\BibitemShut {NoStop}%
\bibitem [{\citenamefont {Lee}\ \emph {et~al.}(2007)\citenamefont {Lee},
  \citenamefont {Cheng},\ and\ \citenamefont {Fleming}}]{Lee-S-2007-1462}%
  \BibitemOpen
  \bibfield  {author} {\bibinfo {author} {\bibfnamefont {H.}~\bibnamefont
  {Lee}}, \bibinfo {author} {\bibfnamefont {Y.-C.}\ \bibnamefont {Cheng}}, \
  and\ \bibinfo {author} {\bibfnamefont {G.~R.}\ \bibnamefont {Fleming}},\
  }\href@noop {} {\bibfield  {journal} {\bibinfo  {journal} {Science}\ }\textbf
  {\bibinfo {volume} {316}},\ \bibinfo {pages} {1462} (\bibinfo {year}
  {2007})}\BibitemShut {NoStop}%
\bibitem [{\citenamefont {Collini}\ \emph {et~al.}(2010)\citenamefont
  {Collini}, \citenamefont {Wong}, \citenamefont {Wilk}, \citenamefont {Curmi},
  \citenamefont {Brumer},\ and\ \citenamefont {Scholes}}]{Collini-N-2010-644}%
  \BibitemOpen
  \bibfield  {author} {\bibinfo {author} {\bibfnamefont {E.}~\bibnamefont
  {Collini}}, \bibinfo {author} {\bibfnamefont {C.~Y.}\ \bibnamefont {Wong}},
  \bibinfo {author} {\bibfnamefont {K.~E.}\ \bibnamefont {Wilk}}, \bibinfo
  {author} {\bibfnamefont {P.~M.~G.}\ \bibnamefont {Curmi}}, \bibinfo {author}
  {\bibfnamefont {P.}~\bibnamefont {Brumer}}, \ and\ \bibinfo {author}
  {\bibfnamefont {G.~D.}\ \bibnamefont {Scholes}},\ }\href@noop {} {\bibfield
  {journal} {\bibinfo  {journal} {Nature}\ }\textbf {\bibinfo {volume} {463}},\
  \bibinfo {pages} {644} (\bibinfo {year} {2010})}\BibitemShut {NoStop}%
\bibitem [{\citenamefont {Panitchayangkoon}\ \emph {et~al.}(2010)\citenamefont
  {Panitchayangkoon}, \citenamefont {Hayes}, \citenamefont {Fransted},
  \citenamefont {Caram}, \citenamefont {Harel}, \citenamefont {Wen},
  \citenamefont {Blankenship},\ and\ \citenamefont
  {Engel}}]{Panitchayangkoon-PNASU-2010-12766}%
  \BibitemOpen
  \bibfield  {author} {\bibinfo {author} {\bibfnamefont {G.}~\bibnamefont
  {Panitchayangkoon}}, \bibinfo {author} {\bibfnamefont {D.}~\bibnamefont
  {Hayes}}, \bibinfo {author} {\bibfnamefont {K.~A.}\ \bibnamefont {Fransted}},
  \bibinfo {author} {\bibfnamefont {J.~R.}\ \bibnamefont {Caram}}, \bibinfo
  {author} {\bibfnamefont {E.}~\bibnamefont {Harel}}, \bibinfo {author}
  {\bibfnamefont {J.}~\bibnamefont {Wen}}, \bibinfo {author} {\bibfnamefont
  {R.~E.}\ \bibnamefont {Blankenship}}, \ and\ \bibinfo {author} {\bibfnamefont
  {G.~S.}\ \bibnamefont {Engel}},\ }\href@noop {} {\bibfield  {journal}
  {\bibinfo  {journal} {Proc. Natl. Acad. Sci. U.S.A.}\ }\textbf {\bibinfo
  {volume} {107}},\ \bibinfo {pages} {12766} (\bibinfo {year}
  {2010})}\BibitemShut {NoStop}%
\bibitem [{\citenamefont {Romero}\ \emph {et~al.}(2014)\citenamefont {Romero},
  \citenamefont {Augulis}, \citenamefont {Novoderezhkin}, \citenamefont
  {Ferretti}, \citenamefont {Thieme}, \citenamefont {Zigmantas},\ and\
  \citenamefont {Van~Grondelle}}]{Romero-NP-2014-676}%
  \BibitemOpen
  \bibfield  {author} {\bibinfo {author} {\bibfnamefont {E.}~\bibnamefont
  {Romero}}, \bibinfo {author} {\bibfnamefont {R.}~\bibnamefont {Augulis}},
  \bibinfo {author} {\bibfnamefont {V.~I.}\ \bibnamefont {Novoderezhkin}},
  \bibinfo {author} {\bibfnamefont {M.}~\bibnamefont {Ferretti}}, \bibinfo
  {author} {\bibfnamefont {J.}~\bibnamefont {Thieme}}, \bibinfo {author}
  {\bibfnamefont {D.}~\bibnamefont {Zigmantas}}, \ and\ \bibinfo {author}
  {\bibfnamefont {R.}~\bibnamefont {van~Grondelle}},\ }\href@noop {} {\bibfield
   {journal} {\bibinfo  {journal} {Nat. Phys.}\ }\textbf {\bibinfo {volume}
  {10}},\ \bibinfo {pages} {676} (\bibinfo {year} {2014})}\BibitemShut
  {NoStop}%
\bibitem [{\citenamefont {Fuller}\ \emph {et~al.}(2014)\citenamefont {Fuller},
  \citenamefont {Pan}, \citenamefont {Gelzinis}, \citenamefont {Butkus},
  \citenamefont {Senlik}, \citenamefont {Wilcox}, \citenamefont {Yocum},
  \citenamefont {Valkunas}, \citenamefont {Abramavicius},\ and\ \citenamefont
  {Ogilvie}}]{Fuller-NC-2014-706}%
  \BibitemOpen
  \bibfield  {author} {\bibinfo {author} {\bibfnamefont {F.~D.}\ \bibnamefont
  {Fuller}}, \bibinfo {author} {\bibfnamefont {J.}~\bibnamefont {Pan}},
  \bibinfo {author} {\bibfnamefont {A.}~\bibnamefont {Gelzinis}}, \bibinfo
  {author} {\bibfnamefont {V.}~\bibnamefont {Butkus}}, \bibinfo {author}
  {\bibfnamefont {S.~S.}\ \bibnamefont {Senlik}}, \bibinfo {author}
  {\bibfnamefont {D.~E.}\ \bibnamefont {Wilcox}}, \bibinfo {author}
  {\bibfnamefont {C.~F.}\ \bibnamefont {Yocum}}, \bibinfo {author}
  {\bibfnamefont {L.}~\bibnamefont {Valkunas}}, \bibinfo {author}
  {\bibfnamefont {D.}~\bibnamefont {Abramavicius}}, \ and\ \bibinfo {author}
  {\bibfnamefont {J.~P.}\ \bibnamefont {Ogilvie}},\ }\href@noop {} {\bibfield
  {journal} {\bibinfo  {journal} {Nat. Chem.}\ }\textbf {\bibinfo {volume}
  {6}},\ \bibinfo {pages} {706} (\bibinfo {year} {2014})}\BibitemShut {NoStop}%
\bibitem [{\citenamefont {Collini}\ and\ \citenamefont
  {Scholes}(2009)}]{Collini-S-2009-369}%
  \BibitemOpen
  \bibfield  {author} {\bibinfo {author} {\bibfnamefont {E.}~\bibnamefont
  {Collini}}\ and\ \bibinfo {author} {\bibfnamefont {G.~D.}\ \bibnamefont
  {Scholes}},\ }\href@noop {} {\bibfield  {journal} {\bibinfo  {journal}
  {Science}\ }\textbf {\bibinfo {volume} {323}},\ \bibinfo {pages} {369}
  (\bibinfo {year} {2009})}\BibitemShut {NoStop}%
\bibitem [{\citenamefont {Karl}(2003)}]{Karl-SM-2003-649}%
  \BibitemOpen
  \bibfield  {author} {\bibinfo {author} {\bibfnamefont {N.}~\bibnamefont
  {Karl}},\ }\href@noop {} {\bibfield  {journal} {\bibinfo  {journal} {Synth.
  Met.}\ }\textbf {\bibinfo {volume} {133-134}},\ \bibinfo {pages} {649}
  (\bibinfo {year} {2003})}\BibitemShut {NoStop}%
\bibitem [{\citenamefont {Jurchescu}\ \emph {et~al.}(2004)\citenamefont
  {Jurchescu}, \citenamefont {Baas},\ and\ \citenamefont
  {Palstra}}]{Jurchescu-APL-2004-3061}%
  \BibitemOpen
  \bibfield  {author} {\bibinfo {author} {\bibfnamefont {O.~D.}\ \bibnamefont
  {Jurchescu}}, \bibinfo {author} {\bibfnamefont {J.}~\bibnamefont {Baas}}, \
  and\ \bibinfo {author} {\bibfnamefont {T.~T.~M.}\ \bibnamefont {Palstra}},\
  }\href@noop {} {\bibfield  {journal} {\bibinfo  {journal} {Appl. Phys.
  Lett.}\ }\textbf {\bibinfo {volume} {84}},\ \bibinfo {pages} {3061} (\bibinfo
  {year} {2004})}\BibitemShut {NoStop}%
\bibitem [{\citenamefont {Podzorov}\ \emph {et~al.}(2005)\citenamefont
  {Podzorov}, \citenamefont {Menard}, \citenamefont {Rogers},\ and\
  \citenamefont {Gershenson}}]{Podzorov-PRL-2005-226601}%
  \BibitemOpen
  \bibfield  {author} {\bibinfo {author} {\bibfnamefont {V.}~\bibnamefont
  {Podzorov}}, \bibinfo {author} {\bibfnamefont {E.}~\bibnamefont {Menard}},
  \bibinfo {author} {\bibfnamefont {J.~A.}\ \bibnamefont {Rogers}}, \ and\
  \bibinfo {author} {\bibfnamefont {M.~E.}\ \bibnamefont {Gershenson}},\
  }\href@noop {} {\bibfield  {journal} {\bibinfo  {journal} {Phys. Rev. Lett.}\
  }\textbf {\bibinfo {volume} {95}},\ \bibinfo {pages} {226601} (\bibinfo
  {year} {2005})}\BibitemShut {NoStop}%
\bibitem [{\citenamefont {Ostroverkhova}\ \emph {et~al.}(2006)\citenamefont
  {Ostroverkhova}, \citenamefont {Cooke}, \citenamefont {Hegmann},
  \citenamefont {Anthony}, \citenamefont {Podzorov}, \citenamefont
  {Gershenson}, \citenamefont {Jurchescu},\ and\ \citenamefont
  {Palstra}}]{Ostroverkhova-APL-2006-162101}%
  \BibitemOpen
  \bibfield  {author} {\bibinfo {author} {\bibfnamefont {O.}~\bibnamefont
  {Ostroverkhova}}, \bibinfo {author} {\bibfnamefont {D.~G.}\ \bibnamefont
  {Cooke}}, \bibinfo {author} {\bibfnamefont {F.~A.}\ \bibnamefont {Hegmann}},
  \bibinfo {author} {\bibfnamefont {J.~E.}\ \bibnamefont {Anthony}}, \bibinfo
  {author} {\bibfnamefont {V.}~\bibnamefont {Podzorov}}, \bibinfo {author}
  {\bibfnamefont {M.~E.}\ \bibnamefont {Gershenson}}, \bibinfo {author}
  {\bibfnamefont {O.~D.}\ \bibnamefont {Jurchescu}}, \ and\ \bibinfo {author}
  {\bibfnamefont {T.~T.~M.}\ \bibnamefont {Palstra}},\ }\href@noop {}
  {\bibfield  {journal} {\bibinfo  {journal} {Appl. Phys. Lett.}\ }\textbf
  {\bibinfo {volume} {88}},\ \bibinfo {pages} {162101} (\bibinfo {year}
  {2006})}\BibitemShut {NoStop}%
\bibitem [{\citenamefont {Cheng}\ \emph {et~al.}(2003)\citenamefont {Cheng},
  \citenamefont {Silbey}, \citenamefont {da~Silva~Filho}, \citenamefont
  {Calbert}, \citenamefont {Cornil},\ and\ \citenamefont
  {Br\'{e}das}}]{Cheng-JCP-2003-3764}%
  \BibitemOpen
  \bibfield  {author} {\bibinfo {author} {\bibfnamefont {Y.~C.}\ \bibnamefont
  {Cheng}}, \bibinfo {author} {\bibfnamefont {R.~J.}\ \bibnamefont {Silbey}},
  \bibinfo {author} {\bibfnamefont {D.~A.}\ \bibnamefont {da~Silva~Filho}},
  \bibinfo {author} {\bibfnamefont {J.~P.}\ \bibnamefont {Calbert}}, \bibinfo
  {author} {\bibfnamefont {J.}~\bibnamefont {Cornil}}, \ and\ \bibinfo {author}
  {\bibfnamefont {J.~L.}\ \bibnamefont {Br\'{e}das}},\ }\href@noop {}
  {\bibfield  {journal} {\bibinfo  {journal} {J. Chem. Phys.}\ }\textbf
  {\bibinfo {volume} {118}},\ \bibinfo {pages} {3764} (\bibinfo {year}
  {2003})}\BibitemShut {NoStop}%
\bibitem [{\citenamefont {Breckenridge}\ and\ \citenamefont
  {Hosler}(1953)}]{Breckenridge-PR-1953-793}%
  \BibitemOpen
  \bibfield  {author} {\bibinfo {author} {\bibfnamefont {R.~G.}\ \bibnamefont
  {Breckenridge}}\ and\ \bibinfo {author} {\bibfnamefont {W.~R.}\ \bibnamefont
  {Hosler}},\ }\href@noop {} {\bibfield  {journal} {\bibinfo  {journal} {Phys.
  Rev.}\ }\textbf {\bibinfo {volume} {91}},\ \bibinfo {pages} {793} (\bibinfo
  {year} {1953})}\BibitemShut {NoStop}%
\bibitem [{\citenamefont {Austin}\ and\ \citenamefont
  {Mott}(1969)}]{Austin-AP-1969-41}%
  \BibitemOpen
  \bibfield  {author} {\bibinfo {author} {\bibfnamefont {I.~G.}~\bibnamefont
  {Austin}}\ and\ \bibinfo {author} {\bibfnamefont {N.~F.}~\bibnamefont {Mott}},\
  }\href@noop {} {\bibfield  {journal} {\bibinfo  {journal} {Adv. Phys.}\
  }\textbf {\bibinfo {volume} {18}},\ \bibinfo {pages} {41} (\bibinfo {year}
  {1969})}\BibitemShut {NoStop}%
\bibitem [{\citenamefont {Tang}\ \emph {et~al.}(1994)\citenamefont {Tang},
  \citenamefont {Prasad}, \citenamefont {Sanjin\`{e}s}, \citenamefont
  {Schmid},\ and\ \citenamefont {L\'{e}vy}}]{Tang-JAP-1994-2042}%
  \BibitemOpen
  \bibfield  {author} {\bibinfo {author} {\bibfnamefont {H.}~\bibnamefont
  {Tang}}, \bibinfo {author} {\bibfnamefont {K.}~\bibnamefont {Prasad}},
  \bibinfo {author} {\bibfnamefont {R.}~\bibnamefont {Sanjin\`{e}s}}, \bibinfo
  {author} {\bibfnamefont {P.~E.}\ \bibnamefont {Schmid}}, \ and\ \bibinfo
  {author} {\bibfnamefont {F.}~\bibnamefont {L\'{e}vy}},\ }\href@noop {}
  {\bibfield  {journal} {\bibinfo  {journal} {J. Appl. Phys.}\ }\textbf
  {\bibinfo {volume} {75}},\ \bibinfo {pages} {2042} (\bibinfo {year}
  {1994})}\BibitemShut {NoStop}%
\bibitem [{\citenamefont {Forro}\ \emph {et~al.}(1994)\citenamefont {Forro},
  \citenamefont {Chauvet}, \citenamefont {Emin}, \citenamefont {Zuppiroli},
  \citenamefont {Berger},\ and\ \citenamefont {L\'{e}vy}}]{Forro-JAP-1994-633}%
  \BibitemOpen
  \bibfield  {author} {\bibinfo {author} {\bibfnamefont {L.}~\bibnamefont
  {Forro}}, \bibinfo {author} {\bibfnamefont {O.}~\bibnamefont {Chauvet}},
  \bibinfo {author} {\bibfnamefont {D.}~\bibnamefont {Emin}}, \bibinfo {author}
  {\bibfnamefont {L.}~\bibnamefont {Zuppiroli}}, \bibinfo {author}
  {\bibfnamefont {H.}~\bibnamefont {Berger}}, \ and\ \bibinfo {author}
  {\bibfnamefont {F.}~\bibnamefont {L\'{e}vy}},\ }\href@noop {} {\bibfield
  {journal} {\bibinfo  {journal} {J. Appl. Phys.}\ }\textbf {\bibinfo {volume}
  {75}},\ \bibinfo {pages} {633} (\bibinfo {year} {1994})}\BibitemShut
  {NoStop}%
\bibitem [{\citenamefont {Yagi}\ \emph {et~al.}(1996)\citenamefont {Yagi},
  \citenamefont {Hasiguti},\ and\ \citenamefont {Aono}}]{Yagi-PRB-1996-7945}%
  \BibitemOpen
  \bibfield  {author} {\bibinfo {author} {\bibfnamefont {E.}~\bibnamefont
  {Yagi}}, \bibinfo {author} {\bibfnamefont {R.~R.}\ \bibnamefont {Hasiguti}},
  \ and\ \bibinfo {author} {\bibfnamefont {M.}~\bibnamefont {Aono}},\
  }\href@noop {} {\bibfield  {journal} {\bibinfo  {journal} {Phys. Rev. B}\
  }\textbf {\bibinfo {volume} {54}},\ \bibinfo {pages} {7945} (\bibinfo {year}
  {1996})}\BibitemShut {NoStop}%
\bibitem [{\citenamefont {Bak}\ \emph {et~al.}(2003)\citenamefont {Bak},
  \citenamefont {Nowotny}, \citenamefont {Rekas},\ and\ \citenamefont
  {Sorrell}}]{Bak-JPCS-2003-1069}%
  \BibitemOpen
  \bibfield  {author} {\bibinfo {author} {\bibfnamefont {T.}~\bibnamefont
  {Bak}}, \bibinfo {author} {\bibfnamefont {J.}~\bibnamefont {Nowotny}},
  \bibinfo {author} {\bibfnamefont {M.}~\bibnamefont {Rekas}}, \ and\ \bibinfo
  {author} {\bibfnamefont {C.~C.}\ \bibnamefont {Sorrell}},\ }\href@noop {}
  {\bibfield  {journal} {\bibinfo  {journal} {J. Phys. Chem. Solids}\ }\textbf
  {\bibinfo {volume} {64}},\ \bibinfo {pages} {1069} (\bibinfo {year}
  {2003})}\BibitemShut {NoStop}%
\bibitem [{\citenamefont {Setvin}\ \emph {et~al.}(2014)\citenamefont {Setvin},
  \citenamefont {Franchini}, \citenamefont {Hao}, \citenamefont {Schmid},
  \citenamefont {Janotti}, \citenamefont {Kaltak}, \citenamefont {Van~de
  Walle}, \citenamefont {Kresse},\ and\ \citenamefont
  {Diebold}}]{Setvin-PRL-2014-86402}%
  \BibitemOpen
  \bibfield  {author} {\bibinfo {author} {\bibfnamefont {M.}~\bibnamefont
  {Setvin}}, \bibinfo {author} {\bibfnamefont {C.}~\bibnamefont {Franchini}},
  \bibinfo {author} {\bibfnamefont {X.}~\bibnamefont {Hao}}, \bibinfo {author}
  {\bibfnamefont {M.}~\bibnamefont {Schmid}}, \bibinfo {author} {\bibfnamefont
  {A.}~\bibnamefont {Janotti}}, \bibinfo {author} {\bibfnamefont
  {M.}~\bibnamefont {Kaltak}}, \bibinfo {author} {\bibfnamefont {C.~G.}\
  \bibnamefont {Van~de Walle}}, \bibinfo {author} {\bibfnamefont
  {G.}~\bibnamefont {Kresse}}, \ and\ \bibinfo {author} {\bibfnamefont
  {U.}~\bibnamefont {Diebold}},\ }\href@noop {} {\bibfield  {journal} {\bibinfo
   {journal} {Phys. Rev. Lett.}\ }\textbf {\bibinfo {volume} {113}},\ \bibinfo
  {pages} {086402} (\bibinfo {year} {2014})}\BibitemShut {NoStop}%
\bibitem [{\citenamefont {Yang}\ \emph {et~al.}(2013)\citenamefont {Yang},
  \citenamefont {Brant}, \citenamefont {Giles},\ and\ \citenamefont
  {Halliburton}}]{Yang-PRB-2013-125201}%
  \BibitemOpen
  \bibfield  {author} {\bibinfo {author} {\bibfnamefont {S.}~\bibnamefont
  {Yang}}, \bibinfo {author} {\bibfnamefont {A.~T.}\ \bibnamefont {Brant}},
  \bibinfo {author} {\bibfnamefont {N.~C.}\ \bibnamefont {Giles}}, \ and\
  \bibinfo {author} {\bibfnamefont {L.~E.}\ \bibnamefont {Halliburton}},\
  }\href@noop {} {\bibfield  {journal} {\bibinfo  {journal} {Phys. Rev. B}\
  }\textbf {\bibinfo {volume} {87}},\ \bibinfo {pages} {125201} (\bibinfo
  {year} {2013})}\BibitemShut {NoStop}%
\bibitem [{\citenamefont {De\'{a}k}\ \emph {et~al.}(2011)\citenamefont
  {De\'{a}k}, \citenamefont {Aradi},\ and\ \citenamefont
  {Frauenheim}}]{Deak-PRB-2011-155207}%
  \BibitemOpen
  \bibfield  {author} {\bibinfo {author} {\bibfnamefont {P.}~\bibnamefont
  {De\'{a}k}}, \bibinfo {author} {\bibfnamefont {B.}~\bibnamefont {Aradi}}, \
  and\ \bibinfo {author} {\bibfnamefont {T.}~\bibnamefont {Frauenheim}},\
  }\href@noop {} {\bibfield  {journal} {\bibinfo  {journal} {Phys. Rev. B}\
  }\textbf {\bibinfo {volume} {83}},\ \bibinfo {pages} {155207} (\bibinfo
  {year} {2011})}\BibitemShut {NoStop}%
\bibitem [{\citenamefont {De\'{a}k}\ \emph {et~al.}(2012)\citenamefont
  {De\'{a}k}, \citenamefont {Aradi},\ and\ \citenamefont
  {Frauenheim}}]{Deak-PRB-2012-195206}%
  \BibitemOpen
  \bibfield  {author} {\bibinfo {author} {\bibfnamefont {P.}~\bibnamefont
  {De\'{a}k}}, \bibinfo {author} {\bibfnamefont {B.}~\bibnamefont {Aradi}}, \
  and\ \bibinfo {author} {\bibfnamefont {T.}~\bibnamefont {Frauenheim}},\
  }\href@noop {} {\bibfield  {journal} {\bibinfo  {journal} {Phys. Rev. B}\
  }\textbf {\bibinfo {volume} {86}},\ \bibinfo {pages} {195206} (\bibinfo
  {year} {2012})}\BibitemShut {NoStop}%
\bibitem [{\citenamefont {Di~Valentin}\ \emph {et~al.}(2009)\citenamefont
  {Di~Valentin}, \citenamefont {Pacchioni},\ and\ \citenamefont
  {Selloni}}]{DiValentin-JPCC-2009-20543}%
  \BibitemOpen
  \bibfield  {author} {\bibinfo {author} {\bibfnamefont {C.}~\bibnamefont
  {Di~Valentin}}, \bibinfo {author} {\bibfnamefont {G.}~\bibnamefont
  {Pacchioni}}, \ and\ \bibinfo {author} {\bibfnamefont {A.}~\bibnamefont
  {Selloni}},\ }\href@noop {} {\bibfield  {journal} {\bibinfo  {journal} {J.
  Phys. Chem. C}\ }\textbf {\bibinfo {volume} {113}},\ \bibinfo {pages} {20543}
  (\bibinfo {year} {2009})}\BibitemShut {NoStop}%
\bibitem [{\citenamefont {Tanimura}\ and\ \citenamefont
  {Kubo}(1989)}]{Tanimura-JPSJ-1989-101}%
  \BibitemOpen
  \bibfield  {author} {\bibinfo {author} {\bibfnamefont {Y.}~\bibnamefont
  {Tanimura}}\ and\ \bibinfo {author} {\bibfnamefont {R.}~\bibnamefont
  {Kubo}},\ }\href@noop {} {\bibfield  {journal} {\bibinfo  {journal} {J. Phys.
  Soc. Jpn.}\ }\textbf {\bibinfo {volume} {58}},\ \bibinfo {pages} {101}
  (\bibinfo {year} {1989})}\BibitemShut {NoStop}%
\bibitem [{\citenamefont {Ishizaki}\ and\ \citenamefont
  {Fleming}(2009)}]{Ishizaki-PNASU-2009-17255}%
  \BibitemOpen
  \bibfield  {author} {\bibinfo {author} {\bibfnamefont {A.}~\bibnamefont
  {Ishizaki}}\ and\ \bibinfo {author} {\bibfnamefont {G.~R.}\ \bibnamefont
  {Fleming}},\ }\href@noop {} {\bibfield  {journal} {\bibinfo  {journal} {Proc.
  Natl. Acad. Sci. U.S.A.}\ }\textbf {\bibinfo {volume} {106}},\ \bibinfo
  {pages} {17255} (\bibinfo {year} {2009})}\BibitemShut {NoStop}%
\bibitem [{\citenamefont {Tanimura}(2020)}]{Tanimura-JCP-2020-20901}%
  \BibitemOpen
  \bibfield  {author} {\bibinfo {author} {\bibfnamefont {Y.}~\bibnamefont
  {Tanimura}},\ }\href@noop {} {\bibfield  {journal} {\bibinfo  {journal} {J.
  Chem. Phys.}\ }\textbf {\bibinfo {volume} {153}},\ \bibinfo {pages} {020901}
  (\bibinfo {year} {2020})}\BibitemShut {NoStop}%
\bibitem [{\citenamefont {Makri}\ and\ \citenamefont
  {Makarov}(1995{\natexlab{a}})}]{Makri-JCP-1995-4600}%
  \BibitemOpen
  \bibfield  {author} {\bibinfo {author} {\bibfnamefont {N.}~\bibnamefont
  {Makri}}\ and\ \bibinfo {author} {\bibfnamefont {D.~E.}\ \bibnamefont
  {Makarov}},\ }\href@noop {} {\bibfield  {journal} {\bibinfo  {journal} {J.
  Chem. Phys.}\ }\textbf {\bibinfo {volume} {102}},\ \bibinfo {pages} {4600}
  (\bibinfo {year} {1995}{\natexlab{a}})}\BibitemShut {NoStop}%
\bibitem [{\citenamefont {Makri}\ and\ \citenamefont
  {Makarov}(1995{\natexlab{b}})}]{Makri-JCP-1995-4611}%
  \BibitemOpen
  \bibfield  {author} {\bibinfo {author} {\bibfnamefont {N.}~\bibnamefont
  {Makri}}\ and\ \bibinfo {author} {\bibfnamefont {D.~E.}\ \bibnamefont
  {Makarov}},\ }\href@noop {} {\bibfield  {journal} {\bibinfo  {journal} {J.
  Chem. Phys.}\ }\textbf {\bibinfo {volume} {102}},\ \bibinfo {pages} {4611}
  (\bibinfo {year} {1995}{\natexlab{b}})}\BibitemShut {NoStop}%
\bibitem [{\citenamefont {Fujita}\ \emph {et~al.}(2012)\citenamefont {Fujita},
  \citenamefont {Brookes}, \citenamefont {Saikin},\ and\ \citenamefont
  {Aspuru-Guzik}}]{Fujita-JPCL-2012-2357}%
  \BibitemOpen
  \bibfield  {author} {\bibinfo {author} {\bibfnamefont {T.}~\bibnamefont
  {Fujita}}, \bibinfo {author} {\bibfnamefont {J.~C.}\ \bibnamefont {Brookes}},
  \bibinfo {author} {\bibfnamefont {S.~K.}\ \bibnamefont {Saikin}}, \ and\
  \bibinfo {author} {\bibfnamefont {A.}~\bibnamefont {Aspuru-Guzik}},\
  }\href@noop {} {\bibfield  {journal} {\bibinfo  {journal} {J. Phys. Chem.
  Lett.}\ }\textbf {\bibinfo {volume} {3}},\ \bibinfo {pages} {2357} (\bibinfo
  {year} {2012})}\BibitemShut {NoStop}%
\bibitem [{\citenamefont {Fujita}\ \emph {et~al.}(2014)\citenamefont {Fujita},
  \citenamefont {Huh}, \citenamefont {Saikin}, \citenamefont {Brookes},\ and\
  \citenamefont {Aspuru-Guzik}}]{Fujita-PR-2014-273}%
  \BibitemOpen
  \bibfield  {author} {\bibinfo {author} {\bibfnamefont {T.}~\bibnamefont
  {Fujita}}, \bibinfo {author} {\bibfnamefont {J.}~\bibnamefont {Huh}},
  \bibinfo {author} {\bibfnamefont {S.~K.}\ \bibnamefont {Saikin}}, \bibinfo
  {author} {\bibfnamefont {J.~C.}\ \bibnamefont {Brookes}}, \ and\ \bibinfo
  {author} {\bibfnamefont {A.}~\bibnamefont {Aspuru-Guzik}},\ }\href@noop {}
  {\bibfield  {journal} {\bibinfo  {journal} {Photosynth. Res.}\ }\textbf
  {\bibinfo {volume} {120}},\ \bibinfo {pages} {273} (\bibinfo {year}
  {2014})}\BibitemShut {NoStop}%
\bibitem [{\citenamefont {Huh}\ \emph {et~al.}(2014)\citenamefont {Huh},
  \citenamefont {Saikin}, \citenamefont {Brookes}, \citenamefont {Valleau},
  \citenamefont {Fujita},\ and\ \citenamefont
  {Aspuru-Guzik}}]{Huh-JACS-2014-2048}%
  \BibitemOpen
  \bibfield  {author} {\bibinfo {author} {\bibfnamefont {J.}~\bibnamefont
  {Huh}}, \bibinfo {author} {\bibfnamefont {S.~K.}\ \bibnamefont {Saikin}},
  \bibinfo {author} {\bibfnamefont {J.~C.}\ \bibnamefont {Brookes}}, \bibinfo
  {author} {\bibfnamefont {S.}~\bibnamefont {Valleau}}, \bibinfo {author}
  {\bibfnamefont {T.}~\bibnamefont {Fujita}}, \ and\ \bibinfo {author}
  {\bibfnamefont {A.}~\bibnamefont {Aspuru-Guzik}},\ }\href@noop {} {\bibfield
  {journal} {\bibinfo  {journal} {J. Am. Chem. Soc.}\ }\textbf {\bibinfo
  {volume} {136}},\ \bibinfo {pages} {2048} (\bibinfo {year}
  {2014})}\BibitemShut {NoStop}%
\bibitem [{\citenamefont {Sawaya}\ \emph {et~al.}(2015)\citenamefont {Sawaya},
  \citenamefont {Huh}, \citenamefont {Fujita}, \citenamefont {Saikin},\ and\
  \citenamefont {Aspuru-Guzik}}]{Sawaya-NL-2015-1722}%
  \BibitemOpen
  \bibfield  {author} {\bibinfo {author} {\bibfnamefont {N.~P.~D.}\
  \bibnamefont {Sawaya}}, \bibinfo {author} {\bibfnamefont {J.}~\bibnamefont
  {Huh}}, \bibinfo {author} {\bibfnamefont {T.}~\bibnamefont {Fujita}},
  \bibinfo {author} {\bibfnamefont {S.~K.}\ \bibnamefont {Saikin}}, \ and\
  \bibinfo {author} {\bibfnamefont {A.}~\bibnamefont {Aspuru-Guzik}},\
  }\href@noop {} {\bibfield  {journal} {\bibinfo  {journal} {Nano Lett.}\
  }\textbf {\bibinfo {volume} {15}},\ \bibinfo {pages} {1722} (\bibinfo {year}
  {2015})}\BibitemShut {NoStop}%
\bibitem [{\citenamefont {Li}\ \emph {et~al.}(2020{\natexlab{a}})\citenamefont
  {Li}, \citenamefont {Buda}, \citenamefont {de~Groot},\ and\ \citenamefont
  {Sevink}}]{Li-JPCB-2020-4026}%
  \BibitemOpen
  \bibfield  {author} {\bibinfo {author} {\bibfnamefont {X.}~\bibnamefont
  {Li}}, \bibinfo {author} {\bibfnamefont {F.}~\bibnamefont {Buda}}, \bibinfo
  {author} {\bibfnamefont {H.~J.~M.}\ \bibnamefont {de~Groot}}, \ and\ \bibinfo
  {author} {\bibfnamefont {G.~J.~A.}\ \bibnamefont {Sevink}},\ }\href@noop {}
  {\bibfield  {journal} {\bibinfo  {journal} {J. Phys. Chem. B}\ }\textbf
  {\bibinfo {volume} {124}},\ \bibinfo {pages} {4026} (\bibinfo {year}
  {2020}{\natexlab{a}})}\BibitemShut {NoStop}%
\bibitem [{\citenamefont {Nan}\ \emph {et~al.}(2009)\citenamefont {Nan},
  \citenamefont {Yang}, \citenamefont {Wang}, \citenamefont {Shuai},\ and\
  \citenamefont {Zhao}}]{Nan-PRB-2009-115203}%
  \BibitemOpen
  \bibfield  {author} {\bibinfo {author} {\bibfnamefont {G.}~\bibnamefont
  {Nan}}, \bibinfo {author} {\bibfnamefont {X.}~\bibnamefont {Yang}}, \bibinfo
  {author} {\bibfnamefont {L.}~\bibnamefont {Wang}}, \bibinfo {author}
  {\bibfnamefont {Z.}~\bibnamefont {Shuai}}, \ and\ \bibinfo {author}
  {\bibfnamefont {Y.}~\bibnamefont {Zhao}},\ }\href@noop {} {\bibfield
  {journal} {\bibinfo  {journal} {Phys. Rev. B}\ }\textbf {\bibinfo {volume}
  {79}},\ \bibinfo {pages} {115203} (\bibinfo {year} {2009})}\BibitemShut
  {NoStop}%
\bibitem [{\citenamefont {Hannewald}\ and\ \citenamefont
  {Bobbert}(2004{\natexlab{a}})}]{Hannewald-PRB-2004-75212}%
  \BibitemOpen
  \bibfield  {author} {\bibinfo {author} {\bibfnamefont {K.}~\bibnamefont
  {Hannewald}}\ and\ \bibinfo {author} {\bibfnamefont {P.~A.}\ \bibnamefont
  {Bobbert}},\ }\href@noop {} {\bibfield  {journal} {\bibinfo  {journal} {Phys.
  Rev. B}\ }\textbf {\bibinfo {volume} {69}},\ \bibinfo {pages} {075212}
  (\bibinfo {year} {2004}{\natexlab{a}})}\BibitemShut {NoStop}%
\bibitem [{\citenamefont {Wang}\ \emph {et~al.}(2007)\citenamefont {Wang},
  \citenamefont {Peng}, \citenamefont {Li},\ and\ \citenamefont
  {Shuai}}]{Wang-JCP-2007-44506}%
  \BibitemOpen
  \bibfield  {author} {\bibinfo {author} {\bibfnamefont {L.~J.}\ \bibnamefont
  {Wang}}, \bibinfo {author} {\bibfnamefont {Q.}~\bibnamefont {Peng}}, \bibinfo
  {author} {\bibfnamefont {Q.~K.}\ \bibnamefont {Li}}, \ and\ \bibinfo {author}
  {\bibfnamefont {Z.}~\bibnamefont {Shuai}},\ }\href@noop {} {\bibfield
  {journal} {\bibinfo  {journal} {J. Chem. Phys.}\ }\textbf {\bibinfo {volume}
  {127}},\ \bibinfo {pages} {044506} (\bibinfo {year} {2007})}\BibitemShut
  {NoStop}%
\bibitem [{\citenamefont {Troisi}\ and\ \citenamefont
  {Orlandi}(2006)}]{Troisi-PRL-2006-86601}%
  \BibitemOpen
  \bibfield  {author} {\bibinfo {author} {\bibfnamefont {A.}~\bibnamefont
  {Troisi}}\ and\ \bibinfo {author} {\bibfnamefont {G.}~\bibnamefont
  {Orlandi}},\ }\href@noop {} {\bibfield  {journal} {\bibinfo  {journal} {Phys.
  Rev. Lett.}\ }\textbf {\bibinfo {volume} {96}},\ \bibinfo {pages} {086601}
  (\bibinfo {year} {2006})}\BibitemShut {NoStop}%
\bibitem [{\citenamefont {Ciuchi}\ \emph {et~al.}(2011)\citenamefont {Ciuchi},
  \citenamefont {Fratini},\ and\ \citenamefont
  {Mayou}}]{Ciuchi-PRB-2011-81202}%
  \BibitemOpen
  \bibfield  {author} {\bibinfo {author} {\bibfnamefont {S.}~\bibnamefont
  {Ciuchi}}, \bibinfo {author} {\bibfnamefont {S.}~\bibnamefont {Fratini}}, \
  and\ \bibinfo {author} {\bibfnamefont {D.}~\bibnamefont {Mayou}},\
  }\href@noop {} {\bibfield  {journal} {\bibinfo  {journal} {Phys. Rev. B}\
  }\textbf {\bibinfo {volume} {83}},\ \bibinfo {pages} {081202} (\bibinfo
  {year} {2011})}\BibitemShut {NoStop}%
\bibitem [{\citenamefont {Ciuchi}\ and\ \citenamefont
  {Fratini}(2012)}]{Ciuchi-PRB-2012-245201}%
  \BibitemOpen
  \bibfield  {author} {\bibinfo {author} {\bibfnamefont {S.}~\bibnamefont
  {Ciuchi}}\ and\ \bibinfo {author} {\bibfnamefont {S.}~\bibnamefont
  {Fratini}},\ }\href@noop {} {\bibfield  {journal} {\bibinfo  {journal} {Phys.
  Rev. B}\ }\textbf {\bibinfo {volume} {86}},\ \bibinfo {pages} {245201}
  (\bibinfo {year} {2012})}\BibitemShut {NoStop}%
\bibitem [{\citenamefont {Fratini}\ \emph {et~al.}(2016)\citenamefont
  {Fratini}, \citenamefont {Mayou},\ and\ \citenamefont
  {Ciuchi}}]{Fratini-AFM-2016-2292}%
  \BibitemOpen
  \bibfield  {author} {\bibinfo {author} {\bibfnamefont {S.}~\bibnamefont
  {Fratini}}, \bibinfo {author} {\bibfnamefont {D.}~\bibnamefont {Mayou}}, \
  and\ \bibinfo {author} {\bibfnamefont {S.}~\bibnamefont {Ciuchi}},\
  }\href@noop {} {\bibfield  {journal} {\bibinfo  {journal} {Adv. Funct.
  Mater.}\ }\textbf {\bibinfo {volume} {26}},\ \bibinfo {pages} {2292}
  (\bibinfo {year} {2016})}\BibitemShut {NoStop}%
\bibitem [{\citenamefont {Fratini}\ \emph {et~al.}(2017)\citenamefont
  {Fratini}, \citenamefont {Ciuchi}, \citenamefont {Mayou}, \citenamefont
  {de~Laissardi\`{e}re},\ and\ \citenamefont {Troisi}}]{Fratini-NM-2017-998}%
  \BibitemOpen
  \bibfield  {author} {\bibinfo {author} {\bibfnamefont {S.}~\bibnamefont
  {Fratini}}, \bibinfo {author} {\bibfnamefont {S.}~\bibnamefont {Ciuchi}},
  \bibinfo {author} {\bibfnamefont {D.}~\bibnamefont {Mayou}}, \bibinfo
  {author} {\bibfnamefont {G.~T.}\ \bibnamefont {de~Laissardi\`{e}re}}, \ and\
  \bibinfo {author} {\bibfnamefont {A.}~\bibnamefont {Troisi}},\ }\href@noop {}
  {\bibfield  {journal} {\bibinfo  {journal} {Nat. Mater.}\ }\textbf {\bibinfo
  {volume} {16}},\ \bibinfo {pages} {998} (\bibinfo {year} {2017})}\BibitemShut
  {NoStop}%
\bibitem [{\citenamefont {Zhong}\ \emph {et~al.}(2014)\citenamefont {Zhong},
  \citenamefont {Zhao},\ and\ \citenamefont {Cao}}]{Zhong-NJP-2014-45009}%
  \BibitemOpen
  \bibfield  {author} {\bibinfo {author} {\bibfnamefont {X.}~\bibnamefont
  {Zhong}}, \bibinfo {author} {\bibfnamefont {Y.}~\bibnamefont {Zhao}}, \ and\
  \bibinfo {author} {\bibfnamefont {J.}~\bibnamefont {Cao}},\ }\href@noop {}
  {\bibfield  {journal} {\bibinfo  {journal} {New J. Phys.}\ }\textbf {\bibinfo
  {volume} {16}},\ \bibinfo {pages} {045009} (\bibinfo {year}
  {2014})}\BibitemShut {NoStop}%
\bibitem [{\citenamefont {Jiang}\ \emph {et~al.}(2016)\citenamefont {Jiang},
  \citenamefont {Zhong}, \citenamefont {Shi}, \citenamefont {Peng},
  \citenamefont {Geng}, \citenamefont {Zhao},\ and\ \citenamefont
  {Shuai}}]{Jiang-NH-2016-53}%
  \BibitemOpen
  \bibfield  {author} {\bibinfo {author} {\bibfnamefont {Y.}~\bibnamefont
  {Jiang}}, \bibinfo {author} {\bibfnamefont {X.}~\bibnamefont {Zhong}},
  \bibinfo {author} {\bibfnamefont {W.}~\bibnamefont {Shi}}, \bibinfo {author}
  {\bibfnamefont {Q.}~\bibnamefont {Peng}}, \bibinfo {author} {\bibfnamefont
  {H.}~\bibnamefont {Geng}}, \bibinfo {author} {\bibfnamefont {Y.}~\bibnamefont
  {Zhao}}, \ and\ \bibinfo {author} {\bibfnamefont {Z.}~\bibnamefont {Shuai}},\
  }\href@noop {} {\bibfield  {journal} {\bibinfo  {journal} {Nanoscale Horiz.}\
  }\textbf {\bibinfo {volume} {1}},\ \bibinfo {pages} {53} (\bibinfo {year}
  {2016})}\BibitemShut {NoStop}%
\bibitem [{\citenamefont {Lian}\ \emph {et~al.}(2019)\citenamefont {Lian},
  \citenamefont {Wang}, \citenamefont {Ke},\ and\ \citenamefont
  {Zhao}}]{Lian-JCP-2019-44115}%
  \BibitemOpen
  \bibfield  {author} {\bibinfo {author} {\bibfnamefont {M.}~\bibnamefont
  {Lian}}, \bibinfo {author} {\bibfnamefont {Y.-C.}\ \bibnamefont {Wang}},
  \bibinfo {author} {\bibfnamefont {Y.}~\bibnamefont {Ke}}, \ and\ \bibinfo
  {author} {\bibfnamefont {Y.}~\bibnamefont {Zhao}},\ }\href@noop {} {\bibfield
   {journal} {\bibinfo  {journal} {J. Chem. Phys.}\ }\textbf {\bibinfo {volume}
  {151}},\ \bibinfo {pages} {044115} (\bibinfo {year} {2019})}\BibitemShut
  {NoStop}%
\bibitem [{\citenamefont {Wang}\ \emph {et~al.}(2010)\citenamefont {Wang},
  \citenamefont {Chen}, \citenamefont {Zheng}, \citenamefont {Wang},\ and\
  \citenamefont {Shi}}]{Wang-JCP-2010-81101}%
  \BibitemOpen
  \bibfield  {author} {\bibinfo {author} {\bibfnamefont {D.}~\bibnamefont
  {Wang}}, \bibinfo {author} {\bibfnamefont {L.}~\bibnamefont {Chen}}, \bibinfo
  {author} {\bibfnamefont {R.}~\bibnamefont {Zheng}}, \bibinfo {author}
  {\bibfnamefont {L.}~\bibnamefont {Wang}}, \ and\ \bibinfo {author}
  {\bibfnamefont {Q.}~\bibnamefont {Shi}},\ }\href@noop {} {\bibfield
  {journal} {\bibinfo  {journal} {J. Chem. Phys.}\ }\textbf {\bibinfo {volume}
  {132}},\ \bibinfo {pages} {081101} (\bibinfo {year} {2010})}\BibitemShut
  {NoStop}%
\bibitem [{\citenamefont {Li}\ \emph {et~al.}(2020{\natexlab{b}})\citenamefont
  {Li}, \citenamefont {Ren},\ and\ \citenamefont {Shuai}}]{Li-JPCL-2020-4930}%
  \BibitemOpen
  \bibfield  {author} {\bibinfo {author} {\bibfnamefont {W.}~\bibnamefont
  {Li}}, \bibinfo {author} {\bibfnamefont {J.}~\bibnamefont {Ren}}, \ and\
  \bibinfo {author} {\bibfnamefont {Z.}~\bibnamefont {Shuai}},\ }\href@noop {}
  {\bibfield  {journal} {\bibinfo  {journal} {J. Phys. Chem. Lett.}\ }\textbf
  {\bibinfo {volume} {11}},\ \bibinfo {pages} {4930} (\bibinfo {year}
  {2020}{\natexlab{b}})}\BibitemShut {NoStop}%
\bibitem [{\citenamefont {Li}\ \emph {et~al.}(2021)\citenamefont {Li},
  \citenamefont {Ren},\ and\ \citenamefont {Shuai}}]{Li-NC-2021-4260}%
  \BibitemOpen
  \bibfield  {author} {\bibinfo {author} {\bibfnamefont {W.}~\bibnamefont
  {Li}}, \bibinfo {author} {\bibfnamefont {J.}~\bibnamefont {Ren}}, \ and\
  \bibinfo {author} {\bibfnamefont {Z.}~\bibnamefont {Shuai}},\ }\href@noop {}
  {\bibfield  {journal} {\bibinfo  {journal} {Nat. Commun.}\ }\textbf {\bibinfo
  {volume} {12}},\ \bibinfo {pages} {4260} (\bibinfo {year}
  {2021})}\BibitemShut {NoStop}%
\bibitem [{\citenamefont {Hendry}\ \emph {et~al.}(2004)\citenamefont {Hendry},
  \citenamefont {Wang}, \citenamefont {Shan}, \citenamefont {Heinz},\ and\
  \citenamefont {Bonn}}]{Hendry-PRB-2004-81101}%
  \BibitemOpen
  \bibfield  {author} {\bibinfo {author} {\bibfnamefont {E.}~\bibnamefont
  {Hendry}}, \bibinfo {author} {\bibfnamefont {F.}~\bibnamefont {Wang}},
  \bibinfo {author} {\bibfnamefont {J.}~\bibnamefont {Shan}}, \bibinfo {author}
  {\bibfnamefont {T.~F.}\ \bibnamefont {Heinz}}, \ and\ \bibinfo {author}
  {\bibfnamefont {M.}~\bibnamefont {Bonn}},\ }\href@noop {} {\bibfield
  {journal} {\bibinfo  {journal} {Phys. Rev. B}\ }\textbf {\bibinfo {volume}
  {69}},\ \bibinfo {pages} {081101} (\bibinfo {year} {2004})}\BibitemShut
  {NoStop}%
\bibitem [{\citenamefont {Persson}\ and\ \citenamefont {Ferreira~da
  Silva}(2005)}]{Persson-APL-2005-231912}%
  \BibitemOpen
  \bibfield  {author} {\bibinfo {author} {\bibfnamefont {C.}~\bibnamefont
  {Persson}}\ and\ \bibinfo {author} {\bibfnamefont {A.}~\bibnamefont
  {Ferreira~da Silva}},\ }\href@noop {} {\bibfield  {journal} {\bibinfo
  {journal} {Appl. Phys. Lett.}\ }\textbf {\bibinfo {volume} {86}},\ \bibinfo
  {pages} {231912} (\bibinfo {year} {2005})}\BibitemShut {NoStop}%
\bibitem [{\citenamefont {Moser}\ \emph {et~al.}(2013)\citenamefont {Moser},
  \citenamefont {Moreschini}, \citenamefont {Ja\'{c}imovi\'{c}}, \citenamefont
  {Bari\v{s}i\'{c}}, \citenamefont {Berger}, \citenamefont {Magrez},
  \citenamefont {Chang}, \citenamefont {Kim}, \citenamefont {Bostwick},
  \citenamefont {Rotenberg}, \citenamefont {Forr\'{o}},\ and\ \citenamefont
  {Grioni}}]{Moser-PRL-2013-196403}%
  \BibitemOpen
  \bibfield  {author} {\bibinfo {author} {\bibfnamefont {S.}~\bibnamefont
  {Moser}}, \bibinfo {author} {\bibfnamefont {L.}~\bibnamefont {Moreschini}},
  \bibinfo {author} {\bibfnamefont {J.}~\bibnamefont {Ja\'{c}imovi\'{c}}},
  \bibinfo {author} {\bibfnamefont {O.~S.}\ \bibnamefont {Bari\v{s}i\'{c}}},
  \bibinfo {author} {\bibfnamefont {H.}~\bibnamefont {Berger}}, \bibinfo
  {author} {\bibfnamefont {A.}~\bibnamefont {Magrez}}, \bibinfo {author}
  {\bibfnamefont {Y.~J.}\ \bibnamefont {Chang}}, \bibinfo {author}
  {\bibfnamefont {K.~S.}\ \bibnamefont {Kim}}, \bibinfo {author} {\bibfnamefont
  {A.}~\bibnamefont {Bostwick}}, \bibinfo {author} {\bibfnamefont
  {E.}~\bibnamefont {Rotenberg}}, \bibinfo {author} {\bibfnamefont
  {L.}~\bibnamefont {Forr\'{o}}}, \ and\ \bibinfo {author} {\bibfnamefont
  {M.}~\bibnamefont {Grioni}},\ }\href@noop {} {\bibfield  {journal} {\bibinfo
  {journal} {Phys. Rev. Lett.}\ }\textbf {\bibinfo {volume} {110}},\ \bibinfo
  {pages} {196403} (\bibinfo {year} {2013})}\BibitemShut {NoStop}%
\bibitem [{\citenamefont {Verdi}\ and\ \citenamefont
  {Giustino}(2015)}]{Verdi-PRL-2015-176401}%
  \BibitemOpen
  \bibfield  {author} {\bibinfo {author} {\bibfnamefont {C.}~\bibnamefont
  {Verdi}}\ and\ \bibinfo {author} {\bibfnamefont {F.}~\bibnamefont
  {Giustino}},\ }\href@noop {} {\bibfield  {journal} {\bibinfo  {journal}
  {Phys. Rev. Lett.}\ }\textbf {\bibinfo {volume} {115}},\ \bibinfo {pages}
  {176401} (\bibinfo {year} {2015})}\BibitemShut {NoStop}%
\bibitem [{\citenamefont {Himmetoglu}\ and\ \citenamefont
  {Janotti}(2016)}]{Himmetoglu-JPCM-2016-65502}%
  \BibitemOpen
  \bibfield  {author} {\bibinfo {author} {\bibfnamefont {B.}~\bibnamefont
  {Himmetoglu}}\ and\ \bibinfo {author} {\bibfnamefont {A.}~\bibnamefont
  {Janotti}},\ }\href@noop {} {\bibfield  {journal} {\bibinfo  {journal} {J.
  Phys.: Condens. Matter}\ }\textbf {\bibinfo {volume} {28}},\ \bibinfo {pages}
  {065502} (\bibinfo {year} {2016})}\BibitemShut {NoStop}%
\bibitem [{\citenamefont {Franchini}\ \emph {et~al.}(2021)\citenamefont
  {Franchini}, \citenamefont {Reticcioli}, \citenamefont {Setvin},\ and\
  \citenamefont {Diebold}}]{Franchini-NRM-2021-1}%
  \BibitemOpen
  \bibfield  {author} {\bibinfo {author} {\bibfnamefont {C.}~\bibnamefont
  {Franchini}}, \bibinfo {author} {\bibfnamefont {M.}~\bibnamefont
  {Reticcioli}}, \bibinfo {author} {\bibfnamefont {M.}~\bibnamefont {Setvin}},
  \ and\ \bibinfo {author} {\bibfnamefont {U.}~\bibnamefont {Diebold}},\
  }\href@noop {} {\bibfield  {journal} {\bibinfo  {journal} {Nat. Rev. Mater.}\
  }\textbf {\bibinfo {volume} {6}},\ \bibinfo {pages} {560} (\bibinfo {year}
  {2021})}\BibitemShut {NoStop}%
\bibitem [{\citenamefont {Deskins}\ and\ \citenamefont
  {Dupuis}(2007)}]{Deskins-PRB-2007-195212}%
  \BibitemOpen
  \bibfield  {author} {\bibinfo {author} {\bibfnamefont {N.~A.}\ \bibnamefont
  {Deskins}}\ and\ \bibinfo {author} {\bibfnamefont {M.}~\bibnamefont
  {Dupuis}},\ }\href@noop {} {\bibfield  {journal} {\bibinfo  {journal} {Phys.
  Rev. B}\ }\textbf {\bibinfo {volume} {75}},\ \bibinfo {pages} {195212}
  (\bibinfo {year} {2007})}\BibitemShut {NoStop}%
\bibitem [{\citenamefont {Spreafico}\ and\ \citenamefont
  {VandeVondele}(2014)}]{Spreafico-PCCP-2014-26144}%
  \BibitemOpen
  \bibfield  {author} {\bibinfo {author} {\bibfnamefont {C.}~\bibnamefont
  {Spreafico}}\ and\ \bibinfo {author} {\bibfnamefont {J.}~\bibnamefont
  {VandeVondele}},\ }\href@noop {} {\bibfield  {journal} {\bibinfo  {journal}
  {Phys. Chem. Chem. Phys.}\ }\textbf {\bibinfo {volume} {16}},\ \bibinfo
  {pages} {26144} (\bibinfo {year} {2014})}\BibitemShut {NoStop}%
\bibitem [{\citenamefont {Mahan}(2013)}]{Mahan--2013-}%
  \BibitemOpen
  \bibfield  {author} {\bibinfo {author} {\bibfnamefont {G.~D.}\ \bibnamefont
  {Mahan}},\ }\href@noop {} {\emph {\bibinfo {title} {Many-particle physics}}}\
  (\bibinfo  {publisher} {Springer Science \& Business Media},\ \bibinfo {year}
  {2013})\BibitemShut {NoStop}%
\bibitem [{\citenamefont {Prokof'ev}\ \emph {et~al.}(1996)\citenamefont
  {Prokof'ev}, \citenamefont {Svistunov},\ and\ \citenamefont
  {Tupitsyn}}]{Prokofev-JL-1996-911}%
  \BibitemOpen
  \bibfield  {author} {\bibinfo {author} {\bibfnamefont {N.~V.}\ \bibnamefont
  {Prokof'ev}}, \bibinfo {author} {\bibfnamefont {B.~V.}\ \bibnamefont
  {Svistunov}}, \ and\ \bibinfo {author} {\bibfnamefont {I.~S.}\ \bibnamefont
  {Tupitsyn}},\ }\href@noop {} {\bibfield  {journal} {\bibinfo  {journal} {JETP
  Lett.}\ }\textbf {\bibinfo {volume} {64}},\ \bibinfo {pages} {911} (\bibinfo
  {year} {1996})}\BibitemShut {NoStop}%
\bibitem [{\citenamefont {Beard}\ and\ \citenamefont
  {Wiese}(1996)}]{Beard-PRL-1996-5130}%
  \BibitemOpen
  \bibfield  {author} {\bibinfo {author} {\bibfnamefont {B.~B.}\ \bibnamefont
  {Beard}}\ and\ \bibinfo {author} {\bibfnamefont {U.-J.}\ \bibnamefont
  {Wiese}},\ }\href@noop {} {\bibfield  {journal} {\bibinfo  {journal} {Phys.
  Rev. Lett.}\ }\textbf {\bibinfo {volume} {77}},\ \bibinfo {pages} {5130}
  (\bibinfo {year} {1996})}\BibitemShut {NoStop}%
\bibitem [{\citenamefont {Prokof'ev}\ and\ \citenamefont
  {Svistunov}(1998)}]{Prokofev-PRL-1998-2514}%
  \BibitemOpen
  \bibfield  {author} {\bibinfo {author} {\bibfnamefont {N.~V.}\ \bibnamefont
  {Prokof'ev}}\ and\ \bibinfo {author} {\bibfnamefont {B.~V.}\ \bibnamefont
  {Svistunov}},\ }\href@noop {} {\bibfield  {journal} {\bibinfo  {journal}
  {Phys. Rev. Lett.}\ }\textbf {\bibinfo {volume} {81}},\ \bibinfo {pages}
  {2514} (\bibinfo {year} {1998})}\BibitemShut {NoStop}%
\bibitem [{\citenamefont {Mishchenko}\ \emph {et~al.}(2000)\citenamefont
  {Mishchenko}, \citenamefont {Prokof'ev}, \citenamefont {Sakamoto},\ and\
  \citenamefont {Svistunov}}]{Mishchenko-PRB-2000-6317}%
  \BibitemOpen
  \bibfield  {author} {\bibinfo {author} {\bibfnamefont {A.~S.}\ \bibnamefont
  {Mishchenko}}, \bibinfo {author} {\bibfnamefont {N.~V.}\ \bibnamefont
  {Prokof'ev}}, \bibinfo {author} {\bibfnamefont {A.}~\bibnamefont {Sakamoto}},
  \ and\ \bibinfo {author} {\bibfnamefont {B.~V.}\ \bibnamefont {Svistunov}},\
  }\href@noop {} {\bibfield  {journal} {\bibinfo  {journal} {Phys. Rev. B}\
  }\textbf {\bibinfo {volume} {62}},\ \bibinfo {pages} {6317} (\bibinfo {year}
  {2000})}\BibitemShut {NoStop}%
\bibitem [{\citenamefont {Mishchenko}\ \emph {et~al.}(2003)\citenamefont
  {Mishchenko}, \citenamefont {Nagaosa}, \citenamefont {Prokof'ev},
  \citenamefont {Sakamoto},\ and\ \citenamefont
  {Svistunov}}]{Mishchenko-PRL-2003-236401}%
  \BibitemOpen
  \bibfield  {author} {\bibinfo {author} {\bibfnamefont {A.~S.}\ \bibnamefont
  {Mishchenko}}, \bibinfo {author} {\bibfnamefont {N.}~\bibnamefont {Nagaosa}},
  \bibinfo {author} {\bibfnamefont {N.~V.}\ \bibnamefont {Prokof'ev}}, \bibinfo
  {author} {\bibfnamefont {A.}~\bibnamefont {Sakamoto}}, \ and\ \bibinfo
  {author} {\bibfnamefont {B.~V.}\ \bibnamefont {Svistunov}},\ }\href@noop {}
  {\bibfield  {journal} {\bibinfo  {journal} {Phys. Rev. Lett.}\ }\textbf
  {\bibinfo {volume} {91}},\ \bibinfo {pages} {236401} (\bibinfo {year}
  {2003})}\BibitemShut {NoStop}%
\bibitem [{\citenamefont {De~Filippis}\ \emph {et~al.}(2006)\citenamefont
  {De~Filippis}, \citenamefont {Cataudella}, \citenamefont {Mishchenko},
  \citenamefont {Perroni},\ and\ \citenamefont
  {Devreese}}]{DeFilippis-PRL-2006-136405}%
  \BibitemOpen
  \bibfield  {author} {\bibinfo {author} {\bibfnamefont {G.}~\bibnamefont
  {De~Filippis}}, \bibinfo {author} {\bibfnamefont {V.}~\bibnamefont
  {Cataudella}}, \bibinfo {author} {\bibfnamefont {A.~S.}\ \bibnamefont
  {Mishchenko}}, \bibinfo {author} {\bibfnamefont {C.~A.}\ \bibnamefont
  {Perroni}}, \ and\ \bibinfo {author} {\bibfnamefont {J.~T.}\ \bibnamefont
  {Devreese}},\ }\href@noop {} {\bibfield  {journal} {\bibinfo  {journal}
  {Phys. Rev. Lett.}\ }\textbf {\bibinfo {volume} {96}},\ \bibinfo {pages}
  {136405} (\bibinfo {year} {2006})}\BibitemShut {NoStop}%
\bibitem [{\citenamefont {Marchand}\ \emph {et~al.}(2010)\citenamefont
  {Marchand}, \citenamefont {De~Filippis}, \citenamefont {Cataudella},
  \citenamefont {Berciu}, \citenamefont {Nagaosa}, \citenamefont {Prokof ’ev},
  \citenamefont {Mishchenko},\ and\ \citenamefont
  {Stamp}}]{Marchand-PRL-2010-266605}%
  \BibitemOpen
  \bibfield  {author} {\bibinfo {author} {\bibfnamefont {D.~J.~J.}\
  \bibnamefont {Marchand}}, \bibinfo {author} {\bibfnamefont {G.}~\bibnamefont
  {De~Filippis}}, \bibinfo {author} {\bibfnamefont {V.}~\bibnamefont
  {Cataudella}}, \bibinfo {author} {\bibfnamefont {M.}~\bibnamefont {Berciu}},
  \bibinfo {author} {\bibfnamefont {N.}~\bibnamefont {Nagaosa}}, \bibinfo
  {author} {\bibfnamefont {N.~V.}\ \bibnamefont {Prokof'v}}, \bibinfo
  {author} {\bibfnamefont {A.~S.}\ \bibnamefont {Mishchenko}}, \ and\ \bibinfo
  {author} {\bibfnamefont {P.~C.~E.}\ \bibnamefont {Stamp}},\ }\href@noop {}
  {\bibfield  {journal} {\bibinfo  {journal} {Phys. Rev. Lett.}\ }\textbf
  {\bibinfo {volume} {105}},\ \bibinfo {pages} {266605} (\bibinfo {year}
  {2010})}\BibitemShut {NoStop}%
\bibitem [{\citenamefont {Goodvin}\ \emph {et~al.}(2011)\citenamefont
  {Goodvin}, \citenamefont {Mishchenko},\ and\ \citenamefont
  {Berciu}}]{Goodvin-PRL-2011-76403}%
  \BibitemOpen
  \bibfield  {author} {\bibinfo {author} {\bibfnamefont {G.~L.}\ \bibnamefont
  {Goodvin}}, \bibinfo {author} {\bibfnamefont {A.~S.}\ \bibnamefont
  {Mishchenko}}, \ and\ \bibinfo {author} {\bibfnamefont {M.}~\bibnamefont
  {Berciu}},\ }\href@noop {} {\bibfield  {journal} {\bibinfo  {journal} {Phys.
  Rev. Lett.}\ }\textbf {\bibinfo {volume} {107}},\ \bibinfo {pages} {076403}
  (\bibinfo {year} {2011})}\BibitemShut {NoStop}%
\bibitem [{\citenamefont {Mishchenko}\ \emph {et~al.}(2015)\citenamefont
  {Mishchenko}, \citenamefont {Nagaosa}, \citenamefont {De~Filippis},
  \citenamefont {de~Candia},\ and\ \citenamefont
  {Cataudella}}]{Mishchenko-PRL-2015-146401}%
  \BibitemOpen
  \bibfield  {author} {\bibinfo {author} {\bibfnamefont {A.~S.}\ \bibnamefont
  {Mishchenko}}, \bibinfo {author} {\bibfnamefont {N.}~\bibnamefont {Nagaosa}},
  \bibinfo {author} {\bibfnamefont {G.}~\bibnamefont {De~Filippis}}, \bibinfo
  {author} {\bibfnamefont {A.}~\bibnamefont {de~Candia}}, \ and\ \bibinfo
  {author} {\bibfnamefont {V.}~\bibnamefont {Cataudella}},\ }\href@noop {}
  {\bibfield  {journal} {\bibinfo  {journal} {Phys. Rev. Lett.}\ }\textbf
  {\bibinfo {volume} {114}},\ \bibinfo {pages} {146401} (\bibinfo {year}
  {2015})}\BibitemShut {NoStop}%
\bibitem [{\citenamefont {De~Filippis}\ \emph {et~al.}(2015)\citenamefont
  {De~Filippis}, \citenamefont {Cataudella}, \citenamefont {Mishchenko},
  \citenamefont {Nagaosa}, \citenamefont {Fierro},\ and\ \citenamefont
  {de~Candia}}]{DeFilippis-PRL-2015-86601}%
  \BibitemOpen
  \bibfield  {author} {\bibinfo {author} {\bibfnamefont {G.}~\bibnamefont
  {De~Filippis}}, \bibinfo {author} {\bibfnamefont {V.}~\bibnamefont
  {Cataudella}}, \bibinfo {author} {\bibfnamefont {A.~S.}\ \bibnamefont
  {Mishchenko}}, \bibinfo {author} {\bibfnamefont {N.}~\bibnamefont {Nagaosa}},
  \bibinfo {author} {\bibfnamefont {A.}~\bibnamefont {Fierro}}, \ and\ \bibinfo
  {author} {\bibfnamefont {A.}~\bibnamefont {de~Candia}},\ }\href@noop {}
  {\bibfield  {journal} {\bibinfo  {journal} {Phys. Rev. Lett.}\ }\textbf
  {\bibinfo {volume} {114}},\ \bibinfo {pages} {086601} (\bibinfo {year}
  {2015})}\BibitemShut {NoStop}%
\bibitem [{\citenamefont {Mishchenko}\ \emph {et~al.}(2018)\citenamefont
  {Mishchenko}, \citenamefont {De~Filippis}, \citenamefont {Cataudella},
  \citenamefont {Nagaosa},\ and\ \citenamefont
  {Fehske}}]{Mishchenko-PRB-2018-45141}%
  \BibitemOpen
  \bibfield  {author} {\bibinfo {author} {\bibfnamefont {A.~S.}\ \bibnamefont
  {Mishchenko}}, \bibinfo {author} {\bibfnamefont {G.}~\bibnamefont
  {De~Filippis}}, \bibinfo {author} {\bibfnamefont {V.}~\bibnamefont
  {Cataudella}}, \bibinfo {author} {\bibfnamefont {N.}~\bibnamefont {Nagaosa}},
  \ and\ \bibinfo {author} {\bibfnamefont {H.}~\bibnamefont {Fehske}},\
  }\href@noop {} {\bibfield  {journal} {\bibinfo  {journal} {Phys. Rev. B}\
  }\textbf {\bibinfo {volume} {97}},\ \bibinfo {pages} {045141} (\bibinfo
  {year} {2018})}\BibitemShut {NoStop}%
\bibitem [{\citenamefont {Mishchenko}\ \emph {et~al.}(2019)\citenamefont
  {Mishchenko}, \citenamefont {Pollet}, \citenamefont {Prokof'ev},
  \citenamefont {Kumar}, \citenamefont {Maslov},\ and\ \citenamefont
  {Nagaosa}}]{Mishchenko-PRL-2019-76601}%
  \BibitemOpen
  \bibfield  {author} {\bibinfo {author} {\bibfnamefont {A.~S.}\ \bibnamefont
  {Mishchenko}}, \bibinfo {author} {\bibfnamefont {L.}~\bibnamefont {Pollet}},
  \bibinfo {author} {\bibfnamefont {N.~V.}\ \bibnamefont {Prokof'ev}}, \bibinfo
  {author} {\bibfnamefont {A.}~\bibnamefont {Kumar}}, \bibinfo {author}
  {\bibfnamefont {D.~L.}\ \bibnamefont {Maslov}}, \ and\ \bibinfo {author}
  {\bibfnamefont {N.}~\bibnamefont {Nagaosa}},\ }\href@noop {} {\bibfield
  {journal} {\bibinfo  {journal} {Phys. Rev. Lett.}\ }\textbf {\bibinfo
  {volume} {123}},\ \bibinfo {pages} {076601} (\bibinfo {year}
  {2019})}\BibitemShut {NoStop}%
\bibitem [{\citenamefont {Gull}\ \emph {et~al.}(2011)\citenamefont {Gull},
  \citenamefont {Millis}, \citenamefont {Lichtenstein}, \citenamefont
  {Rubtsov}, \citenamefont {Troyer},\ and\ \citenamefont
  {Werner}}]{Gull-RMP-2011-349}%
  \BibitemOpen
  \bibfield  {author} {\bibinfo {author} {\bibfnamefont {E.}~\bibnamefont
  {Gull}}, \bibinfo {author} {\bibfnamefont {A.~J.}\ \bibnamefont {Millis}},
  \bibinfo {author} {\bibfnamefont {A.~I.}\ \bibnamefont {Lichtenstein}},
  \bibinfo {author} {\bibfnamefont {A.~N.}\ \bibnamefont {Rubtsov}}, \bibinfo
  {author} {\bibfnamefont {M.}~\bibnamefont {Troyer}}, \ and\ \bibinfo {author}
  {\bibfnamefont {P.}~\bibnamefont {Werner}},\ }\href@noop {} {\bibfield
  {journal} {\bibinfo  {journal} {Rev. Mod. Phys.}\ }\textbf {\bibinfo {volume}
  {83}},\ \bibinfo {pages} {349} (\bibinfo {year} {2011})}\BibitemShut
  {NoStop}%
\bibitem [{\citenamefont {Giustino}(2017)}]{Giustino-RMP-2017-105003}%
  \BibitemOpen
  \bibfield  {author} {\bibinfo {author} {\bibfnamefont {F.}~\bibnamefont
  {Giustino}},\ }\href@noop {} {\bibfield  {journal} {\bibinfo  {journal} {Rev.
  Mod. Phys.}\ }\textbf {\bibinfo {volume} {89}},\ \bibinfo {pages} {015003}
  (\bibinfo {year} {2017})}\BibitemShut {NoStop}%
\bibitem [{\citenamefont {Marzari}\ and\ \citenamefont
  {Vanderbilt}(1997)}]{Marzari-PRB-1997-12847}%
  \BibitemOpen
  \bibfield  {author} {\bibinfo {author} {\bibfnamefont {N.}~\bibnamefont
  {Marzari}}\ and\ \bibinfo {author} {\bibfnamefont {D.}~\bibnamefont
  {Vanderbilt}},\ }\href@noop {} {\bibfield  {journal} {\bibinfo  {journal}
  {Phys. Rev. B}\ }\textbf {\bibinfo {volume} {56}},\ \bibinfo {pages} {12847}
  (\bibinfo {year} {1997})}\BibitemShut {NoStop}%
\bibitem [{\citenamefont {Marzari}\ \emph {et~al.}(2012)\citenamefont
  {Marzari}, \citenamefont {Mostofi}, \citenamefont {Yates}, \citenamefont
  {Souza},\ and\ \citenamefont {Vanderbilt}}]{Marzari-RMP-2012-1419}%
  \BibitemOpen
  \bibfield  {author} {\bibinfo {author} {\bibfnamefont {N.}~\bibnamefont
  {Marzari}}, \bibinfo {author} {\bibfnamefont {A.~A.}\ \bibnamefont
  {Mostofi}}, \bibinfo {author} {\bibfnamefont {J.~R.}\ \bibnamefont {Yates}},
  \bibinfo {author} {\bibfnamefont {I.}~\bibnamefont {Souza}}, \ and\ \bibinfo
  {author} {\bibfnamefont {D.}~\bibnamefont {Vanderbilt}},\ }\href@noop {}
  {\bibfield  {journal} {\bibinfo  {journal} {Rev. Mod. Phys.}\ }\textbf
  {\bibinfo {volume} {84}},\ \bibinfo {pages} {1419} (\bibinfo {year}
  {2012})}\BibitemShut {NoStop}%
\bibitem [{\citenamefont {Stockburger}\ and\ \citenamefont
  {Grabert}(2002)}]{Stockburger-PRL-2002-170407}%
  \BibitemOpen
  \bibfield  {author} {\bibinfo {author} {\bibfnamefont {J.~T.}\ \bibnamefont
  {Stockburger}}\ and\ \bibinfo {author} {\bibfnamefont {H.}~\bibnamefont
  {Grabert}},\ }\href@noop {} {\bibfield  {journal} {\bibinfo  {journal} {Phys.
  Rev. Lett.}\ }\textbf {\bibinfo {volume} {88}},\ \bibinfo {pages} {170407}
  (\bibinfo {year} {2002})}\BibitemShut {NoStop}%
\bibitem [{\citenamefont {Moix}\ \emph {et~al.}(2012)\citenamefont {Moix},
  \citenamefont {Zhao},\ and\ \citenamefont {Cao}}]{Moix-PRB-2012-115412}%
  \BibitemOpen
  \bibfield  {author} {\bibinfo {author} {\bibfnamefont {J.~M.}\ \bibnamefont
  {Moix}}, \bibinfo {author} {\bibfnamefont {Y.}~\bibnamefont {Zhao}}, \ and\
  \bibinfo {author} {\bibfnamefont {J.}~\bibnamefont {Cao}},\ }\href@noop {}
  {\bibfield  {journal} {\bibinfo  {journal} {Phys. Rev. B}\ }\textbf {\bibinfo
  {volume} {85}},\ \bibinfo {pages} {115412} (\bibinfo {year}
  {2012})}\BibitemShut {NoStop}%
\bibitem [{\citenamefont {Mishchenko}()}]{Mishchenko--2012-}%
  \BibitemOpen
  \bibfield  {author} {\bibinfo {author} {\bibfnamefont {A.~S.}\ \bibnamefont
  {Mishchenko}},\ }\bibinfo {note} {\textit{Stochastic Optimization Method for
  Analytic Continuation} in Correlated Electrons: From Models to Materials
  Modeling and Simulation, edited by E. Pavarini, E. Koch, F. Anders, and M.
  Jarrell (Verlag des Forschungszentrum, Julich, 2012), Vol. 2.}\BibitemShut
  {Stop}%
\bibitem [{\citenamefont {Hannewald}\ and\ \citenamefont
  {Bobbert}(2004{\natexlab{b}})}]{Hannewald-PRB-2004-75212a}%
  \BibitemOpen
  \bibfield  {author} {\bibinfo {author} {\bibfnamefont {K.}~\bibnamefont
  {Hannewald}}\ and\ \bibinfo {author} {\bibfnamefont {P.~A.}\ \bibnamefont
  {Bobbert}},\ }\href@noop {} {\bibfield  {journal} {\bibinfo  {journal} {Phys.
  Rev. B}\ }\textbf {\bibinfo {volume} {69}},\ \bibinfo {pages} {075212}
  (\bibinfo {year} {2004}{\natexlab{b}})}\BibitemShut {NoStop}%
\bibitem [{\citenamefont {Baroni}\ \emph {et~al.}(2001)\citenamefont {Baroni},
  \citenamefont {de~Gironcoli}, \citenamefont {Dal~Corso},\ and\ \citenamefont
  {Giannozzi}}]{Baroni-RMP-2001-515}%
  \BibitemOpen
  \bibfield  {author} {\bibinfo {author} {\bibfnamefont {S.}~\bibnamefont
  {Baroni}}, \bibinfo {author} {\bibfnamefont {S.}~\bibnamefont
  {de~Gironcoli}}, \bibinfo {author} {\bibfnamefont {A.}~\bibnamefont
  {Dal~Corso}}, \ and\ \bibinfo {author} {\bibfnamefont {P.}~\bibnamefont
  {Giannozzi}},\ }\href@noop {} {\bibfield  {journal} {\bibinfo  {journal}
  {Rev. Mod. Phys.}\ }\textbf {\bibinfo {volume} {73}},\ \bibinfo {pages} {515}
  (\bibinfo {year} {2001})}\BibitemShut {NoStop}%
\bibitem [{\citenamefont {Giustino}\ \emph {et~al.}(2007)\citenamefont
  {Giustino}, \citenamefont {Cohen},\ and\ \citenamefont
  {Louie}}]{Giustino-PRB-2007-165108}%
  \BibitemOpen
  \bibfield  {author} {\bibinfo {author} {\bibfnamefont {F.}~\bibnamefont
  {Giustino}}, \bibinfo {author} {\bibfnamefont {M.~L.}\ \bibnamefont {Cohen}},
  \ and\ \bibinfo {author} {\bibfnamefont {S.~G.}\ \bibnamefont {Louie}},\
  }\href@noop {} {\bibfield  {journal} {\bibinfo  {journal} {Phys. Rev. B}\
  }\textbf {\bibinfo {volume} {76}},\ \bibinfo {pages} {165108} (\bibinfo
  {year} {2007})}\BibitemShut {NoStop}%
\bibitem [{\citenamefont {Yates}\ \emph {et~al.}(2007)\citenamefont {Yates},
  \citenamefont {Wang}, \citenamefont {Vanderbilt},\ and\ \citenamefont
  {Souza}}]{Yates-PRB-2007-195121}%
  \BibitemOpen
  \bibfield  {author} {\bibinfo {author} {\bibfnamefont {J.~R.}\ \bibnamefont
  {Yates}}, \bibinfo {author} {\bibfnamefont {X.}~\bibnamefont {Wang}},
  \bibinfo {author} {\bibfnamefont {D.}~\bibnamefont {Vanderbilt}}, \ and\
  \bibinfo {author} {\bibfnamefont {I.}~\bibnamefont {Souza}},\ }\href@noop {}
  {\bibfield  {journal} {\bibinfo  {journal} {Phys. Rev. B}\ }\textbf {\bibinfo
  {volume} {75}},\ \bibinfo {pages} {195121} (\bibinfo {year}
  {2007})}\BibitemShut {NoStop}%
\bibitem [{\citenamefont {Feng}\ \emph {et~al.}(2020)\citenamefont {Feng},
  \citenamefont {Wang}, \citenamefont {Ke}, \citenamefont {Liang},\ and\
  \citenamefont {Zhao}}]{Feng-JCP-2020-34116}%
  \BibitemOpen
  \bibfield  {author} {\bibinfo {author} {\bibfnamefont {S.}~\bibnamefont
  {Feng}}, \bibinfo {author} {\bibfnamefont {Y.-C.}\ \bibnamefont {Wang}},
  \bibinfo {author} {\bibfnamefont {Y.}~\bibnamefont {Ke}}, \bibinfo {author}
  {\bibfnamefont {W.~Z.}~\bibnamefont {Liang}}, \ and\ \bibinfo {author}
  {\bibfnamefont {Y.}~\bibnamefont {Zhao}},\ }\href@noop {} {\bibfield
  {journal} {\bibinfo  {journal} {J. Chem. Phys.}\ }\textbf {\bibinfo {volume}
  {153}},\ \bibinfo {pages} {034116} (\bibinfo {year} {2020})}\BibitemShut
  {NoStop}%
\bibitem [{\citenamefont {Feng}\ \emph {et~al.}(2021)\citenamefont {Feng},
  \citenamefont {Wang}, \citenamefont {Liang},\ and\ \citenamefont
  {Zhao}}]{Feng-JPCA-2021-2932}%
  \BibitemOpen
  \bibfield  {author} {\bibinfo {author} {\bibfnamefont {S.}~\bibnamefont
  {Feng}}, \bibinfo {author} {\bibfnamefont {Y.-C.}\ \bibnamefont {Wang}},
  \bibinfo {author} {\bibfnamefont {W.}~\bibnamefont {Liang}}, \ and\ \bibinfo
  {author} {\bibfnamefont {Y.}~\bibnamefont {Zhao}},\ }\href@noop {} {\bibfield
   {journal} {\bibinfo  {journal} {J. Phys. Chem. A}\ }\textbf {\bibinfo
  {volume} {125}},\ \bibinfo {pages} {2932} (\bibinfo {year}
  {2021})}\BibitemShut {NoStop}%
\bibitem [{\citenamefont {Barclay}\ \emph {et~al.}(2014)\citenamefont
  {Barclay}, \citenamefont {Constantopoulos},\ and\ \citenamefont
  {Matisons}}]{Barclay-CR-2014-10217}%
  \BibitemOpen
  \bibfield  {author} {\bibinfo {author} {\bibfnamefont {T.~G.}\ \bibnamefont
  {Barclay}}, \bibinfo {author} {\bibfnamefont {K.}~\bibnamefont
  {Constantopoulos}}, \ and\ \bibinfo {author} {\bibfnamefont {J.}~\bibnamefont
  {Matisons}},\ }\href@noop {} {\bibfield  {journal} {\bibinfo  {journal}
  {Chem. Rev.}\ }\textbf {\bibinfo {volume} {114}},\ \bibinfo {pages} {10217}
  (\bibinfo {year} {2014})}\BibitemShut {NoStop}%
\bibitem [{\citenamefont {Vlaming}\ \emph {et~al.}(2009)\citenamefont
  {Vlaming}, \citenamefont {Augulis}, \citenamefont {Stuart}, \citenamefont
  {Knoester},\ and\ \citenamefont {van Loosdrecht}}]{Vlaming-JPCB-2009-2273}%
  \BibitemOpen
  \bibfield  {author} {\bibinfo {author} {\bibfnamefont {S.~M.}\ \bibnamefont
  {Vlaming}}, \bibinfo {author} {\bibfnamefont {R.}~\bibnamefont {Augulis}},
  \bibinfo {author} {\bibfnamefont {M.~C.~A.}\ \bibnamefont {Stuart}}, \bibinfo
  {author} {\bibfnamefont {J.}~\bibnamefont {Knoester}}, \ and\ \bibinfo
  {author} {\bibfnamefont {P.~H.~M.}\ \bibnamefont {van Loosdrecht}},\
  }\href@noop {} {\bibfield  {journal} {\bibinfo  {journal} {J. Phys. Chem. B}\
  }\textbf {\bibinfo {volume} {113}},\ \bibinfo {pages} {2273} (\bibinfo {year}
  {2009})}\BibitemShut {NoStop}%
\bibitem [{\citenamefont {Eisele}\ \emph {et~al.}(2012)\citenamefont {Eisele},
  \citenamefont {Cone}, \citenamefont {Bloemsma}, \citenamefont {Vlaming},
  \citenamefont {van~der Kwaak}, \citenamefont {Silbey}, \citenamefont
  {Bawendi}, \citenamefont {Knoester}, \citenamefont {Rabe},\ and\
  \citenamefont {Vanden~Bout}}]{Eisele-NC-2012-655}%
  \BibitemOpen
  \bibfield  {author} {\bibinfo {author} {\bibfnamefont {D.~M.}\ \bibnamefont
  {Eisele}}, \bibinfo {author} {\bibfnamefont {C.~W.}\ \bibnamefont {Cone}},
  \bibinfo {author} {\bibfnamefont {E.~A.}\ \bibnamefont {Bloemsma}}, \bibinfo
  {author} {\bibfnamefont {S.~M.}\ \bibnamefont {Vlaming}}, \bibinfo {author}
  {\bibfnamefont {C.~G.~F.}\ \bibnamefont {van~der Kwaak}}, \bibinfo {author}
  {\bibfnamefont {R.~J.}\ \bibnamefont {Silbey}}, \bibinfo {author}
  {\bibfnamefont {M.~G.}\ \bibnamefont {Bawendi}}, \bibinfo {author}
  {\bibfnamefont {J.}~\bibnamefont {Knoester}}, \bibinfo {author}
  {\bibfnamefont {J.~P.}\ \bibnamefont {Rabe}}, \ and\ \bibinfo {author}
  {\bibfnamefont {D.~A.}\ \bibnamefont {Vanden~Bout}},\ }\href@noop {}
  {\bibfield  {journal} {\bibinfo  {journal} {Nat. Chem.}\ }\textbf {\bibinfo
  {volume} {4}},\ \bibinfo {pages} {655} (\bibinfo {year} {2012})}\BibitemShut
  {NoStop}%
\bibitem [{\citenamefont {Doria}\ \emph {et~al.}(2018)\citenamefont {Doria},
  \citenamefont {Sinclair}, \citenamefont {Klein}, \citenamefont {Bennett},
  \citenamefont {Chuang}, \citenamefont {Freyria}, \citenamefont {Steiner},
  \citenamefont {Foggi}, \citenamefont {Nelson}, \citenamefont {Cao},
  \citenamefont {Aspuru-Guzik}, \citenamefont {Lloyd}, \citenamefont {Caram},\
  and\ \citenamefont {Bawendi}}]{Doria-AN-2018-4556}%
  \BibitemOpen
  \bibfield  {author} {\bibinfo {author} {\bibfnamefont {S.}~\bibnamefont
  {Doria}}, \bibinfo {author} {\bibfnamefont {T.~S.}\ \bibnamefont {Sinclair}},
  \bibinfo {author} {\bibfnamefont {N.~D.}\ \bibnamefont {Klein}}, \bibinfo
  {author} {\bibfnamefont {D.~I.~G.}\ \bibnamefont {Bennett}}, \bibinfo
  {author} {\bibfnamefont {C.}~\bibnamefont {Chuang}}, \bibinfo {author}
  {\bibfnamefont {F.~S.}\ \bibnamefont {Freyria}}, \bibinfo {author}
  {\bibfnamefont {C.~P.}\ \bibnamefont {Steiner}}, \bibinfo {author}
  {\bibfnamefont {P.}~\bibnamefont {Foggi}}, \bibinfo {author} {\bibfnamefont
  {K.~A.}\ \bibnamefont {Nelson}}, \bibinfo {author} {\bibfnamefont
  {J.}~\bibnamefont {Cao}}, \bibinfo {author} {\bibfnamefont {A.}~\bibnamefont
  {Aspuru-Guzik}}, \bibinfo {author} {\bibfnamefont {S.}~\bibnamefont {Lloyd}},
  \bibinfo {author} {\bibfnamefont {J.~R.}\ \bibnamefont {Caram}}, \ and\
  \bibinfo {author} {\bibfnamefont {M.~G.}\ \bibnamefont {Bawendi}},\
  }\href@noop {} {\bibfield  {journal} {\bibinfo  {journal} {ACS Nano}\
  }\textbf {\bibinfo {volume} {12}},\ \bibinfo {pages} {4556} (\bibinfo {year}
  {2018})}\BibitemShut {NoStop}%
\bibitem [{\citenamefont {Mostofi}\ \emph {et~al.}(2008)\citenamefont
  {Mostofi}, \citenamefont {Yates}, \citenamefont {Lee}, \citenamefont {Souza},
  \citenamefont {Vanderbilt},\ and\ \citenamefont
  {Marzari}}]{Mostofi-CPC-2008-685}%
  \BibitemOpen
  \bibfield  {author} {\bibinfo {author} {\bibfnamefont {A.~A.}\ \bibnamefont
  {Mostofi}}, \bibinfo {author} {\bibfnamefont {J.~R.}\ \bibnamefont {Yates}},
  \bibinfo {author} {\bibfnamefont {Y.-S.}\ \bibnamefont {Lee}}, \bibinfo
  {author} {\bibfnamefont {I.}~\bibnamefont {Souza}}, \bibinfo {author}
  {\bibfnamefont {D.}~\bibnamefont {Vanderbilt}}, \ and\ \bibinfo {author}
  {\bibfnamefont {N.}~\bibnamefont {Marzari}},\ }\href@noop {} {\bibfield
  {journal} {\bibinfo  {journal} {Comput. Phys. Commun.}\ }\textbf {\bibinfo
  {volume} {178}},\ \bibinfo {pages} {685} (\bibinfo {year}
  {2008})}\BibitemShut {NoStop}%
\end{thebibliography}
%

\clearpage
\end{document}